\documentclass[a4paper,fleqn,usenatbib]{mnras}

\usepackage[normalem]{ulem}

\usepackage{newtxtext,newtxmath}
\usepackage{siunitx}
\usepackage[T1]{fontenc}
\usepackage{ae,aecompl}

\usepackage{graphicx}	
\usepackage{amsmath}	
\usepackage{amssymb}	
\usepackage{pdflscape}
\usepackage{hyperref}

\defcitealias{2019MNRAS.490.3806N}{Paper\,I}

\title[PATHOS\,II: candidates in OCs]{A PSF-based Approach to TESS High
  quality data Of Stellar clusters (PATHOS) - II. Search for exoplanets
  in open clusters of the southern ecliptic hemisphere and their frequency.}

\author[D.\ Nardiello et al.]{D.\ Nardiello$^{1,2}$\thanks{E-mail: domenico.nardiello@lam.fr}, 
  G.\ Piotto$^{2,3}$,
  M.\ Deleuil$^{1}$,
  L.\ Malavolta$^{3}$,
  M.\ Montalto$^{2,3}$,
  \newauthor
  L.\ R.\ Bedin$^{2}$,
  L.\ Borsato$^{2}$,
  V.\ Granata$^{2,3}$,
  M.\ Libralato$^{4}$,
  E.\ E.\ Manthopoulou$^{2,3}$\\
$^{1}$Aix Marseille Univ, CNRS, CNES, LAM, Marseille, France \\
$^{2}$Istituto Nazionale di Astrofisica - Osservatorio Astronomico di Padova, Vicolo dell'Osservatorio 5, IT-35122, Padova, Italy \\
$^{3}$Dipartimento di Fisica e Astronomia ``Galileo Galilei'', Universit\`a di Padova, Vicolo dell'Osservatorio 3, IT-35122, Padova, Italy  \\
$^{4}$Space Telescope Science Institute, 3800 San Martin Drive, Baltimore, MD 21218, USA \\
}

\date{Accepted 2020 May 20. Received 2020 May 20; in original form 2020 April 22.}

\pubyear{2020}
\begin{document}
\label{firstpage}
\pagerange{\pageref{firstpage}--\pageref{lastpage}}
\maketitle

\begin{abstract}
The scope of the project ``A PSF-based Approach to {\it TESS} High
Quality data Of Stellar clusters'' (PATHOS) is the extraction and
analysis of high-precision light curves of stars in stellar clusters
and young associations for the identification of candidate exoplanets
and variable stars. The cutting-edge tools used in this project allow
us to measure the real flux of stars in dense fields, minimising the
effects due to contamination by neighbour sources. We extracted about
200\,000 light curves of stars in 645 open clusters located in the
southern ecliptic hemisphere and observed by {\it TESS} during the
first year of its mission. We searched for transiting signals and we
found 33 objects of interest, 11 of them are strong candidate
exoplanets.  Because of the limited S/N, we did not find
any Earth or super-Earth. We identified two Neptune-size planets
orbiting stars with $R_{\star}<1.5\,R_{\sun}$, implying a frequency
$f_{\star}=1.34 \pm 0.95\,\%$, consistent with the frequency around
field stars. The 7 Jupiter candidates around stars with
$R_{\star}<\,1.5R_{\sun}$ imply a frequency $f_{\star}=0.19\pm
0.07\,\%$, smaller than in the field. A more complete estimate of the
survey completeness and false positive rate is needed to confirm these
results. Light curves used in this work will be made available to the
astronomical community on the Mikulski Archive for Space Telescope
under the project PATHOS.
\end{abstract}




\section{Introduction}

In our Galaxy there are more than 1500 among globular clusters, open
clusters and stellar associations. These objects can comprise up to
several hundreds thousand stars covering all ranges of spectral
classes and stellar evolutionary stages and have ages that span from
few million years to almost the age of the Universe. Their chemical
properties reflect the chemical evolution of the Milky Way. For these
stars we can extract with high accuracy essential pieces of
information, like the radius, the mass, the chemical content, and the
age, i.e. parameters that suffer usually of large uncertainties for
many field stars.  The possibility of measuring stellar host
parameters with high precision makes the search for exoplanets around
cluster stars, and their frequency, of particular interest, especially
if we want to comprehend how they formed and evolved in different
environments.

Despite the large number of stars in stellar clusters and
associations, only a small fraction of known (candidate) exoplanets
orbits these stars. These objects have been found by looking for
transits in the light curves of the cluster/association members and/or
analysing their radial velocities. To date, 28 between confirmed and
candidate exoplanets have been discovered in stellar clusters and
associations. The first exoplanet discovered in a stellar cluster, the
Hyades, was found by \citet{2007ApJ...661..527S}; since then, other
three exoplanets and an exoplanetary system have been found in this
open cluster, both by using radial velocities
(\citealt{2014ApJ...787...27Q}) and {\it Kepler/K2}
(\citealt{2014PASP..126..398H}) light curves
(\citealt{2016AJ....151..112D,2016ApJ...818...46M,2018AJ....155...10C,2018AJ....155....4M,2018AJ....156...46V})
The first two transiting exoplanets in an open cluster, NGC\,6811,
were discovered by \citet{2013Natur.499...55M} using {\it Kepler} main
mission data (\citealt{2010Sci...327..977B}).The Praesepe cluster
hosts 8 candidate/confirmed exoplanets and an exoplanetary system
(\citealt{2012ApJ...756L..33Q,2016A&A...594A.100B,2016MNRAS.463.1780L,
  2016A&A...588A.118M,2016AJ....152..223O,2016MNRAS.461.3399P,2017AJ....153...64M,2017AJ....153..177P,2020arXiv200312940G}).
A transiting exoplanet has been discovered by
\citet{2018AJ....155..173C} in the 3 Gyr old open cluster
Ruprecht\,147.  No transiting exoplanets around members of the old
open cluster M\,67 have been reported by \citet{2016MNRAS.463.1831N},
despite
\citet{2014A&A...561L...9B,2016A&A...592L...1B,2017A&A...603A..85B}
found an excess of hot Jupiters by using radial velocity measurements.
Noteworthy, the unusual null detection of exoplanets around Pleiades
members has been reported by \citet{2017MNRAS.464..850G} and by
\citet{2017AJ....153...64M}.  Some exoplanets have also been found in
young associations and moving groups. The first one is K2-33b,
orbiting a pre-main sequence star that belongs to the Upper Scorpius
OB association
(\citealt{2016Natur.534..658D,2016AJ....152...61M}). Exoplanets were
also found in the associations Cas-Tau (EPIC\,247267267\,b,
\citealt{2018AJ....156..302D}) and Tuc-Hor (DS\,Tuc\,A\,b,
\citealt{2019ApJ...880L..17N,2019A&A...630A..81B}).

\begin{figure*}
\includegraphics[bb=15 19 565 257, width=0.85\textwidth]{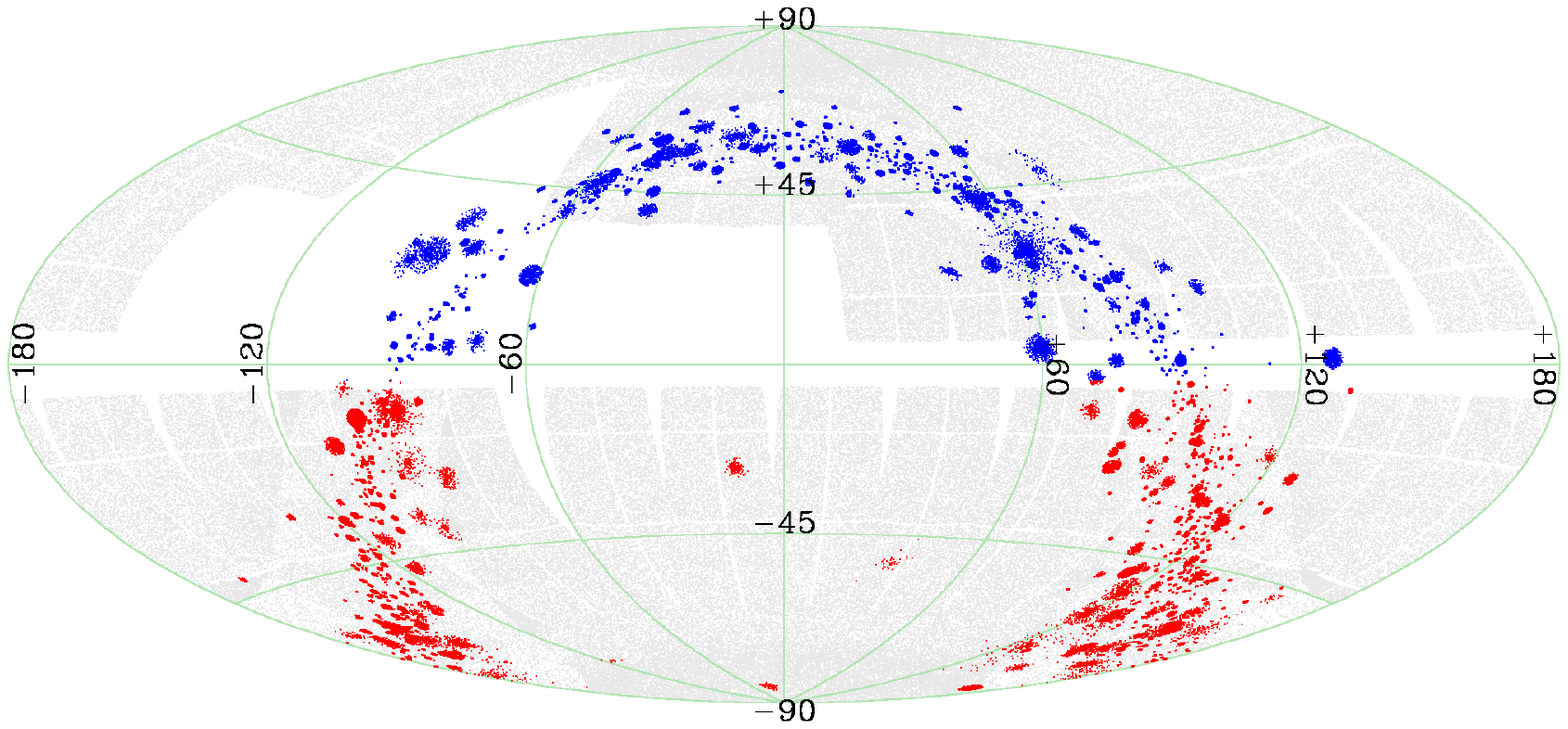}
\caption{Aitoff projection in ecliptic coordinates of the {\it TESS}
  observations: grey points represent the sources observed in 2-min
  cadence mode in Sectors 1-24; red/blue points are the open cluster
  members for which we propose to extract the light curves in our
  project; red points are the stars in the input catalogue used in this
  work. \label{fig:1}}
\end{figure*}

Among the previously mentioned exoplanets in stellar clusters and
associations, 18 are transiting exoplanets and, excluding
DS\,Tuc\,A\,b, they were all discovered by using data collected with
{\it Kepler}. But only a handful of clusters and associations were
observed during the {\it Kepler} main and {\it K2} missions, and the
number of studied cluster members is limited to few thousand stars.
The opportunity that the {\it Transiting Exoplanet Survey Satellite}
({\it TESS}, \citealt{2015JATIS...1a4003R}) is giving us is without equal:
at the end of its 2-years nominal mission, this satellite will have
observed more than 80\,\% of the sky, thus including almost all the
known clusters and associations.

In {\it TESS} observations, the large part of the cluster members fall
in 30-minute cadence stacked frames, called Full Frame Images (FFIs),
collected during 27-days campaigns; each campaign covers a sector of
the sky of $24 \times 96$\,deg$^2$. One of the major problems to be
addressed in the extraction of high precision light curves of cluster
members from {\it TESS} FFIs is related to their low angular
resolution ($\sim 21$\,arcsec\,pix$^{-1}$).  In fact, even low density
stellar clusters appear crowded in {\it TESS} images, where their
members are likely blended among each others or with field stars.  In
these conditions, the extraction of light curves using simple aperture
photometry is not reliable, because contamination and blending effects
prevail.

In the last years, many approaches have been developed for the
extraction of high precision light curves of stars in crowded regions,
based on the use of the Point Spread Functions (PSFs,
\citealt{2015MNRAS.447.3536N,2016MNRAS.456.1137L}) or the Difference
Imaging Analysis
(\citealt{1998ApJ...503..325A,2015AJ....149..135C,2017PASP..129d4501S,2019ApJS..245...13B,2019ApJS..244...12W,2020ApJS..246...15S}).
In the first paper of this series we presented our project ``A
PSF-based Approach to {\it TESS} High Quality data Of Stellar
clusters'' (PATHOS, \citealt[hereafter
  Paper\,I]{2019MNRAS.490.3806N}), aimed at the extraction of high
precision light curves for stars in stellar clusters (open and
globular clusters) and young associations. Our approach of light curve
extraction, already widely tested on ground-based, {\it Kepler} and
also {\it TESS} data, is based on the use of empirical PSFs and an
input catalogue to measure the flux of a given star after having
suppressed the contribution of its neighbours. In this way we are able
to measure the real flux of the target star, minimise the dilution
effects due to the contamination by neighbour sources (essential in
the search for exoplanetary transits), and extract high precision
photometry even for stars close to the {\it TESS} limiting magnitude
($T\sim17$).

In \citetalias{2019MNRAS.490.3806N} we successfully applied our method
to an extreme case of crowded region, a field containing the globular
cluster 47\,Tuc, which includes Galactic and Small Magellanic Cloud
stars. We searched for exoplanets and variable stars among all the
extracted light curves, finding one candidate exoplanet (a hot Jupiter
around a field star, PATHOS-1) and many new variable stars along the
red and asymptotic giant branch sequences of 47\,Tuc. For this work,
we applied our PSF-based approach to extract high-precision light
curves for stars belonging to open clusters in the southern ecliptic
hemisphere and listed in the catalogue published by
\citet[Sect.~\ref{sec:dr}]{2018A&A...618A..93C}. We analysed these
light curves searching for transit signals, we selected the most
promising transiting objects and validated them with a series of
vetting tests (Sect.~\ref{sec:search}).  We modelled the transits of
the candidate exoplanets to derive their parameters
(Sect.~\ref{sec:modelling}), and analysed their dependence on cluster
properties (Sects.~\ref{sec:modelling} and \ref{sec:results}). We also
estimated the observed and expected occurrence rate of exoplanets and
discussed our results in Section~\ref{sec:conclusions}.

\section{Observations and data reduction}
\label{sec:dr}

This work is focused on the open clusters located in the southern ecliptic
hemisphere and that were observed by {\it TESS} in Sectors 1-13.  In
this work we used observations carried out by {\it TESS} in a period
of $\sim 1$ year (357 days), between 2018 July 25 and 2019 July 17;
the total number of FFIs used for the extraction
of the light curves is 245\,576, that are all the downloadable FFIs
for the Sectors
1-13\footnote{\url{https://archive.stsci.edu/tess/bulk_downloads}-\url{/bulk_downloads_ffi-tp-lc-dv.html}}.

The light curve extraction has been performed using an improved
version of \texttt{img2lc}, the software described in
\citetalias{2019MNRAS.490.3806N}. The three main ingredients of our
approach are: (i) Full Frame Images, (ii) PSFs, and (iii) input catalogue. In the following, we give a short
description of the input catalogue adopted and of our pipeline.

\subsection{The input catalogue}
\label{sec:input_cat}

Our pipeline extracts the light curves of all the sources which
coordinates are listed in an input catalogue. For this work we used as
input the catalogue published by \citet{2018A&A...618A..93C}.  This
catalogue contains, for 1229 clusters, a list of stellar cluster
members, whose membership probabilities are calculated by using Gaia
DR2 proper motions and parallaxes
(\citealt{2018A&A...616A...1G}). From this catalogue we selected a
sub-sample of stars having magnitude $G<17.5$ and (a conservative)
ecliptic latitude $\beta < -4 \degr$, that corresponds to the fields
covered by {\it TESS} in the Sectors 1-13. In this way we selected a
total of 189\,090 cluster members, although $\sim 13.9\,\%$ of them
are not observed by {\it TESS}, because they fall between CCD/Camera
gaps or between sector gaps, or because they are outside the field of view
of {\it TESS} observations, as shown in Figure~\ref{fig:1}. On the
other hand, $\sim 1/4$ of the stars in the input catalogue are
observed in more than one sector. Finally, we extracted a total of 219\,256
light curves of stars in 645 open clusters.  In the following analysis
we excluded the members of NGC\,1901, an open cluster in the {\it
  TESS} continuous viewing zone; this cluster is the only one observed
in all the 13 Sectors, and will be the subject of a future work
(Manthopoulou et al., in preparation).

\begin{figure*}
\includegraphics[bb=1 47 1366 495, width=0.99\textwidth]{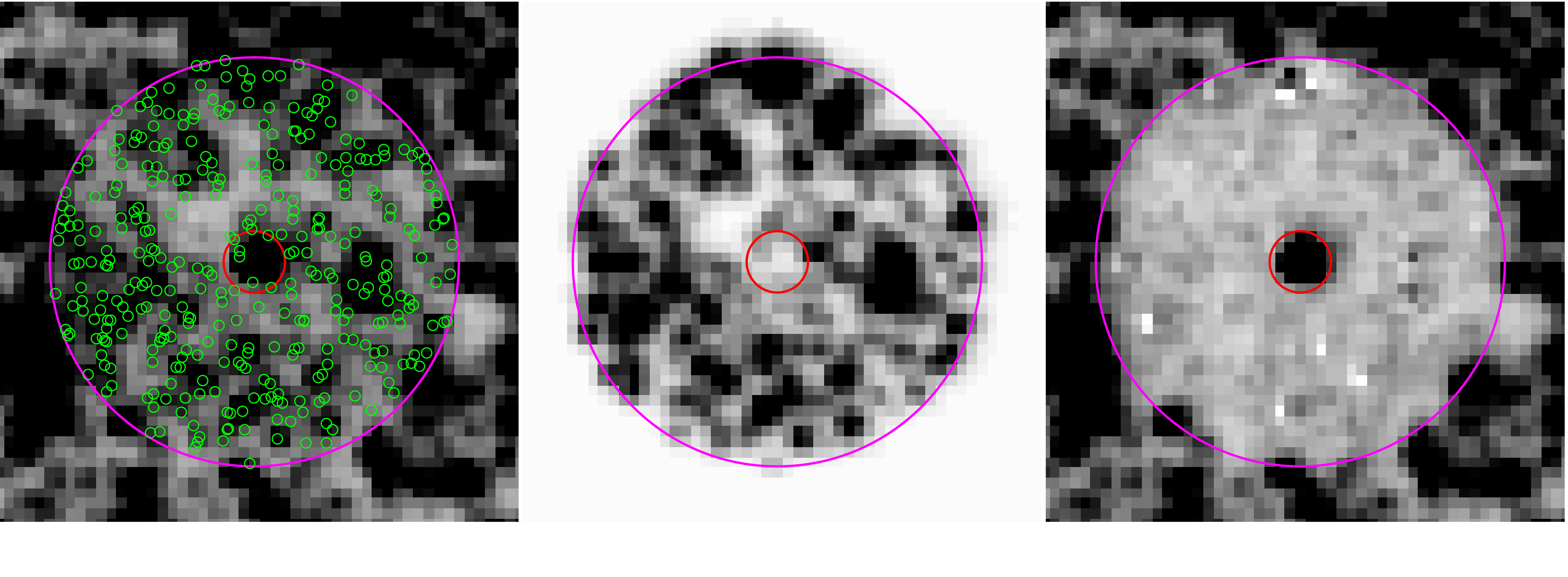}
\includegraphics[bb=0 262 554 551, width=0.99\textwidth]{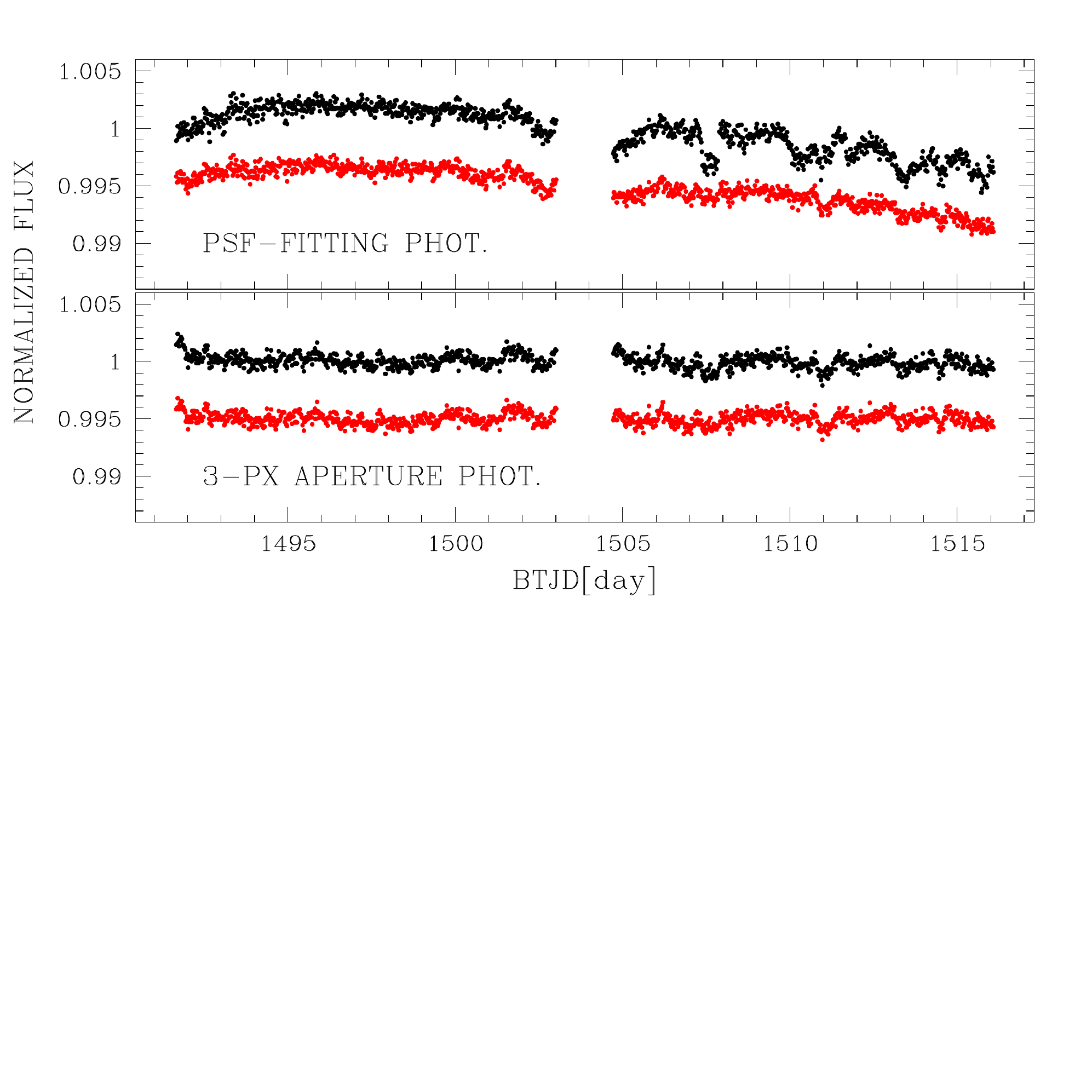}
\caption{From FFI to the final light curve. {\it Top Panels.} Example
  of neighbour subtraction. Left-hand panel shows the original FFI
  (\texttt{tess2019009222936-s0007-1-1-0131-s\_ffic.fits}) centred on
  the star Gaia DR2\,3111381928222301184, a possible cluster member of
  Gulliver\,47; red circle is the photometric aperture used to analyse
  this star, green circles are the neighbours located within a radius
  of 20 pixels (magenta circle) from the target star. Middle panel
  shows the models of the neighbours subtracted from the FFI;
  right-hand panel is the FFI after neighbour subtraction. {\it Bottom
    Panels.} The PSF-fitting and 3-pixel aperture photometry light
  curves of the star Gaia DR2\,3111381928222301184 before (black
  points) and after (red points) the systematic
  correction. \label{fig:2}}
\end{figure*}

\begin{figure*}
\includegraphics[width=0.27\textwidth]{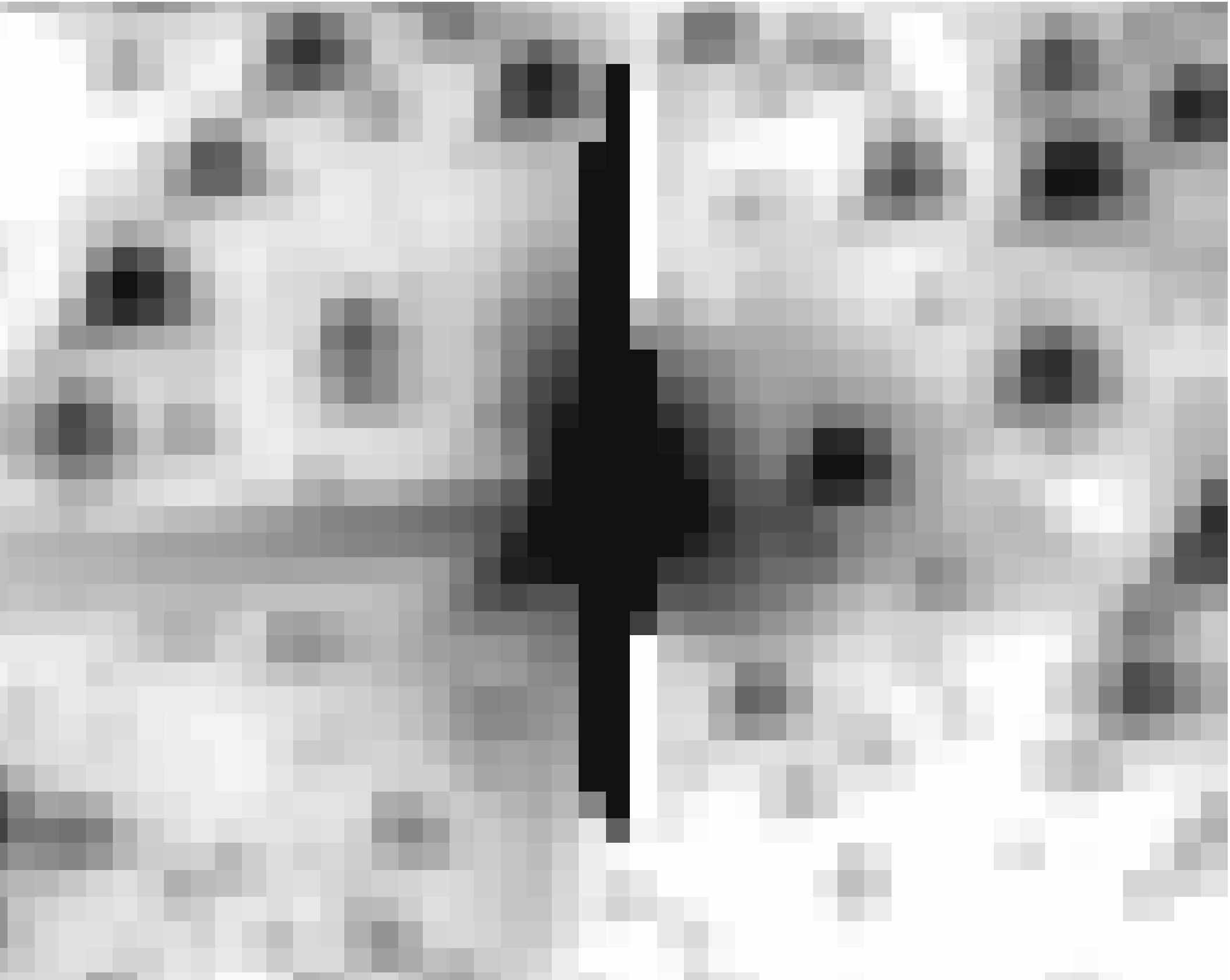}
\includegraphics[bb=3 363 566 556, width=0.65\textwidth]{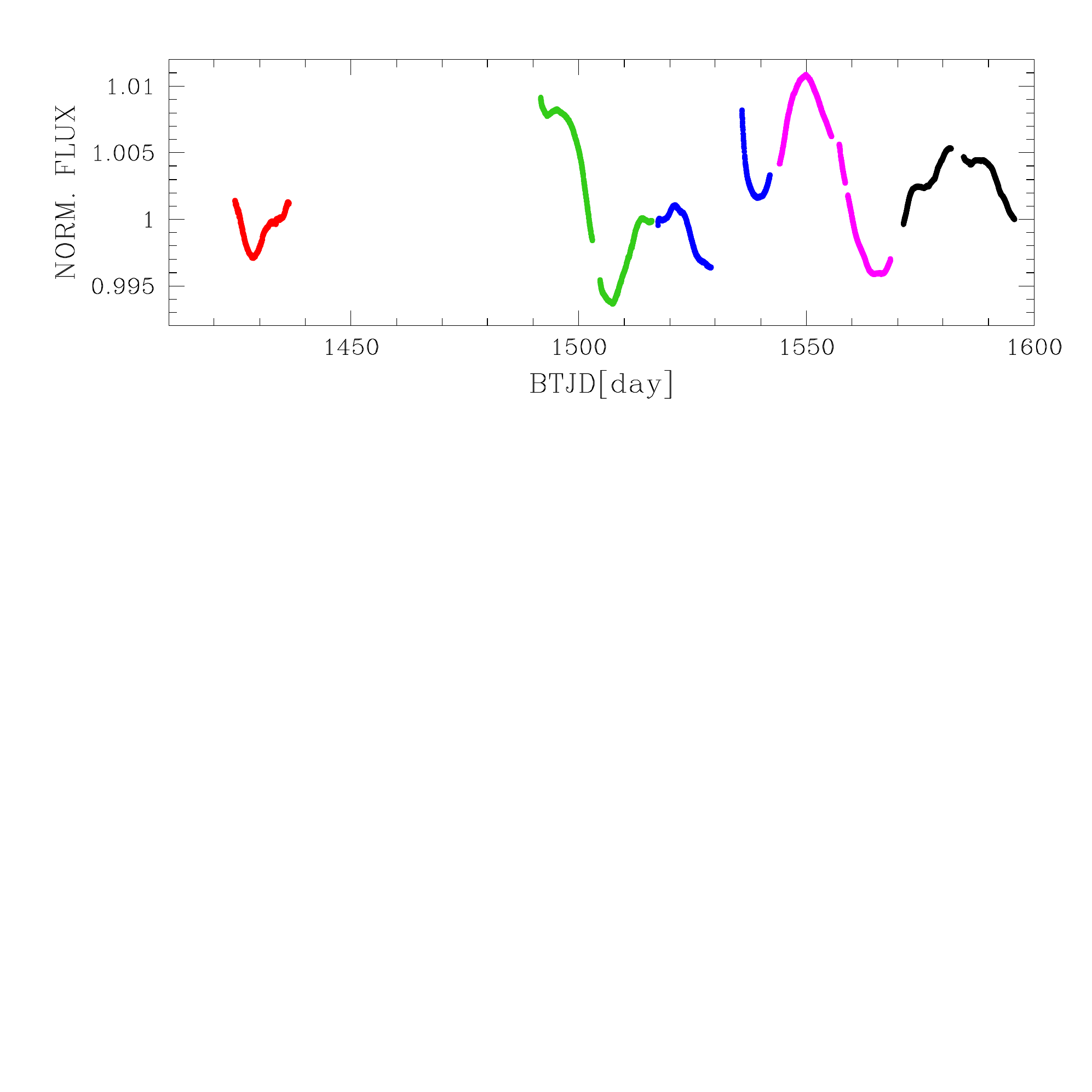}
\caption{Left-hand panel shows the saturated star Gaia
  DR2\,05290818758516639104 on the FFI
  \texttt{tess2019010175936-s0007-4-1-0131-s\_ffic.fits}. Right-hand
  panel is its light curve extracted from Sectors 4 (red), 7 (green),
  8 (blue), 9 (magenta), and 10 (black). \label{fig:3}}
\end{figure*}

\subsection{Light curve extraction}

For the extraction of the light curves from {\it TESS} FFIs, we
adopted the PSF-based approach developed by
\citet{2015MNRAS.447.3536N,2016MNRAS.455.2337N} for the ground-based
data collected with the Asiago Schmidt Telescope 67/92 cm. This
pipeline was also adapted to {\it Kepler/K2} space-based data-set by
\citet[see also \citealt{2016MNRAS.463.1780L,2016MNRAS.463.1831N}]{
  2016MNRAS.456.1137L} and finally to {\it TESS} data in
\citetalias{2019MNRAS.490.3806N}.

As a first step, the routine transforms the positions and the
luminosities of the stars in the input catalogue into the reference
system of a single image.  For all the stars in the input catalogue,
it transforms the ($\alpha,\delta$)-coordinates into the image
reference system using the transformation coefficients listed in the
FITS header of the single FFI.

In a second step, the software calculates the photometric zero-point
between the calibrated and instrumental {\it TESS} magnitudes as
follows: for the 200 brightest, not saturated stars of each FFI, it
extracts the PSF-fitting magnitudes $T_{\rm inst}$, and calculates the
3.5-$\sigma$-clipped mean value of $T^i_{\rm cal} - T^i_{\rm inst}$,
with $i=1,...,200$ and $T_{\rm cal}$ the calibrated {\it TESS}
magnitudes calculated by using Gaia DR2 magnitudes and the equation in
\citet{2019AJ....158..138S}.

Finally, for each star in the input catalogue, the routine searches in
the Gaia DR2 catalogue (\citealt{2018A&A...616A...1G}) all the
neighbours located between 0.5 and 20 {\it TESS} pixels from the
target star and having $T<17.5$, and transforms their positions and
luminosities in the FFI reference system, as described above. The
software models the neighbour sources by using the local PSF and the
transformed positions and luminosities, and it subtracts these models
to the original FFI, as shown in Fig.~\ref{fig:2}. After the
neighbour-subtraction, our software performs PSF-fitting and four
different aperture photometries (1-, 2-, 3-, and 4-pixels radius) of the
target star.

\subsection{Systematic effects correction}

Variations of the spacecraft, detector and environment conditions
affect the quality of the light curves, introducing systematic
artifacts. These systematic trends are common to all the stars of a
given {\it TESS} Camera/CCD/Sector, and can be corrected using
orthonormal functions called cotrending basis vectors (CBVs).

To extract the CBVs, we combined two samples of light curves: (i) the
light curves of the stars in the input catalogue described in
Sect.~\ref{sec:input_cat} and (ii) the light curves of the stars
also observed in 2-minute cadence mode (grey points of
Fig.~\ref{fig:1}) and extracted from FFIs with our pipeline. We
extracted the second sample of light curves to increase the number of
bright stars in our list and extract better CBVs associated to 3- and
4-pixel aperture photometries.

To extract the CBVs, we followed a procedure similar to that described
in \citetalias{2019MNRAS.490.3806N}: for each star and for each
photometric method, we calculated the raw
\texttt{RMS}\footnote{Defined as the 68.27th-percentile of the sorted
  residual from the median value of the light curve}. On the basis of
the \texttt{RMS} distributions, we identified the magnitude interval
where each photometric method works better and we selected the best
measured stars on the basis of their \texttt{RMS} values as follows:
for each photometric method, we divided the \texttt{RMS}
  distributions in bins of width 0.75 $T$ magnitude and we estimated
  in each magnitude interval the 3.5$\sigma$-clipped mean value of
  the \texttt{RMS}. We interpolated the binned points with a spline:
  for each photometric method, we selected the magnitude interval
  where the mean \texttt{RMS} distribution is lower compared to the
  other distributions.  For the stars that passed the first selection
criteria, we calculated the median of the correlation coefficients
between a given light curve and the light curves of the other
stars. We kept all the stars with a median correlation coefficient
$>50\,\%$, and we iterated other 2 times, restarting from the
calculation of the median correlation coefficient, and considering at
each iteration only the stars that passed the selection criteria of
the previous iteration. With the surviving stars, we extracted 20 CBVs
for each photometric method using the single value decomposition. On
average, CBVs are obtained by using between 200 and 300 stars.

To correct the light curves, we used the Levenberg-Marquardt method
(\citealt{MINPACK-1}) to find the coefficients $A_i$ that minimised
the expression:
\begin{equation}
F^j_{\rm raw}-\sum_i(A_i \cdot {\rm CBV}_i^j)
\end{equation}
with $F^j_{\rm raw}$ the raw flux of the light curve at the epoch $j$,
and ${\rm CBV}_i $ the $i$-th CBV, with $i=1,...,20$. We changed the
number of CBVs applied to obtain the correction from 1 to 20, and, for
each light curve and photometric method, we selected the number of
CBVs whose correction gives the lower Akaike's information criterion
(AIC) score.

Bottom panels of Fig.~\ref{fig:2} show the light curves of Gaia
DR2\,3111381928222301184 before (black points) and after (red points)
 correction, in the case of PSF-fitting photometry and 3-pixel
aperture photometry: the improvement of the light curve in the case of
PSF-fitting photometry is clear, and the \texttt{RMS} of the light curve passes from 2.1 mmag to 1.6 mmag.

\begin{figure*}
\includegraphics[bb=0 187 563 556, width=0.9\textwidth]{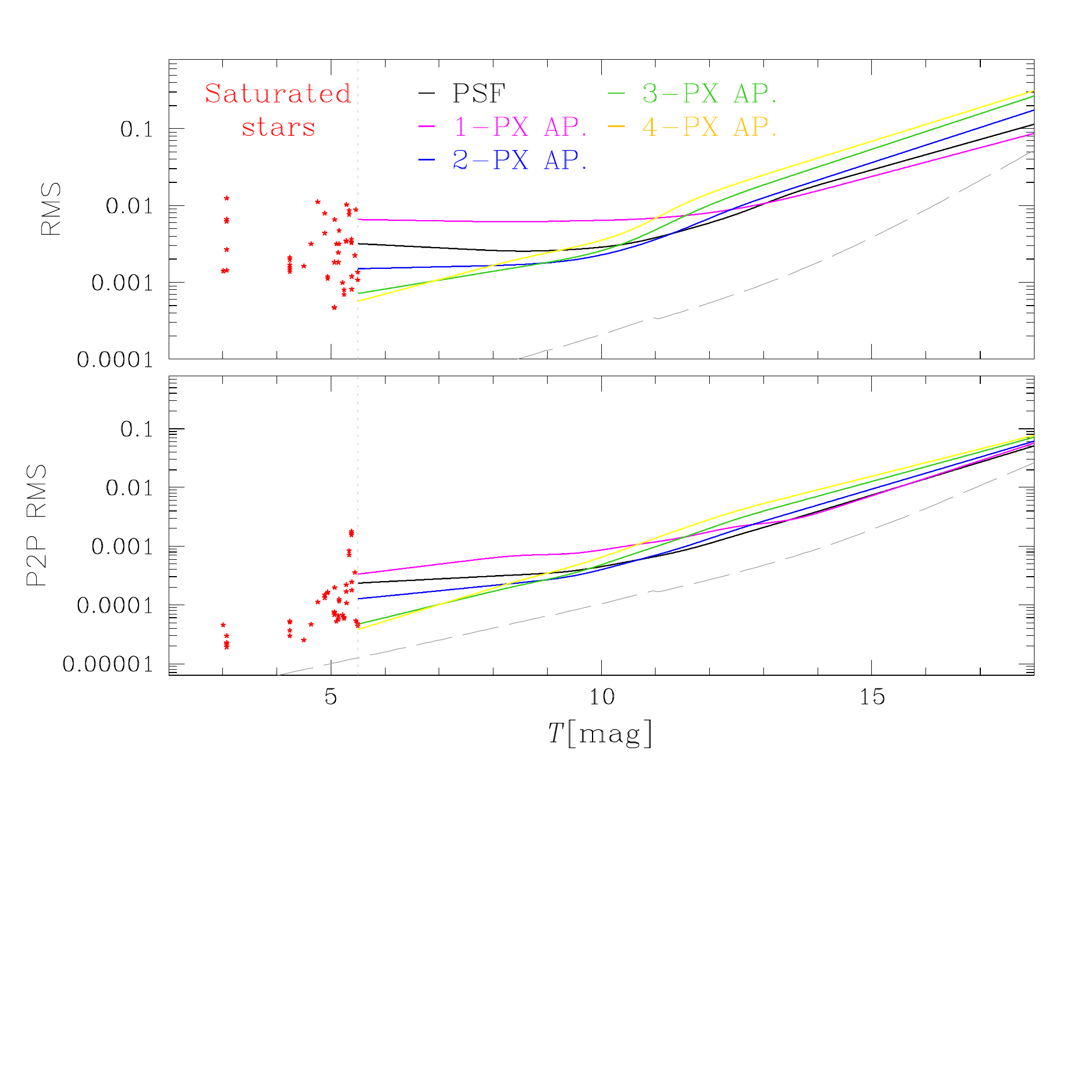}
\caption{Photometric \texttt{RMS} (top panel) and \texttt{P2P RMS}
  (bottom panel) mean trends as a function of the {\it TESS} magnitude
  for all the light curves extracted in this work, colour-coded on the
  basis of the photometric method adopted: black for PSF-fitting
  photometry, magenta, blue, green, and yellow for 1-pixel, 2-pixel,
  3-pixel, and 4-pixel aperture photometries, respectively. Red points
  represent the saturated stars. The grey dashed lines represent the
  theoretical limit.
  \label{fig:4}}
\end{figure*}

\subsection{Saturated stars}
\label{sec:sat}

Stars with $T\lesssim 5.5$ are saturated on FFIs, and form a bleeding
column. As explained in the {\it TESS} Instrument Handbook, the CCDs
conserve the electrons even when saturation causes their moving along
the columns neighbouring the pixel where they were generated. To
recover the charge along the bleeding column we used the technique
described in \citet{2004acs..rept...17G} and \citet{2010wfc..rept...10G}, and already tested in many works based on Hubble
Space Telescope data (see, e.g.,
\citealt{2008AJ....135.2055A,2017ApJ...842....6B,2018MNRAS.481.3382N}).
Briefly, our routine checks if the central pixel and three of the
neighbouring pixels have counts $>80000\,e^{-}/s$; in this case, it
checks the presence of a bleeding column and retraces its
shape. Finally, it adds the contribution that comes
from the bleeding columns to the flux calculated inside an aperture of
radius 5.5 pixels centred on the star.  In the case of saturated stars, the
routine did not perform neighbour subtraction and we did not
correct for systematic trends, because the sample of saturated (and
not variable) stars is too small to extract CBVs. Figure~\ref{fig:3}
shows the saturated star Gaia DR2\,05290818758516639104 ($T \sim 3.1$)
on a FFI (left panel) and its light curve extracted from 5 sectors
(right panel). The standard deviation from the mean value of the light
curve, after removing the contribution of the variability, is $\sim
26$\,ppm.

\subsection{Data release}
All the light curves extracted in this work are released on the
Mikulski Archive for Space Telescopes (MAST) as a High Level Science
Product (HLSP) under the project
PATHOS\footnote{\url{https://archive.stsci.edu/hlsp/pathos}} (DOI:
10.17909/t9-es7m-vw14). Each light curve (in \texttt{fits} and
\texttt{ascii} format) contains the epoch in TESS Barycentric Julian
Day (BTJD), the five extracted raw and corrected photometries
(PSF-fitting, 1-pixel, 2-pixel, 3-pixel, 4-pixel aperture), the value
of the local sky, the position (x, y) on the image, and the data
quality flag \texttt{DQUALITY} (see TESS Science Data Products
Description Document for details). All the main information on the
star extracted from the Gaia DR2 catalogue
(\citealt{2018A&A...616A...1G}) and on the observations are reported
in the header of each light curve.

\subsection{Photometric precision}

Figure~\ref{fig:4} shows the two quality parameters adopted also in
\citetalias{2019MNRAS.490.3806N} and previous works of our group: (i)
the \texttt{RMS}, calculated using the cotrended light curves and
defined as the 68.27th-percentile of the sorted residual from the
median value, obtained clipping-out the outliers in 10 iterations; (ii) the
\texttt{P2P RMS} (point-to-point), that is not sensitive to intrinsic
stellar variability, and obtained calculating the 68.27th percentile
of the distribution of the sorted residual from the median value of
$\delta F_j=F_j-F_{j+1}$, with $F_j$ and $F_{j+1}$ the flux of the
light curve at the epochs $j$ and $j+1$.

\begin{figure*}
\includegraphics[bb=0 31 566 567, width=0.8\textwidth]{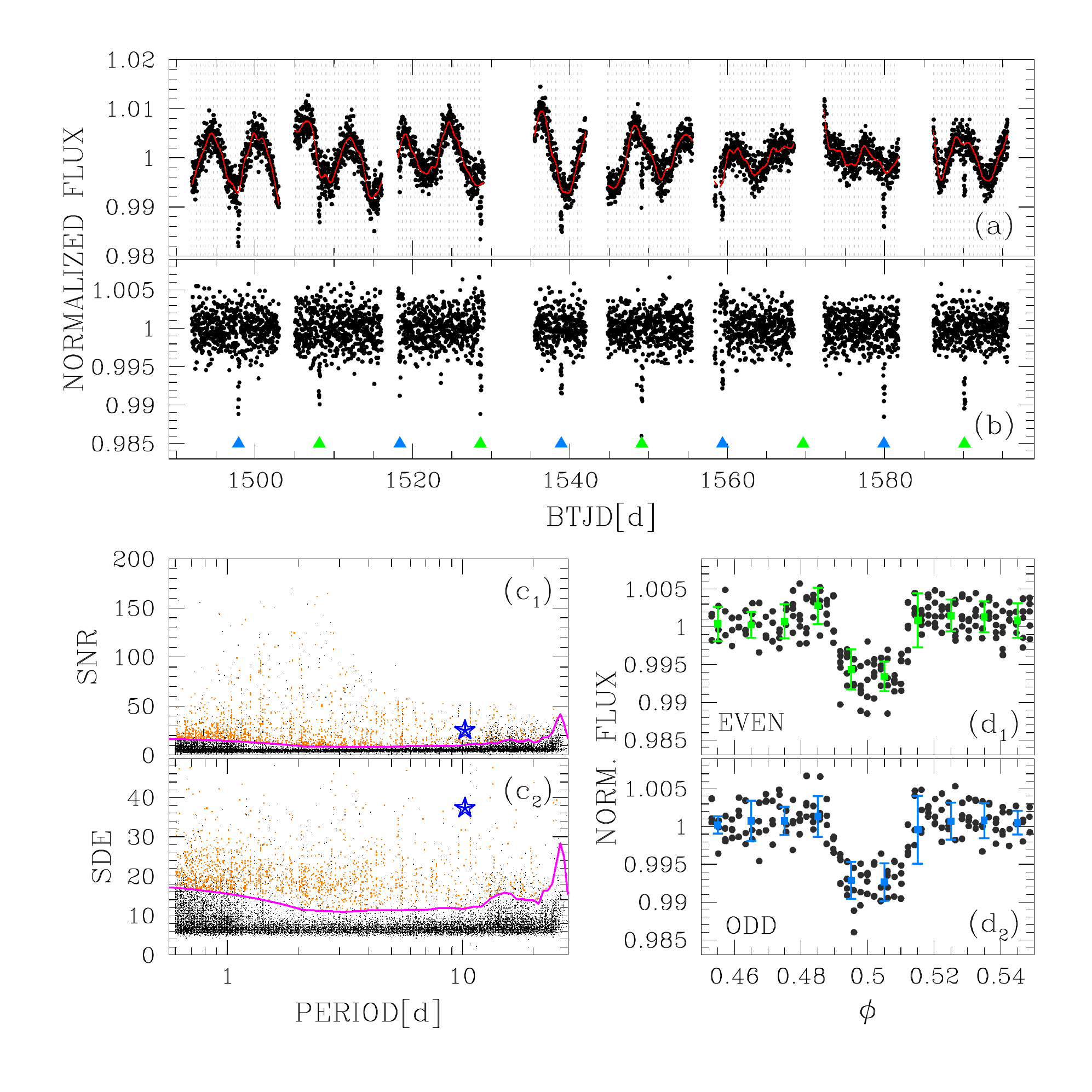}
\caption{Overview of the successive steps of the procedure adopted to
  select candidate transiting objects: panel (a) shows the normalised
  light curve of Gaia DR2\,5290850609994130560 from Sectors 7, 8, 9,
  and 10. The model (in red) is obtained interpolating a 5th order
  spline on a grid of knots spaced every 13-hours (grey lines). Panel
  (b) is the same light curve after removing the variability of the
  star: azure/green arrows are the odd/even transit events, found
  extracting the TLS periodogram. Panels (c) illustrate the procedure
  of selection of stars on the basis of the parameters extracted with
  the TLS routine: black points are all the stars observed in more
  than 1 sector, magenta lines are the limits above which stars pass
  the selection based on SDE/SNR parameters, orange points are the
  stars that pass all the selections
  (SDE/SNR/depth-of-transit/$\sigma_{\rm odd-even}$), and blue star is
  Gaia DR2\,5290850609994130560. Panels (d$_1$) and (d$_2$) are the
  phased even and odd transits, respectively. Green and azure points
  are the means (and their standard deviations) of the phased points
  calculated in bins of width 0.01. Comparison of even/odd transit
  depths is useful to discriminate between candidate exoplanets and
  eclipsing binaries (see text for details).
  \label{fig:5}}
\end{figure*}

For each photometric method, we derived, for all the extracted light
curves, the mean trends of the \texttt{RMS} and \texttt{P2P RMS} as a
function of the $T$ magnitude as follows: we divided each parameter
distribution in intervals of 0.75 $T$ magnitude and we calculated in
each bin the $3.5\sigma$-clipped average of the parameter. Finally, we
interpolated the mean values with a cubic spline. Figure~\ref{fig:4}
shows these mean trends both for the \texttt{RMS} (top panel) and for
the \texttt{P2P RMS} (bottom panel); in the plot, different colours
are associated to different photometric methods: black, magenta, blue,
green, and yellow lines correspond to PSF-fitting, 1-, 2-, 3-, and
4-pixel aperture photometries, respectively. Red stars are the
saturated stars, measured as described in Sect.~\ref{sec:sat}.

We used the \texttt{P2P RMS} mean trends to select the best
photometric method for each light curve: given a star of magnitude
$T_\star$, we have chosen to analyse the light curve extracted with
the photometric method that in $T_\star$ have the lower mean
\texttt{P2P RMS}. On average, for not saturated stars with $T\lesssim
7.0$, we used the 4-pixel aperture photometry; for stars with $7.0
\lesssim T \lesssim 9.0$ and with $9.0 \lesssim T \lesssim 10.5$ the
best photometric methods are the 3-pixel and 2-pixel aperture
photometries, respectively. PSF-fitting photometry works well for
stars having $10.5 \lesssim T \lesssim 13.5$; finally, in the faint
regime of magnitudes ($T \gtrsim 13.5$) the 1-pixel aperture
photometry gives the best results. Gray dashed line in
Fig.~\ref{fig:4} is the theoretical limit obtained taking into account
all the sources of noise (shot noise, sky, readout-noise RON, and dark
current DC), adopting the average values of RON$\sim 9\,e^{-}\,{\rm
  px}^{-1}$, DC$\sim 1\,e^{-}\,s^{-1}\,{\rm px}^{-1}$, sky$\sim
150\,e^{-}\,s^{-1}$, and using in different intervals of magnitudes
the aperture radius where the light curve shows the lower scatter, on
the basis of the considerations done previously. Theoretical limit
distributions are, as expected, lower then the mean observed trends;
this is mainly due to a combination of different effects, like stellar
variability (that mainly affects the \texttt{RMS} distributions),
variations of the background in the light curves, contamination by and
blending with not-subtracted sources, etc.

Before of the analysis, we excluded all the light curves for which
instrumental magnitude $T_{\rm instr}$ is too different from the
expected calibrated magnitude $T_{\rm calib}$, by using the following
procedure: for each photometric method (and for each Sector), we
analysed the $\delta T= T_{\rm instr}-T_{\rm calib}$ distribution, and
we excluded all the light curves whose $\delta T$ value deviates more
than $4 \sigma$ from the mean value. In this way we excluded all the
light curves of stars strongly contaminated by other sources that were
not subtracted during the light curve extraction (e.g., close bleeding
columns, hot pixels, background galaxies or other sources that were
not in the Gaia DR2 catalogue). We also excluded all the light curves
that have $< 75 \%$ of good points (i.e. \texttt{DQUALITY}$=0$ and
\texttt{FLUX}$\neq 0$). Finally, we analysed 196\,231 light curves
associated to 147\,702 stars.

\begin{figure*}
\includegraphics[bb=0 130 484 554, width=0.85\textwidth]{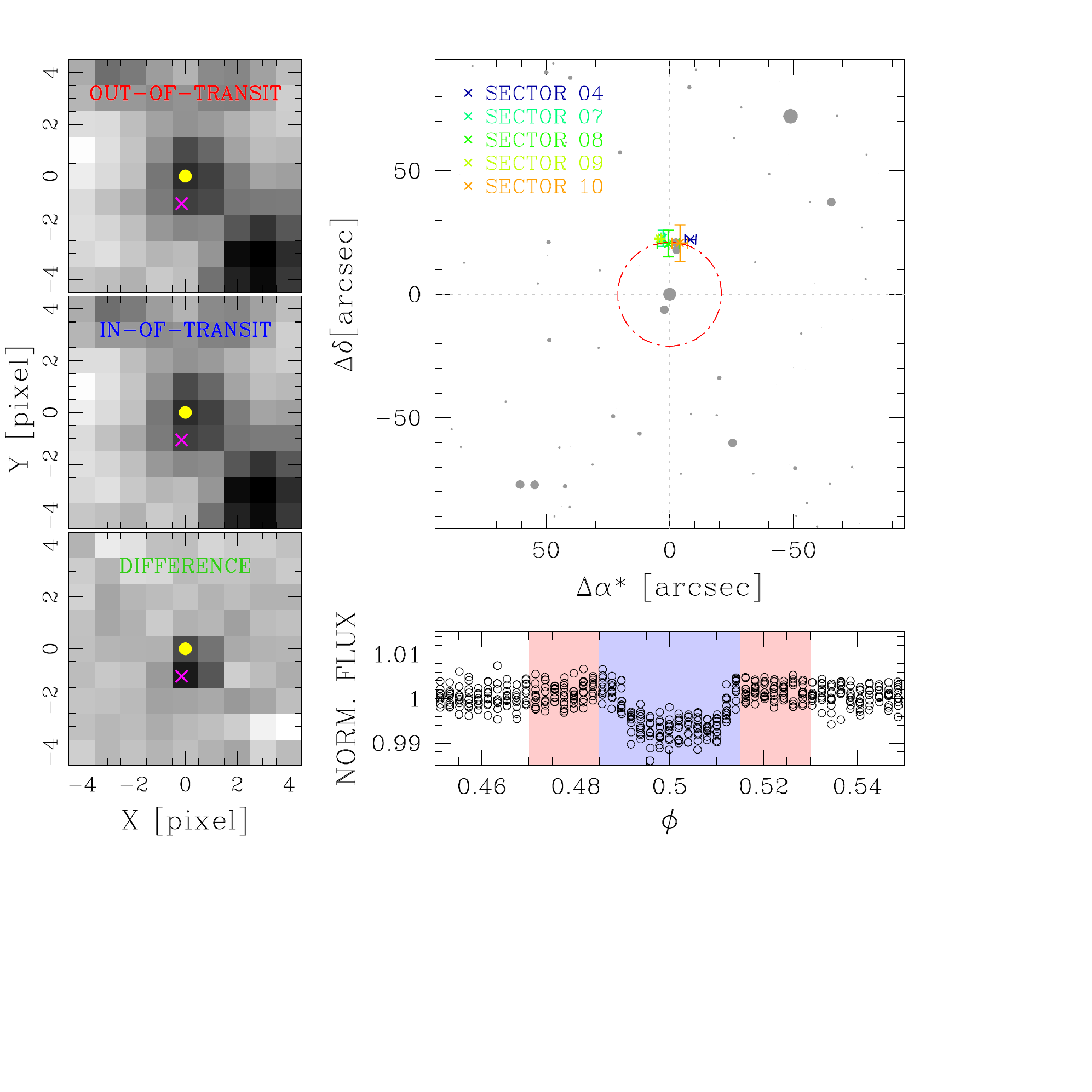}
\caption{Vetting procedure of the candidate Gaia
  DR2\,5290850609994130560 based on the position of the
  out-/in-of-transit difference image centroid. Left panels show the
  average of the images corresponding to out-of-transit (top panel)
  and in-of-transit (middle panel) points, and the difference between
  these two images (bottom panel); yellow circle is the position of
  the target star as reported in the Gaia DR2 catalogue, while magenta
  cross is the photocenter calculated on the difference image.  Top
  right-hand panel shows the $95\times 95$\,arcsec$^2$ finding chart
  centred on the target star and based on the Gaia DR2 catalogue. The
  crosses, colour-coded as in the legenda, are the mean offsets
  calculated for each sector in which the star is observed; red circle
  represents the photometric aperture adopted to analyse the light
  curve. Bottom right-hand panel is the phased light curve of the
  target star: light red and blue regions contain the out- and
  in-of-transit points, respectively.
  \label{fig:6}}
\end{figure*}

\section{Candidate exoplanets selection}
\label{sec:search}

In order to select candidate transiting exoplanets orbiting stellar
cluster members, we performed a series of analyses and tests on the
light curves selected as described in the previous sections, to
identify and remove the large part of false positive events.

As a first step, we removed the variability of the stars. We flattened
the light curves using 5th order splines defined on $N_{knots}$
knots. We calculated two different grid of knots: (i) a knot every
6.5-hours, and (ii) a knot every 13-hours. In this way, we were able
to better model short- and long-period variable stars, but also to
prevent the flattening of the transits whose duration is longer than
6.5 hours. In this phase we also ``\textit{cleaned}'' each light
curve, excluding: (i) points whose quality flag was
\texttt{DQUALITY}$>0$; (ii) all the outliers above $3.5\sigma$ the
median flux; (iii) the photometric points associated to local sky
background higher than 5$\sigma_{\rm sky}$ above median value of the
sky. Panels (a) and (b) of Fig.~\ref{fig:5} show the procedure of
flattening of the light curve of the star Gaia
DR2\,5290850609994130560: in this case we defined a knot every
13-hours (grey lines in panel (a)).

We performed the procedure that we are going to describe considering
each sector independent from the others; in a second step we
performed again the same procedure for that stars observed in more than
one sector and considering the entire light curve.

For the flattened light curves, we extracted the Transit-fitting Least
Squares (TLS) periodograms\footnote{TLS v.1.0.24,
  \url{https://github.com/hippke/tls}}(\citealt{2019A&A...623A..39H}),
searching for transiting objects having period between $0.6\,{\rm d}
\leq P \leq 0.5\times T_{\rm LC}$, where $T_{\rm LC}$ is the maximum
temporal baseline of the light curve (usually, for a single sector
$T_{\rm LC} \sim 27$\,days). In addition to the periodograms, the
routine extracted many parameters useful to discriminate between light
curves with/without transit signals, such as: the signal detection
efficiency (SDE), the signal-to-noise ratio (SNR), the depth of the
transit, and the significance between odd and even transit depths
($\sigma_{\rm odd-even}$). We used these parameters to perform a first
selection of candidates: we divided the SDE and SNR distributions in
intervals $\delta P = 0.5$\,d, and within each of them we calculated
their $3.5\sigma$-clipped mean (${\bar {\rm SDE}}$ and ${\bar {\rm
    SNR}}$) and standard deviation ($\sigma_{\bar {\rm SDE}}$ and
$\sigma_{\bar {\rm SNR}}$). We interpolated ${\bar {\rm
    SDE}}+3.5\sigma_{\bar {\rm SDE}}$ and ${\bar {\rm SNR}}+3.5
\sigma_{\bar {\rm SNR}}$ with splines, and we saved all the light
curves which SDE and SNR values are above these splines, depth
$<10\,\%$, and $\sigma_{\rm odd-even}< 2.5$. This selection procedure
is shown in panels (c$_1$) and (c$_2$) of Fig.~\ref{fig:5}: black
points are all the stars for which we extracted the periodograms and
that are observed in more than one sector, magenta line is the lower
boundary for SNR and SDE selections, orange points are the stars that
passed all the above described selections, and blue star is the
candidate Gaia DR2\,5290850609994130560.  On average, the number of
stars that passed these selection is $\sim 3\,\%$ of the total number
of analysed light curves. After this selection, in order to avoid false
positive due to blends, we identified the groups of stars which
  present similar period within 0.05\,d and they are separated by $<5$
  pixels, and we exclude all these stars if one of them shows a
  signal that can not be associated to a transiting planet
  (e.g., a too deep transit or the presence of a secondary eclipse). 
  Finally, we
visually inspected the remaining light curves to check the odd/even
transit depths (like in panels (d) of Fig.~\ref{fig:5}), the presence
of artifacts that generate false transits, and the presence of
secondary eclipses. On average, the final list of candidate transiting
object contains $\sim 0.1\,\%$ of the analysed objects; these
candidates went through of a detailed vetting procedure to exclude
possible remaining false positives.

\subsection{Vetting of the candidates}
We verified that the objects that survived to the previously described
selections are genuine candidates, checking if the transit signals are
due to a neighbour source. In order to verify this hypothesis, we
performed three different tests.

For the first test, we verified if the transit depth of the candidate
object changes considering light curves obtained with different
photometric methods. For example, if the transit events are due to a
neighbour eclipsing binary, we expect that large photometric apertures
give deeper transit events. Moreover, as already demonstrated in
\citetalias{2019MNRAS.490.3806N}, PSF-fitting photometry is less
affected by contamination than aperture photometry and allows us to
discriminate between true and false positives. To perform this test,
we estimated the mean value (and its standard deviation $\sigma$) of
the transit depth for each photometric method; if the transit
associated to a photometric method is $2 \sigma$ deeper than the mean
transit depth associated to the analysed light curve, we considered
the star as false positive.

The second test consisted in checking the binned phased light curve
with a period equal to $0.5 \times$, $1 \times$ and $2 \times$ the
period found by TLS, and searching for secondary eclipses outside the
primary transit.  A star is considered false positive if the mean
  depths of even/odd transits differ $>2 \sigma$.

Finally, we computed the in/out-of-transit difference centroid to
check if the transit events are associated to the target or to a close
star. To calculate the centroid, we followed the procedure illustrated
in Fig.~\ref{fig:6} in the case of the star Gaia
DR2\,5290850609994130560: we identified the FFIs corresponding to the
in-of-transit and out-of-transit points of the light curve, (light
blue and light red shaded regions in the bottom-right panel of
Fig.~\ref{fig:6}, respectively). For each transit, we calculated the
mean out-of-transit image, the mean in-of-transit image, and their
difference (top, middle, and bottom left-hand panels of
Fig.~\ref{fig:6}, respectively). For each transit, we then calculated
the photocenter on the difference image (magenta cross) and its offset
relative to the Gaia DR2 position of the target star (yellow circle).
Finally, for each sector, we calculated the mean value and its
standard deviation of all the offsets. Top right-hand panel shows the
finding chart ($95\times 95$\,arcsec$^2$) based on the Gaia DR2
catalogue and centred on Gaia DR2\,5290850609994130560: the average
offsets for all the sectors in which the star is observed are all
located at a distance $\sim 20$\,arcsec from the target star, in a
region corresponding to two close stars of magnitude $\sim 14.5$ (Gaia
DR2\,5290850609994130944 and Gaia
DR2\,5290850605693476608). Therefore, we discarded this candidate
because the transit signals are not associated to the cluster member,
but to a background star close to our target.

\subsection{{\it TESS} Objects of Interest}

We cross-matched our catalogue of analysed cluster members with the
list of the {\it TESS} Objects of
Interest\footnote{\url{https://tess.mit.edu/toi-releases/go-to-alerts/}}
(TOIs) released by the {\it TESS} team. We found 5 stars in common:
three of them (TOI-496, TOI-681, and TOI-837) are also detected by our
pipeline (PATHOS-6, PATHOS-25, and PATHOS-30). The other two TOIs were
initially detected in the first selection of candidates we did, and
then excluded by our vetting: (1) TOI-517 resulted to be an eccentric
eclipsing binary with period $P\sim 9.885$\,d (as also reported in the
public comments of the TOI); (2) from the analysis of the
in-/out-of-transit centroid of TOI-861 we found that the signal is
associated to a close star (TIC\,0372913432) at $\sim 40$ arcsec from
the target, and that this star is catalogued as binary by
\citet{2018MNRAS.475.1609B}.

\section{Modelling of the transits}
\label{sec:modelling}

In our final catalogue of genuine transiting objects there are 33
sources associated to 28 stellar clusters.  For the extraction of the
physical parameters (that can be obtained from a light curve analyses)
of the transiting objects we need to know the stellar parameters.  As
stars in open clusters are located at the same distance and have the
same chemical composition and age, we can use theoretical models to
estimate their main parameters including temperature, mass and radius.
In the next section, we describe how we extracted these, that will be
then used as priors for the modelling of the transits
(Sect.~\ref{sec:mod}).

\subsection{Stellar parameters}

We extracted the information on the cluster members that host
candidate transiting exoplanets by fitting isochrones to the
colour-magnitude diagrams (CMDs) of each stellar cluster and deriving
the stellar parameters (mass, effective temperature, luminosity, and
radius) interpolating colour and magnitude of each star on the
isochrone.  In this work we used the last release of BaSTI (``a Bag of
Stellar Tracks and Isochrones'') models
(\citealt{2018ApJ...856..125H}) to fit isochrones on the CMDs.  In
order to perform this fitting, it is necessary to know the age,
metallicity, distance, and reddening of the stellar clusters.

Unfortunately, metallicity measurements are not available for the
large part of the stellar clusters in our sample. For this reason, we
considered a constant metallicity [Fe/H]$=0.0$ for all the clusters
and, at the end, we added to the final error on stellar parameters,
the contribution of considering a wrong [Fe/H] in the isochrone
fitting procedure. The corresponding uncertainty has been estimated
using 3 different isochrones with age of 300 Myr (the average age of
the studied clusters), but with metallicity [Fe/H]=$-0.3$,
[Fe/H]$=0.0$, and [Fe/H]$=+0.3$, that is the range of metallicities
spanned by the Galactic stellar clusters for which there are [Fe/H]
measurements. We found that, for main sequence stars with mass
$0.5\,M_{\sun}<M_\star<1.5\,M_{\sun}$, the mean differences on the
stellar radius $R_\star$, mass $M_\star$, and effective temperature
$T_{\rm eff}$ between isochrones with [Fe/H]$=0.0$ and [Fe/H]$=\pm0.3$
are $\sim 0.03$~R$_{\sun}$, $\sim 0.05$~M$_{\sun}$, and $\sim 200$~K,
respectively.

We extracted the information on the distance of the clusters from
\citet{2018A&A...618A..93C}, while ages and reddenings are extracted
from two different catalogues: for 13 stellar clusters we used the
cluster parameters and the associated errors found by
\citet{2019A&A...623A.108B} analysing the Gaia\,DR2 CMDs; for the
other clusters we used the information given by
\citet{2016A&A...585A.101K}. In the latter catalogue the errors on age
and reddening are not given. In this case, we adopted an error equal
to $10\%$ of the measurement. Muzzio\,1 is the only cluster that is
not present in both catalogues; for this very young cluster we used
the information given by \citet{2011A&A...530A..32B}.

Table~\ref{tab:1} lists the cluster parameters.  Thanks to isochrone
fits, we were able to extract the stellar parameters of the cluster
members that host candidate transiting exoplanets. These parameters
were then used as priors in the phase of modelling described in the
next section. Isochrone fits are shown in Figs.~\ref{fig:7a},
\ref{fig:7b}, and \ref{fig:7c}.

\begin{table}
  \caption{Cluster parameters}
  \resizebox{0.5\textwidth}{!}{
      \begin{tabular}{l  S[table-format=4.0(3)]   S[table-format=5.1(3)]  r  c c}
  \hline
  Cluster Name & {Age}   & {Distance} & $E(B-V)$ & $\pi^{\rm (a)}$ & Reference \\
               & {(Myr)} &   {(pc)}   &          &  (mas)   &    \\ 
  \hline
  Alessi\,8      &  137 \pm  35 &  663.7 \pm  9.5 & $0.10 \pm 0.01$ &  $1.479\pm0.046$    & (2) \\
  ASCC\,85       &   32 \pm   3 &  870.0 \pm  5.1 & $0.52 \pm 0.05$ &  $1.119\pm0.085$    & (2) \\
  ASCC\,88       &   20 \pm   2 &  883.3 \pm  5.5 & $0.73 \pm 0.07$ &  $1.097\pm0.060$    & (2) \\
  Collinder\,292 &  498 \pm  52 & 1461.5 \pm 16.9 & $0.29 \pm 0.01$ &  $0.528\pm0.045$    & (1) \\ 
  Haffner\,14    &  195 \pm  10 & 4359.1 \pm 38.4 & $0.61 \pm 0.01$ &  $0.226\pm0.041$    & (1) \\
  Harvard\,5     &   65 \pm  34 & 1203.9 \pm 15.6 & $0.22 \pm 0.01$ &  $0.764\pm0.044$    & (1) \\
  IC\,2602       &   35 \pm   2 &  152.3 \pm  1.0 & $0.03 \pm 0.01$ &  $6.561\pm0.157$    & (1) \\
  Melotte\,101   &  125 \pm  13 & 2134.7 \pm  9.2 & $0.45 \pm 0.05$ &  $0.439\pm0.048$    & (2) \\
  Muzzio\,1      &    5 \pm   4 & 1824.8 \pm 24.3 & $1.02 \pm 0.56$ &  $0.519\pm0.023$    & (3) \\
  NGC\,2112      & 2065 \pm 207 & 1103.6 \pm  3.6 & $0.63 \pm 0.06$ &  $0.877\pm0.066$    & (2) \\
  NGC\,2318      &  560 \pm  56 & 1342.9 \pm  8.5 & $0.44 \pm 0.04$ &  $0.716\pm0.042$    & (2) \\
  NGC\,2323      &   95 \pm  10 &  921.3 \pm  6.8 & $0.15 \pm 0.01$ &  $0.997\pm0.057$    & (2) \\
  NGC\,2437      &  220 \pm  22 & 1582.3 \pm  3.9 & $0.15 \pm 0.01$ &  $0.603\pm0.060$    & (2) \\
  NGC\,2516      &  251 \pm   5 &  415.1 \pm  1.0 & $0.09 \pm 0.01$ &  $2.417\pm0.045$    & (1) \\
  NGC\,2527      &  830 \pm  48 &  572.0 \pm  5.6 & $0.06 \pm 0.01$ &  $1.536\pm0.070$    & (1) \\
  NGC\,2548      &  525 \pm  53 &  758.8 \pm  2.0 & $0.04 \pm 0.01$ &  $1.289\pm0.065$    & (2) \\
  NGC\,2669      &   82 \pm   3 & 1124.1 \pm 10.4 & $0.20 \pm 0.01$ &  $0.845\pm0.046$    & (1) \\
  NGC\,2671      &  338 \pm  34 & 1396.5 \pm  7.2 & $0.98 \pm 0.10$ &  $0.687\pm0.031$    & (2) \\
  NGC\,3114      &  200 \pm  20 & 1017.3 \pm  1.1 & $0.08 \pm 0.01$ &  $0.954\pm0.048$    & (2) \\
  NGC\,3532      &  399 \pm  10 &  485.3 \pm  1.0 & $0.02 \pm 0.01$ &  $2.066\pm0.062$    & (1) \\
  NGC\,5316      &  152 \pm   5 & 1437.5 \pm 12.6 & $0.25 \pm 0.01$ &  $0.665\pm0.055$    & (1) \\
  NGC\,5925      &  180 \pm  18 & 1413.2 \pm  6.2 & $0.75 \pm 0.07$ &  $0.679\pm0.050$    & (1) \\
  Ruprecht\,48   &  398 \pm  40 & 3784.4 \pm 79.5 & $0.28 \pm 0.03$ &  $0.235\pm0.038$    & (2) \\
  Ruprecht\,82   &  353 \pm  47 & 2206.0 \pm 24.5 & $0.33 \pm 0.02$ &  $0.430\pm0.038$    & (1) \\
  Ruprecht\,94   &   20 \pm   2 & 2511.5 \pm 29.5 & $0.05 \pm 0.05$ &  $0.369\pm0.031$    & (2) \\
  Ruprecht\,151  &  411 \pm  25 & 1179.8 \pm 25.2 & $0.19 \pm 0.01$ &  $0.867\pm0.056$    & (1) \\
  SAI\,91        &  676 \pm  68 & 3571.6 \pm 90.7 & $0.44 \pm 0.08$ &  $0.251\pm0.040$    & (2) \\
  Trumpler\,22   &   25 \pm  15 & 2364.8 \pm 24.1 & $0.59 \pm 0.01$ &  $0.386\pm0.028$    & (1) \\
  \hline
  \multicolumn{5}{l}{{(1)~\citet{2019A&A...623A.108B}; (2)~\citet{2016A&A...585A.101K}; (3)~\citet{2011A&A...530A..32B}}} \\
  \multicolumn{5}{l}{{$ ^{\rm (a)}$\,Mean cluster parallax from \citet{2018A&A...618A..93C}}} \\
\end{tabular}

}
      \label{tab:1}

\end{table}

\subsection{Transit modelling}
\label{sec:mod}
For the transit modelling and analysis, we made use of
\texttt{PyORBIT}\footnote{\url{https://github.com/LucaMalavolta/PyORBIT}}
(\citealt{2016A&A...588A.118M,2018AJ....155..107M}; see also, e.g.,
\citealt{2019A&A...630A..81B}). The routine is a wrapper for the
transit modelling code \texttt{batman} (\citealt{2015PASP..127.1161K})
and the affine invariant Markov chain Monte Carlo (MCMC) sampler
\texttt{emcee} (\citealt{2013PASP..125..306F}) combined with global
optimization algorithm
\texttt{PyDE}\footnote{\url{https://github.com/hpparvi/PyDE}}.  We
included in the transit model: the central time of the first transit
($T_0$), the period ($P$), the impact parameter ($b$), the planetary
to stellar radius ratio ($R_{\rm p}/R_\star$), and the stellar density
($\rho_\star$). For the modelling, we adopted a Keplerian orbit with
null eccentricity. In Table~\ref{tab:2} we listed the priors used for
the fit. We used the $T_{\rm eff}$ and $\log(g)$ values obtained with
isochrones fitting to extract, through a bilinear interpolation, the
priors on the limb darkening (LD) coefficients from the grid of values
published by \citet{2018A&A...618A..20C}; for the modelling we adopted
the LD parametrisation used by \citet{2013MNRAS.435.2152K}.  In the
modelling process, we took into account of the 30-minute cadence of
the \textit{TESS} time-series (\citealt{2010MNRAS.408.1758K}). The
routine explored all the parameters in linear space. For our fit, we
made use of a number of walkers $N_{\rm walkers}=110$, equal to 10
times the number of free parameters. We ran, for each model, the
sampler for 50\,000 steps, removing the first 15\,000 steps as burn-in
and using a thinning factor of 100.

\begin{table*}
  \caption{Star parameters and priors for the modelling}
  \label{tab:2}
    \resizebox{0.97\textwidth}{!}{
      \begin{tabular}{l c l S[table-format=4.0(4)] S[table-format=4.0(4)] S[table-format=3.0(4)] S[table-format=3.0(1)] c c c l c c c}
  \hline
  TIC & PATHOS & Cluster & {$\alpha$} & {$\delta$} & {$\pi$}   & {$T$} & $R_\star$ & $M_\star$ & $\rho_\star$ & Period & $T_0$ & LD$_{c1}$ & LD$_{c2}$  \\
  &        &            & {(deg.)}   &  {(deg.)}  &  {(mas)} &{(mag.) }   & ($R_{\sun}$) & ($M_{\sun}$) & ($\rho_{\sun}$) & (d)    & (BTJD) &         &           \\
  \hline
0030654608 &   2   &      Muzzio\,1 &   134.3845     &      -47.8477 & 0.5191 & 10.4 &    $10.0\pm2.00$ & $17.0 \pm 2.00$   & $\mathcal{N}(0.01, 0.10)$ &  $\mathcal{U}( 2.0,  3.0)$     &  $\mathcal{U}(1519.0, 1520.0)$   &  $\mathcal{N}(0.10, 0.50)$    &  $\mathcal{N}(0.10, 0.50)$    \\
0039291805 &   3   &      NGC\,2112 &    88.7164     &       -0.0484 & 0.8964 & 15.2 &    $1.12\pm0.03$ & $1.13 \pm 0.05$   & $\mathcal{N}(0.80, 0.10)$ &  $\mathcal{U}( 7.0,  7.5)$     &  $\mathcal{U}(1468.9, 1469.2)$   &  $\mathcal{N}(0.31, 0.05)$    &  $\mathcal{N}(0.29, 0.05)$    \\
0042524156 &   4   &       ASCC\,88 &   256.8882     &      -35.5036 & 1.1502 & 10.5 &    $2.94\pm0.03$ & $5.18 \pm 0.05$   & $\mathcal{N}(0.20, 0.10)$ &  $\mathcal{U}( 9.8, 10.2)$     &  $\mathcal{U}(1631.0, 1632.0)$   &  $\mathcal{N}(0.15, 0.13)$    &  $\mathcal{N}(0.15, 0.13)$    \\
0080317933 &   5   & Collinder\,292 &   237.9156     &      -57.2797 & 0.6320 & 12.9 &    $1.83\pm0.03$ & $1.91 \pm 0.05$   & $\mathcal{N}(0.32, 0.10)$ &  $\mathcal{U}( 3.0,  3.5)$     &  $\mathcal{U}(1625.0, 1626.0)$   &  $\mathcal{N}(0.19, 0.05)$    &  $\mathcal{N}(0.23, 0.05)$    \\
0088977253 &   6   &      NGC\,2548 &   123.0977     &       -5.7687 & 1.2358 & 12.3 &    $1.42\pm0.03$ & $1.42 \pm 0.05$   & $\mathcal{N}(0.49, 0.10)$ &  $\mathcal{U}( 2.5,  3.0)$     &  $\mathcal{U}(1492.0, 1493.0)$   &  $\mathcal{N}(0.27, 0.05)$    &  $\mathcal{N}(0.28, 0.05)$    \\
0092835691 &   7   &        SAI\,91 &   129.1193     &      -50.1497 & 0.2347 & 14.6 &    $2.24\pm0.05$ & $2.02 \pm 0.07$   & $\mathcal{N}(0.18, 0.10)$ &  $\mathcal{U}(22.0, 23.0)$     &  $\mathcal{U}(1527.0, 1528.0)$   &  $\mathcal{N}(0.19, 0.05)$    &  $\mathcal{N}(0.23, 0.05)$    \\
0094589619 &   8   &      NGC\,2437 &   115.7338     &      -14.5810 & 0.6238 & 13.8 &    $1.46\pm0.03$ & $1.55 \pm 0.05$   & $\mathcal{N}(0.49, 0.10)$ &  $\mathcal{U}(11.8, 12.5)$     &  $\mathcal{U}(1496.0, 1497.0)$   &  $\mathcal{N}(0.24, 0.05)$    &  $\mathcal{N}(0.26, 0.05)$    \\
0125414447 &   9   &      NGC\,2323 &   105.7303     &       -8.4369 & 0.9320 & 14.2 &    $1.00\pm0.03$ & $1.10 \pm 0.05$   & $\mathcal{N}(1.10, 0.10)$ &  $\mathcal{U}( 3.5,  4.0)$     &  $\mathcal{U}(1493.0, 1494.0)$   &  $\mathcal{N}(0.32, 0.05)$    &  $\mathcal{N}(0.30, 0.05)$    \\
0126600730 &  10   &    Haffner\,14 &   116.0772     &      -28.3253 & 0.2145 & 15.4 &    $2.11\pm0.03$ & $2.56 \pm 0.05$   & $\mathcal{N}(0.27, 0.10)$ &  $\mathcal{U}( 6.0,  6.5)$     &  $\mathcal{U}(1497.0, 1498.0)$   &  $\mathcal{N}(0.15, 0.05)$    &  $\mathcal{N}(0.20, 0.05)$    \\
0144995073 &  11   &      NGC\,2669 &   131.4641     &      -52.9284 & 0.8289 & 13.7 &    $1.36\pm0.03$ & $1.41 \pm 0.05$   & $\mathcal{N}(0.55, 0.10)$ &  $\mathcal{U}(19.5, 20.5)$     &  $\mathcal{U}(1518.5, 1519.5)$   &  $\mathcal{N}(0.27, 0.05)$    &  $\mathcal{N}(0.28, 0.05)$    \\
0147069011 &  12   &      Alessi\,8 &   232.8684     &      -51.1783 & 1.5017 & 12.5 &    $1.25\pm0.03$ & $1.30 \pm 0.05$   & $\mathcal{N}(0.66, 0.10)$ &  $\mathcal{U}( 7.0,  7.5)$     &  $\mathcal{U}(1628.0, 1629.0)$   &  $\mathcal{N}(0.28, 0.05)$    &  $\mathcal{N}(0.29, 0.05)$    \\
0147426828 &  13   &      NGC\,2318 &   104.8768     &      -13.2535 & 0.6934 & 14.1 &    $1.49\pm0.03$ & $1.49 \pm 0.05$   & $\mathcal{N}(0.45, 0.10)$ &  $\mathcal{U}( 1.0,  1.7)$     &  $\mathcal{U}(1492.0, 1493.0)$   &  $\mathcal{N}(0.25, 0.05)$    &  $\mathcal{N}(0.27, 0.05)$    \\
0153734545 &  14   &      NGC\,2527 &   121.2402     &      -28.1462 & 1.5019 &  9.9 &    $2.72\pm0.03$ & $2.05 \pm 0.05$   & $\mathcal{N}(0.11, 0.10)$ &  $\mathcal{U}( 3.0,  3.5)$     &  $\mathcal{U}(1493.0, 1494.0)$   &  $\mathcal{N}(0.22, 0.05)$    &  $\mathcal{N}(0.25, 0.05)$    \\
0153735144 &  15   &      NGC\,2527 &   121.1537     &      -28.2948 & 1.5432 & 11.6 &    $1.51\pm0.03$ & $1.46 \pm 0.05$   & $\mathcal{N}(0.43, 0.10)$ &  $\mathcal{U}( 3.2,  4.0)$     &  $\mathcal{U}(1493.0, 1494.0)$   &  $\mathcal{N}(0.26, 0.05)$    &  $\mathcal{N}(0.27, 0.05)$    \\
0159059181 &  16   &      NGC\,2112 &    88.3711     &       +0.4239 & 0.8356 & 14.2 &    $1.52\pm0.03$ & $1.33 \pm 0.05$   & $\mathcal{N}(0.38, 0.10)$ &  $\mathcal{U}( 5.0,  5.5)$     &  $\mathcal{U}(1470.9, 1471.4)$   &  $\mathcal{N}(0.28, 0.05)$    &  $\mathcal{N}(0.28, 0.05)$    \\
0181602717 &  17   &      NGC\,2671 &   131.7366     &      -41.9773 & 0.5930 & 15.2 &    $1.47\pm0.03$ & $1.53 \pm 0.05$   & $\mathcal{N}(0.48, 0.10)$ &  $\mathcal{U}( 2.0,  2.7)$     &  $\mathcal{U}(1517.0, 1518.0)$   &  $\mathcal{N}(0.24, 0.05)$    &  $\mathcal{N}(0.25, 0.05)$    \\
0236084210 &  18   &       ASCC\,85 &   251.5388     &      -45.3214 & 1.1433 & 11.5 &    $1.84\pm0.03$ & $2.66 \pm 0.05$   & $\mathcal{N}(0.42, 0.10)$ &  $\mathcal{U}( 5.0,  5.5)$     &  $\mathcal{U}(1631.5, 1632.5)$   &  $\mathcal{N}(0.15, 0.05)$    &  $\mathcal{N}(0.18, 0.05)$    \\
0300362600 &  19   &      NGC\,5316 &   208.3975     &      -62.0764 & 0.6984 & 12.2 &    $2.14\pm0.03$ & $2.70 \pm 0.05$   & $\mathcal{N}(0.28, 0.10)$ &  $\mathcal{U}(11.0, 12.0)$     &  $\mathcal{U}(1601.9, 1602.5)$   &  $\mathcal{N}(0.15, 0.05)$    &  $\mathcal{N}(0.19, 0.05)$    \\
0306385801 &  20   &      NGC\,3532 &   166.6553     &      -58.8608 & 2.0510 &  9.5 &    $2.22\pm0.03$ & $2.29 \pm 0.05$   & $\mathcal{N}(0.21, 0.10)$ &  $\mathcal{U}(13.8, 14.5)$     &  $\mathcal{U}(1574.8, 1575.3)$   &  $\mathcal{N}(0.16, 0.05)$    &  $\mathcal{N}(0.21, 0.05)$    \\
0308538095 &  21   &      NGC\,2516 &   120.8514     &      -60.6655 & 2.4333 & 11.6 &    $1.29\pm0.03$ & $1.32 \pm 0.05$   & $\mathcal{N}(0.62, 0.10)$ &  $\mathcal{U}(11.0, 12.5)$     &  $\mathcal{U}(1420.0, 1421.0)$   &  $\mathcal{N}(0.29, 0.05)$    &  $\mathcal{N}(0.29, 0.05)$    \\
0317536999 &  22   &   Ruprecht\,94 &   172.5414     &      -63.1323 & 0.3921 & 14.1 &    $1.86\pm0.03$ & $2.76 \pm 0.05$   & $\mathcal{N}(0.43, 0.10)$ &  $\mathcal{U}( 5.5,  6.0)$     &  $\mathcal{U}(1575.5, 1576.5)$   &  $\mathcal{N}(0.15, 0.05)$    &  $\mathcal{N}(0.17, 0.05)$    \\
0372913337 &  23   &      NGC\,2516 &   119.4918     &      -60.8459 & 2.3454 &  8.9 &    $2.49\pm0.03$ & $2.74 \pm 0.05$   & $\mathcal{N}(0.18, 0.10)$ &  $\mathcal{U}( 1.5,  2.0)$     &  $\mathcal{U}(1325.0, 1326.0)$   &  $\mathcal{N}(0.15, 0.05)$    &  $\mathcal{N}(0.19, 0.05)$    \\
0389927567 &  24   &   Melotte\,101 &   160.2669     &      -65.5215 & 0.4321 & 13.0 &    $2.26\pm0.03$ & $2.95 \pm 0.05$   & $\mathcal{N}(0.25, 0.10)$ &  $\mathcal{U}( 6.5,  7.0)$     &  $\mathcal{U}(1575.0, 1576.0)$   &  $\mathcal{N}(0.15, 0.05)$    &  $\mathcal{N}(0.17, 0.05)$    \\
0410450228 &  25   &      NGC\,2516 &   117.8949     &      -60.4124 & 2.3447 & 10.6 &    $1.51\pm0.03$ & $1.67 \pm 0.05$   & $\mathcal{N}(0.48, 0.10)$ &  $\mathcal{U}(15.5, 16.0)$     &  $\mathcal{U}(1420.0, 1421.0)$   &  $\mathcal{N}(0.22, 0.05)$    &  $\mathcal{N}(0.25, 0.05)$    \\
0413809436 &  26   &      NGC\,5925 &   232.0979     &      -54.5586 & 0.6814 & 14.1 &    $1.59\pm0.03$ & $1.88 \pm 0.05$   & $\mathcal{N}(0.47, 0.10)$ &  $\mathcal{U}( 2.5,  3.0)$     &  $\mathcal{U}(1627.0, 1628.0)$   &  $\mathcal{N}(0.18, 0.05)$    &  $\mathcal{N}(0.23, 0.05)$    \\
0419091401 &  27   &   Ruprecht\,48 &   120.6352     &      -31.9951 & 0.2379 & 13.7 &    $3.40\pm0.20$ & $2.71 \pm 0.10$   & $\mathcal{N}(0.07, 0.10)$ &  $\mathcal{U}(12.5, 13.0)$     &  $\mathcal{U}(1499.0, 1500.0)$   &  $\mathcal{N}(0.15, 0.10)$    &  $\mathcal{N}(0.21, 0.10)$    \\
0432564189 &  28   &   Ruprecht\,82 &   146.5229     &      -53.9787 & 0.4632 & 13.6 &    $2.12\pm0.05$ & $2.28 \pm 0.05$   & $\mathcal{N}(0.23, 0.10)$ &  $\mathcal{U}( 3.5,  4.0)$     &  $\mathcal{U}(1546.0, 1547.0)$   &  $\mathcal{N}(0.16, 0.05)$    &  $\mathcal{N}(0.21, 0.05)$    \\
0450610413 &  29   &     Harvard\,5 &   186.7427     &      -60.6906 & 0.6932 & 12.2 &    $1.72\pm0.05$ & $2.30 \pm 0.05$   & $\mathcal{N}(0.45, 0.10)$ &  $\mathcal{U}(10.0, 11.0)$     &  $\mathcal{U}(1599.8, 1600.8)$   &  $\mathcal{N}(0.15, 0.05)$    &  $\mathcal{N}(0.20, 0.05)$    \\
0460205581 &  30   &       IC\,2602 &   157.0373     &      -64.5052 & 6.9893 &  9.9 &    $1.07\pm0.03$ & $1.15 \pm 0.05$   & $\mathcal{N}(0.93, 0.10)$ &  $\mathcal{U}( 8.0,  8.5)$     &  $\mathcal{U}(1574.0, 1575.0)$   &  $\mathcal{N}(0.31, 0.05)$    &  $\mathcal{N}(0.30, 0.05)$    \\
0460950389 &  31   &       IC\,2602 &   159.1580     &      -64.7982 & 6.6229 & 11.6 &    $0.84\pm0.03$ & $0.81 \pm 0.05$   & $\mathcal{N}(1.37, 0.10)$ &  $\mathcal{U}( 2.5,  3.0)$     &  $\mathcal{U}(1572.0, 1573.0)$   &  $\mathcal{N}(0.43, 0.05)$    &  $\mathcal{N}(0.38, 0.05)$    \\
0462004618 &  32   &      NGC\,3114 &   149.9120     &      -59.9032 & 1.0387 & 11.3 &    $2.01\pm0.03$ & $2.43 \pm 0.05$   & $\mathcal{N}(0.29, 0.10)$ &  $\mathcal{U}( 2.0,  2.5)$     &  $\mathcal{U}(1544.0, 1545.0)$   &  $\mathcal{N}(0.15, 0.05)$    &  $\mathcal{N}(0.20, 0.05)$    \\
0748919024 &  33   &  Ruprecht\,151 &   115.1315     &      -16.3216 & 0.7387 & 13.4 &    $1.46\pm0.03$ & $1.49 \pm 0.05$   & $\mathcal{N}(0.48, 0.10)$ &  $\mathcal{U}( 2.0,  3.0)$     &  $\mathcal{U}(1493.8, 1494.5)$   &  $\mathcal{N}(0.25, 0.05)$    &  $\mathcal{N}(0.27, 0.05)$    \\
1036769612 &  34   &   Trumpler\,22 &   217.5064     &      -61.3249 & 0.4244 & 15.8 &    $1.43\pm0.05$ & $1.58 \pm 0.10$   & $\mathcal{N}(0.54, 0.20)$ &  $\mathcal{U}( 7.1,  7.7)$     &  $\mathcal{U}(1599.0, 1600.0)$   &  $\mathcal{N}(0.24, 0.05)$    &  $\mathcal{N}(0.27, 0.05)$    \\
  \hline
\end{tabular}

      }
\end{table*}

In Appendix~\ref{app:cand} are all results: in Table~\ref{tab:3}
are listed the outputs of our fit and the clusters associated to each
transiting object of interest; Figs.~\ref{fig:7a}, \ref{fig:7b}, and
\ref{fig:7c} show, for each transiting object, the position of the
star on the Gaia\,DR2 CMD of the cluster to which it is associated
(left-hand panel), the folded light curve with the overimposed model
of the transit (top right-hand panel), the difference between the
observed points and the model (middle right-hand panel), the finding
chart and the vector-point diagram of the target, and its surrounding
sources from the Gaia\,DR2 catalogue (bottom right-hand
panels).

\section{Exoplanets in open clusters: results}
\label{sec:results}

From our dataset, we extracted 33 objects of interest whose light
curves show transit signals. In this section we isolate the more
significant candidate exoplanets and compare the candidate exoplanets
and cluster parameters.

From our final list, we excluded all stars that host candidate planets
with a radius $R_{\rm P} \geq 3.0\,R_{\rm J}$ (15 objects), because of
their doubtful planet nature. We also excluded all those stars that
have a parallax that differs $>3 \sigma$ from the mean parallax of the
cluster (see Table~\ref{tab:1}), stars that are not located on the
main sequence in the CMD of the cluster (i.e., $>2 \sigma$ from the
mean colour of the main sequence), and stars whose proper motion
differs by $>4 \sigma$ from the mean proper motion of the cluster. In
this way, we were left with 11 candidates (PATHOS-3, 6, 8, 9, 15, 20,
21, 23, 25, 30, 31) in 8 open clusters, 7 of them (NGC\,2112,
NGC\,2437, NGC\,2516, NGC\,2527, NGC\,2548, NGC\,3532, and IC\,2602)
having a solar metallicity (\citealt{2016A&A...585A.150N}), while for
NGC\,2323 there are no metallicity measurements; 9 of these candidate
exoplanets orbit stars with radii $R_{\star}\lesssim 1.5\,R_{\sun}$,
while PATHOS-20 and PATHOS-23 orbit stars with $R_{\star}\sim
2.2\,R_{\sun}$ and $R_{\star} \sim 2.5\,R_{\sun}$, respectively.

\begin{figure*}
\includegraphics[bb=0 0 573 571 , width=0.75\textwidth]{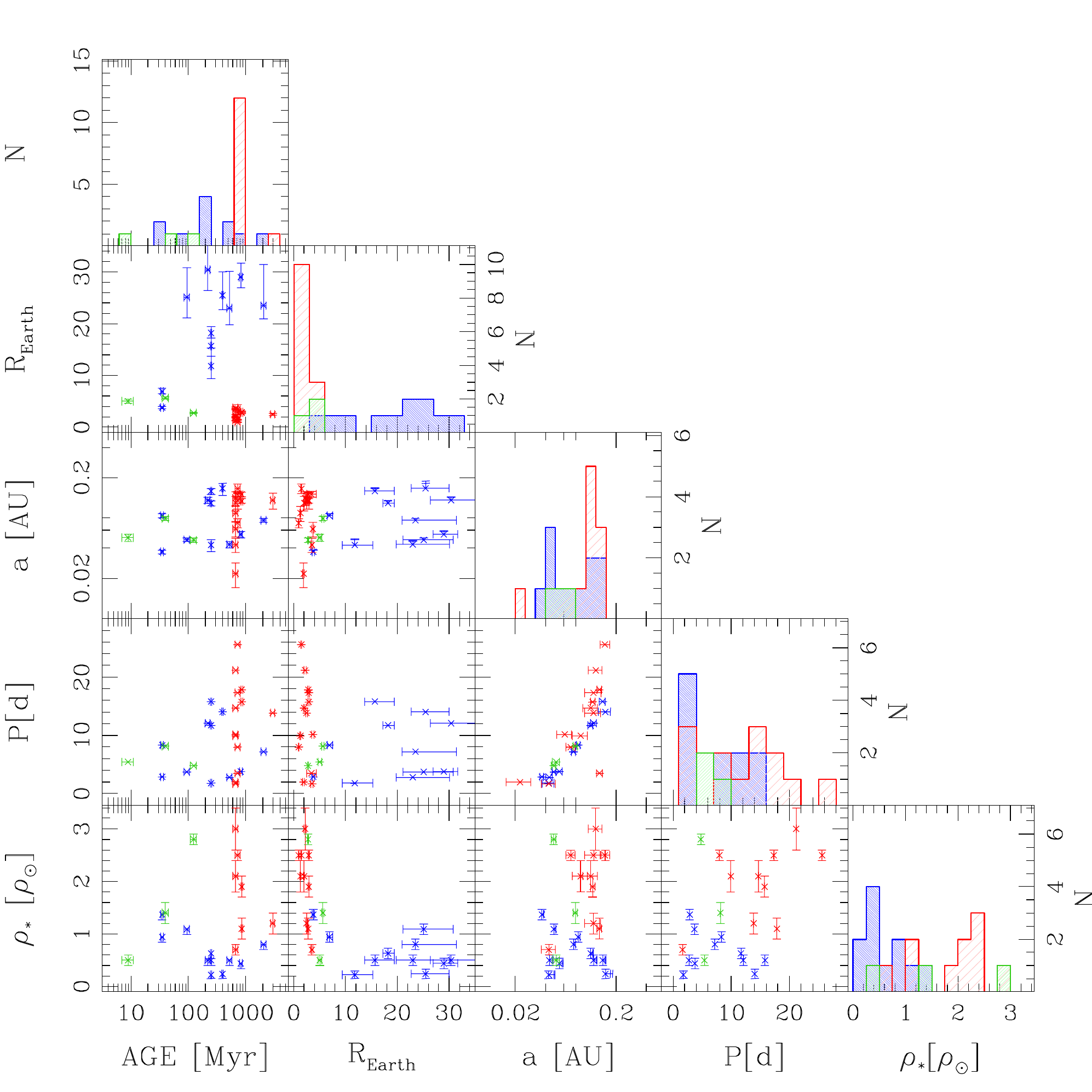}
\caption{Correlations between cluster, stellar, and exoplanets
  properties. Blue points and histograms are the results of this work;
  red points and histograms are the results obtained using data of
  open clusters collected by {\it Kepler}; green points and
  histograms relative to  stars in associations.
  \label{fig:8}}
\end{figure*}

Figure~\ref{fig:8} shows the correlations between cluster age and
stellar density and some candidate exoplanet parameters, like period,
semi-major axis, and radius.  In this analysis we also included the
candidate and confirmed transiting exoplanets in open clusters and
young association found by {\it Kepler/K2} and {\it TESS}, present in
literature. In particular, we considered the first two discovered
exoplanets (Kepler-66b and Kepler-67b, \citealt{2013Natur.499...55M})
in NGC\,6811 ($\sim 863$\,Myr), the 4 exoplanets K2-25b, K2-136Ab,c,d
(\citealt{2018AJ....155...10C,2018AJ....155....4M}) in the Hyades
($\sim 730$\,Myr), the 6 exoplanets in M\,44 ($\sim 670$\,Myr), e.g.,
K2-95b, K2-100b, K2-101b, K2-102b, K2-103b, K2-104b
(\citealt{2016A&A...594A.100B,2016MNRAS.463.1780L,2016MNRAS.461.3399P,2016AJ....152..223O,2017AJ....153..177P,2017AJ....153...64M}),
and the Ruprecht\,147 ($\sim 3$\,Gyr) member K2-213b
(\citealt{2018AJ....155..173C}). In our analysis, we also considered
the exoplanets K2-33b in the young association Upper Scorpius ($\sim
10$\,Myr, \citealt{2016Natur.534..658D,2016AJ....152...61M}),
EPIC\,247267267\,b in the Cas-Tau group ($\sim 120$\,Myr,
\citealt{2018AJ....156..302D}) and the recently discovered exoplanets
DS\,Tuc\,Ab (\citealt{2019ApJ...880L..17N,2019A&A...630A..81B}),
discovered by {\it TESS} and member of the Tucana-Horologium young
association ($\sim 40$\,Myr).

Blue points and histograms in Fig.~\ref{fig:8} correspond to the
objects identified in this work; red points and histograms are the
planets from {\it Kepler} observations; green points and histograms
represent stars in young associations. Considering all the points,
there is no evident correlations among cluster/stellar and (candidate)
exoplanet parameters.  Figure~\ref{fig:8} illustrates the difference
between the type of exoplanets in stellar clusters detected by {\it
  Kepler} and {\it TESS}: the large part of the candidates found in
this work are Jupiter-size objects orbiting stars with
$0.7\,R_{\sun}\lesssim R_{\star} \lesssim 2.5\,R_{\sun}$, while the
transiting objects found with {\it Kepler} and {\it K2} data of open
clusters are Neptune- and Earth-size exoplanets hosted by
$R_{\star}\lesssim 1\,R_{\sun}$ stars. This is mainly due to the
combination of two observational biases: (i) {\it Kepler} data
limiting magnitude is about 2-3 magnitudes fainter than {\it TESS} one,
and therefore, at a given magnitude, {\it Kepler} photometric
precision is higher than {\it TESS} one, allowing us to find smaller
exoplanets around fainter main sequence stars; (ii) 3 of the 4 open
clusters (Hyades, M\,44, and Ruprecht\,147) observed by {\it Kepler}
and whose members host known exoplanets are ``close'' clusters ($d
\lesssim 300$\,pc, while NGC\,6811 is at $\sim 1250$\,pc), while
distances of the open clusters studied in this work are between
100\,pc and 15\,kpc, with 50\,\% of the open cluster having $d \gtrsim
3$\,kpc.

\begin{figure*}
\includegraphics[bb=16 222 559 572, width=0.6\textwidth]{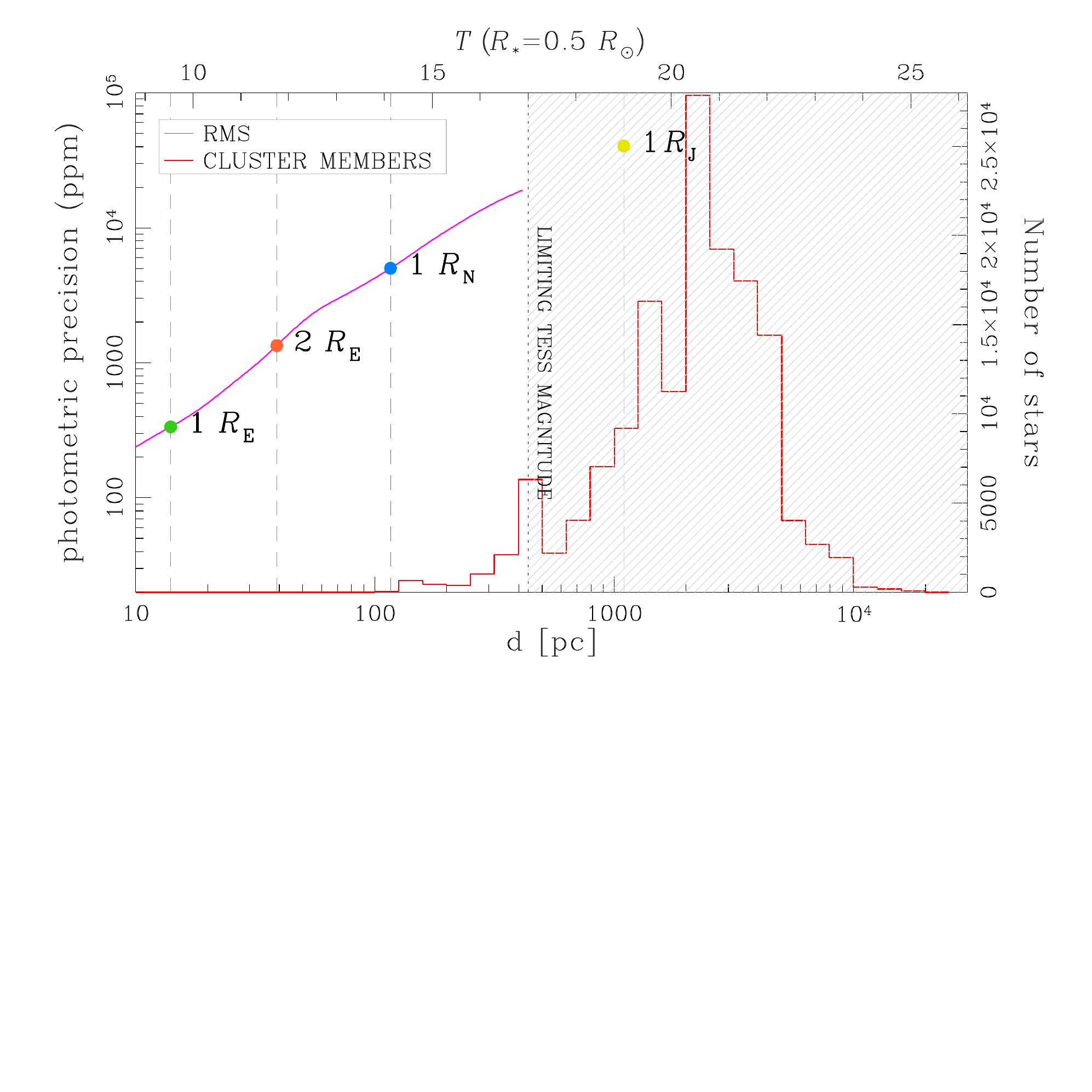}
\includegraphics[bb=16 222 559 572, width=0.6\textwidth]{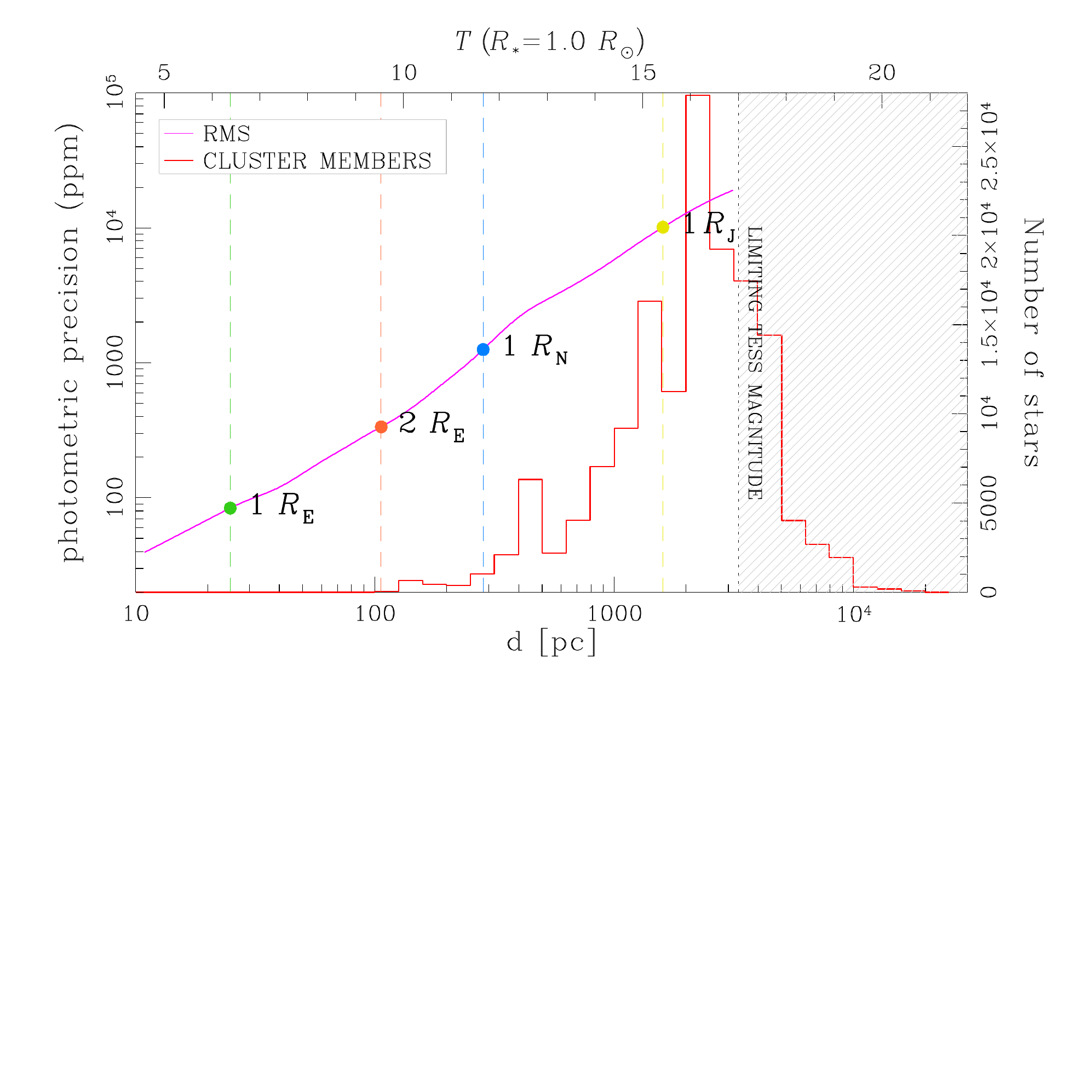}
\includegraphics[bb=16 222 559 572, width=0.6\textwidth]{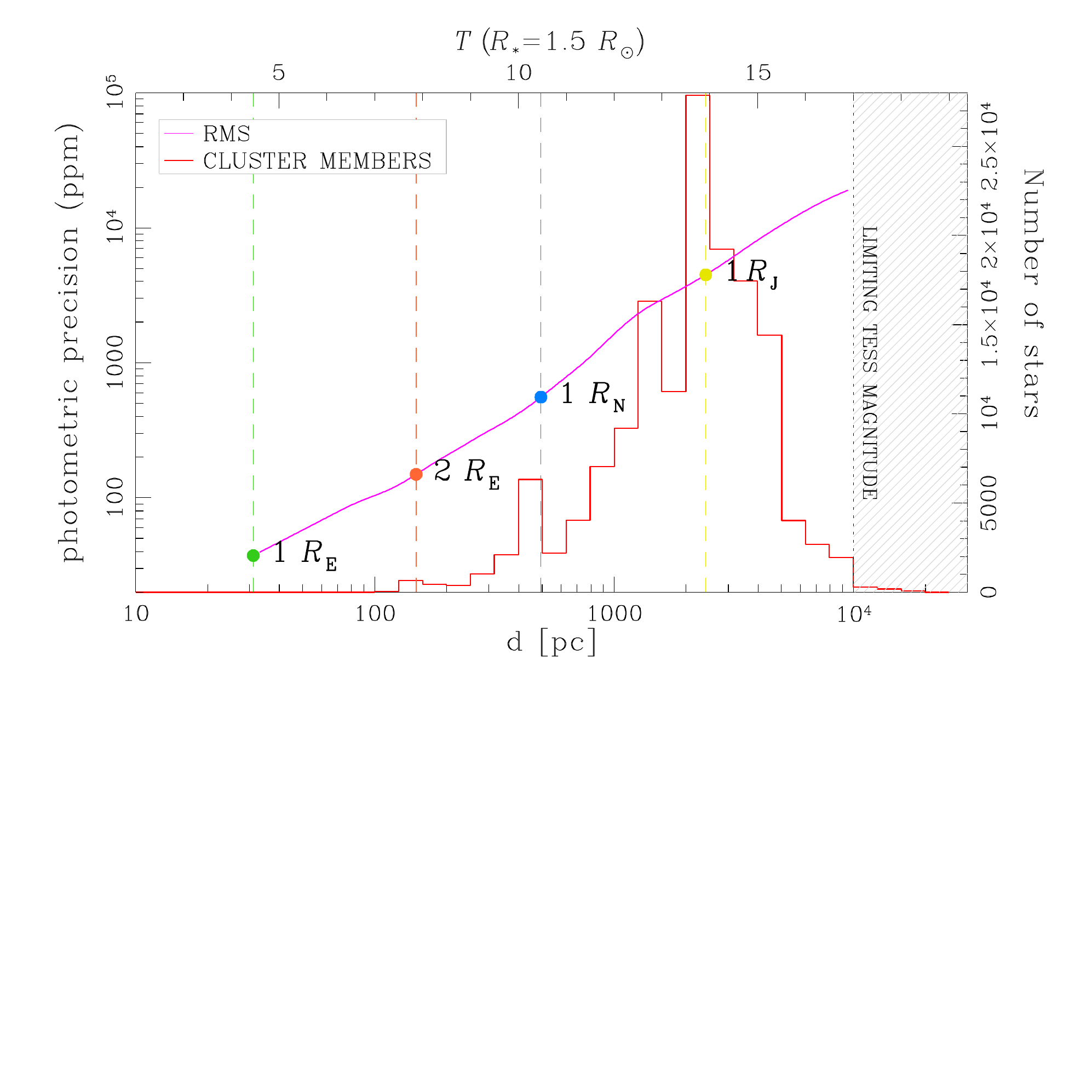}
\caption{Detectability limits of transits of exoplanets of size
  $R_{P}=1\,R_{\rm E}$ (green dot), $R_{P}=2\,R_{\rm E}$ (orange dot),
  $R_{P}=1\,R_{\rm N}$ (light blue dot), and $R_{P}=1\,R_{\rm J}$
  (yellow dot) orbiting stars with radius $R_{\star}=0.5\,R_{\sun}$
  (top panel), $R_{\star}=1.0\,R_{\sun}$ (middle panel), and
  $R_{\star}=1.5\,R_{\sun}$ (bottom panel). Magenta line represents
  the average \texttt{RMS} distribution as a function of the magnitude
  obtained from flattened light curves. Red histogram is the distance
  distribution of the cluster members studied in this work. Grey zone
  represents the range of magnitudes outside the range of detection of
  {\it TESS}. The upper axis of each panel gives the apparent
  magnitude of a star with $R_{\star}=X\,R_{\sun}$, with
  $X=0.5,\,1.0,\,1.5$, at the corresponding distance of the lower
  axis. 
  \label{fig:9}}
\end{figure*}

Figure~\ref{fig:9} shows how the distance of the open clusters affects
our results. We analysed what kind of planets we are able to detect in
the light curves of stars having a radius $R_{\star}=0.5\,R_{\sun}$
(top panel), $R_{\star}=1.0\,R_{\sun}$ (middle panel), and
$R_{\star}=1.5\,R_{\sun}$ (bottom panel), and located at distances
between 10 and 15\,000\,pc. Using isochrones, we calculated the
average absolute {\it TESS} magnitude of these stars and then we
calculated their apparent magnitudes for different distances. The
upper axis of each panel illustrates the apparent magnitude $T$ of a
star at the corresponding distance of the lower axis. The grey zone
represents the magnitude range out of reach of {\it TESS}.  Red
histogram is the distribution of the distances of the stars in the
clusters studied in this work: the peak of the distribution is at $d
\sim 3000$\,pc. Magenta line represents the mean distribution of the
\texttt{RMS} calculated using all the analysed flattened light curves;
the \texttt{RMS} gives us an idea about the detectability of an
exoplanet of radius $R_{\rm p}$ hosted by a star of radius $R_{\star}$
located at distance $d$.

\begin{figure}
\includegraphics[bb=1 180 309 555, width=0.45\textwidth]{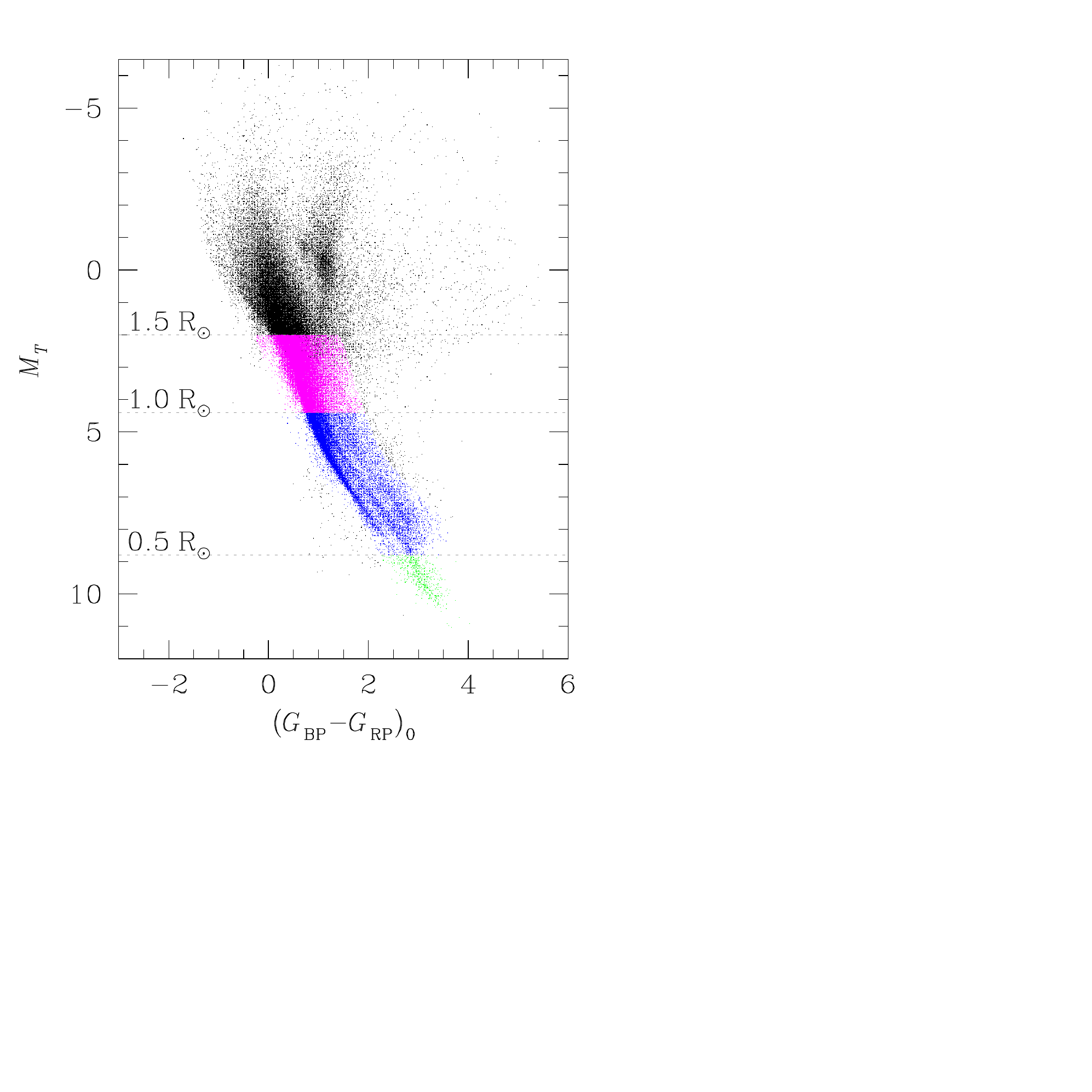}
\caption{The $M_{T}$ versus $(G_{\rm BP}-G_{\rm RP})_0$ CMD for the analysed stars in this work. Green, blue, and magenta points are the main sequence stars in the radii intervals   $R_{\star}\lesssim 0.5\,R_{\sun}$, $0.5\,R_{\sun} \lesssim
  R_{\star}\lesssim 1.0\,R_{\sun}$ and $1.0\,R_{\sun} \lesssim
  R_{\star}\lesssim 1.5\,R_{\sun}$, respectively.
  \label{fig:10}}
\end{figure}

We analysed four different types of exoplanets orbiting stars into the
three $R_\star$ groups: Earth-size planet ($R_{\rm p}=1\,R_{\rm E}$,
green dot in Fig.~\ref{fig:9}), super-Earth-size planet ($R_{\rm
  p}=2\,R_{\rm E}$, orange dot in Fig.~\ref{fig:9}), Neptune-size
planet ($R_{\rm p}=1\,R_{\rm N}$, light blue dot in Fig.~\ref{fig:9}),
and Jupiter-size planet ($R_{\rm p}=1\,R_{\rm J}$, yellow dot in
Fig.~\ref{fig:9}). We calculated the expected photometric decrement in
the light curve $\delta_{\rm phot} = (R_{\rm P}/R_{\star})^2$ due to
the exoplanet transit and, by using the photometric precision
distribution, we calculated the maximum distance at which the transit
of a planet of radius $R_{\rm P}$ is still detectable (at 1$\sigma$ )
in the light curve of a star with radius $R_{\star}$.  Finally, we
used this distance, combined with the CMD $M_{T}$ versus $(G_{\rm
  BP}-G_{\rm RP})_0$, to extract the number ($N_{\star}$) of cluster
members having radius $R_{\star}$ for which this kind of transits can
be detected in their light curves. This CMD is shown in
Fig.~\ref{fig:10}.  For the reddening correction, we used the
tridimensional extinction map of \citet{2018A&A...616A.132L} in
conjunction with \citet{2018AJ....156...58B} distances to estimate the
amount of interstellar reddening to the target stars.  We evaluated
the reddening $E(B-V)$ on a sample of $N$, equally spaced, nodal
points along each target direction by considering the weighted average
of the 27 surrounding voxels reddening ($3 \times 3 \times 3$ voxels
cube around the nodal point).  The reddening was weighted by the
distance of the nodal point from each one of the neighbouring voxels'
centers while the distance between two adjacent nodal points was
2\,pc. We therefore summed up the reddening along the boresight to
derive the integrated reddening to the target stars. For stars falling
outside the map we added a correction to account for the extinction
from the edge of the map to the target position. The correction was
calculated using the dust model described in
\citet{2014MNRAS.437..351B}. The entire procedure will be further
described in Montalto et al.~(in preparation).  For each cluster, we
calculated the 3.5$\sigma$-clipped mean $E(B-V)$ value and we used the
equations reported by \citet{2019AJ....158..138S} to correct the
colours and the magnitudes of the stars.\\ We selected the main
sequence stars as follows:  first, we excluded the evolved stars
  of each open cluster visually inspecting each CMD. In a second step,
  we determined the fiducial line of all the main sequence stars by
using the naive estimator (\citealt{1986desd.book.....S}, see
\citealt{2015MNRAS.451..312N} for a detailed description of the
method), and we selected all the stars whose colours are within
$2\,\sigma$ from the mean colour of the fiducial line. Green, blue,
and magenta points in Fig.~\ref{fig:10} are the main sequence stars
with radii $R_{\star}\lesssim 0.5\,R_{\sun}$, $0.5\,R_{\sun} \lesssim
R_{\star}\lesssim 1.0\,R_{\sun}$ and $1.0\,R_{\sun} \lesssim
R_{\star}\lesssim 1.5\,R_{\sun}$, respectively, and that will be used
in the following analysis.

The three panels of Fig.~\ref{fig:9} show that we are not able to
detect transiting exoplanets with $R_{\rm P} \leq 1\,R_{\rm E}$ around
cluster members with $R_{\star}\gtrsim 0.5\,R_{\sun}$: indeed, in
order to be able to detect this kind of exoplanets it would be
necessary that the hosting star is at a distance $d \lesssim 30$\,pc,
but no open clusters studied in this work satisfies this condition. In
the case of super-Earths ($R_{\rm P} \lesssim 2\,R_{\rm E}$), we have no possibility to
find them around cluster members with $R_{\star}<0.5\,R_{\sun}$, but
we are able to detect this kind of exoplanets in the light curves of
stars with $0.5\,R_{\sun} \lesssim R_{\star}\lesssim 1.5\,R_{\sun}$
and located at $d \lesssim 150$\,pc. For the two intervals of stellar
radii $0.5\,R_{\sun} \lesssim R_{\star}\lesssim 1.0\,R_{\sun}$ and
$1.0\,R_{\sun} \lesssim R_{\star}\lesssim 1.5\,R_{\sun}$, from the CMD
of Fig.~\ref{fig:10} combined with information on the distance from
Fig.~\ref{fig:9}, we extracted a total of $N_{\star}=23 \pm 5$ and
$N_{\star}=20 \pm 4$ stars with these radii, respectively.  In our
survey we detected no super-Earth transits. This result is in
agreement with the frequency of field exoplanets from the {\it Kepler}
survey, as tabulated by \citet{2013ApJ...766...81F}. \\ Following
\citet{2018AJ....155..173C}, we also calculated the number of expected
transiting super-Earths in our sample, by using the equation
  \begin{equation}
  N_{\rm planet} = N_{\star} \times f_{\star}\times {\rm Pr}_{\rm  transit}
  \label{eq:2}
\end{equation}
where $f_{\star}$=4.82\,\% is the percentage of stars with at least
one exoplanet for the period range 0.8-10\,d estimated by
\citet{2013ApJ...766...81F}, ${\rm Pr}_{\rm transit} \simeq
R_{\star}/a$ is the transit probability, and $a$ is calculated
assuming an average period $P=10$\,d. The result of this estimate is
$N_{\rm planet} \simeq 0$ for stars with $R_{\rm \star}=1.0\,R_{\sun}$
and $R_{\rm \star}=1.5\,R_{\sun}$, consistent with our null
detection.

From Figs.~\ref{fig:9} and \ref{fig:10}, transiting exoplanets having
a radius $R_{\rm P} \sim 1\,R_{\rm N}$ and orbiting stars with
$R_{\star} \lesssim 0.5\,R_{\sun}$ can be detected in the light curve
of $N_{\star} = 23 \pm 5$ stars.  \citet{2013ApJ...766...81F}
estimated that $f_{\star}=2.61$\,\% of field stars host Neptune-size
exoplanets with periods $<10$\,d. Correspondingly, we expect a null
detection (0.02 planets) of Neptune-size exoplanets transiting low
mass main sequence stars that are members of the studied clusters.
\\ In the case the hosting star has radius $0.5\,R_{\sun} \lesssim
R_{\star}\lesssim 1.0\,R_{\sun}$ and $1.0\,R_{\sun} \lesssim
R_{\star}\lesssim 1.5\,R_{\sun}$, we are able to detect Neptune-size
transits in the light curves of $N_{\star}=1086 \pm 33$ and
$N_{\star}=1868 \pm 43$ stars. We found 1 Neptune-size exoplanet
around a $R_{\star}\sim 1.1\,R_{\sun}$ star (PATHOS-30), while
PATHOS-31 is a candidate Neptune orbiting a $R_{\star}\sim
0.8\,R_{\sun}$ star. By using these information, we calculated the
frequency of transiting Neptune exoplanets for stars with
$0.5\,R_{\sun} \lesssim R_{\star}\lesssim 1.5\,R_{\sun}$, $f_{\star}$,
from the equation~(\ref{eq:2}). We found that for $0.5\,R_{\sun}
\lesssim R_{\star}\lesssim 1.0\,R_{\sun}$, $f_{\star}=1.84\pm
1.84$\,\%, while for $1.0\,R_{\sun} \lesssim R_{\star}\lesssim
1.5\,R_{\sun}$ we obtained $f_{\star}=0.67\pm 0.67$\,\%; if we
consider all the stars with $R \lesssim 1.5\,R_{\sun}$ we have a
frequency of transiting Neptunes $f_{\star}=1.34\pm 0.95$\,\%, in
agreement with \citet{2013ApJ...766...81F}, who
found. $f_{\star}=2.61\pm 0.18$\,\%, within $\sim 1\sigma$.

In our survey, Jupiter-size exoplanets orbiting stars of radius
$R_{\star}\lesssim 0.5\,R_{\sun}$ can be detected in the light curves
of $N_{\star}=544 \pm 23$ cluster members, while, for stars with
$0.5\,R_{\sun} \lesssim R_{\star}\lesssim 1.0\,R_{\sun}$ and
$1.0\,R_{\sun} \lesssim R_{\star}\lesssim 1.5\,R_{\sun}$, we were able
to detect transits of this kind of exoplanets in the light curves of
$N_{\star}=17\,583 \pm 133$ and $N_{\star}=26\,776 \pm 164$ cluster
stars, respectively.  We did not detect any Jupiter-size exoplanet
around stars with $R_{\star}\lesssim 0.5\,R_{\sun}$; it is expected
assuming a frequency of giant planets ($\sim 0.43\,\%$) as in
\citet{2013ApJ...766...81F}.  We calculated the observed frequency of
transiting Jupiters in the other two ranges of stellar radii,
following the procedure already adopted for Neptune-size planets. We
found 1 giant exoplanets orbiting $0.5\,R_{\sun} \lesssim
R_{\star}\lesssim 1.0\,R_{\sun}$, with a corresponding frequency
$f_{\star}=0.11 \pm 0.11$\,\%; 6 Jupiter-size exoplanets were
detected for stars with radii $1.0\,R_{\sun} \lesssim
R_{\star}\lesssim 1.5\,R_{\sun}$, with a frequency of $f_{\star}=0.28
\pm 0.11$\,\%. Considering all the stars with $R_{\star}\lesssim
1.5\,R_{\sun}$ for which we are able to detect Jupiter-size transiting
exoplanets, we obtain a frequency $f_{\star}=0.19\pm0.07\,\%$, that is
significantly lower than that found by
\citet[$f_{\star}=(0.43\pm0.05)\,\%$]{2013ApJ...766...81F}.

We want to emphasise that, in the statistical analysis we performed,
there two effects that partially compensate each other and that can
affect our estimated frequencies: (i) our detection method might miss
some transit, i.e., the sample of detected candidate exoplanets might
be not complete. Including the completeness correction might result in a
higher value of $f_{\star}$; (ii) though we applied a validation check
to our targets, we could not identify all false positives, estimated
to be $\sim 40$\,\% by \citet{2015ApJ...809...77S}, and $\sim 33$\,\%
by \citet{2019AJ....158...81C}. Because of our false positive check,
the fraction of false positives in our sample is surely smaller, but
not null, and this fact would decrease $f_{\star}$.

\section{Summary and Conclusion}
\label{sec:conclusions}

In the second work of our project PATHOS, we applied our PSF-based
approach to {\it TESS} FFIs in order to extract high-precision light
curves of stars that belong to open clusters observed by {\it TESS}
during the first year of its mission (Sectors 1-13).  We extracted and
analysed 219\,256 light curves of 162\,901 stars located in 645 open
clusters (51\,475 stars are observed in more than one {\it TESS}
sector). These open clusters span a wide range of ages (from few tens
Myr up to $\sim 3$\,Gyr) and distances ($\sim $100 -- $10^4$\,pc),
allowing us to probe stars born at different times and in different
environmental conditions. The light curves will be publicly
  available as HLSP on the PATHOS project
  webpage\footnote{\url{https://archive.stsci.edu/hlsp/pathos}} (DOI:
  10.17909/t9-es7m-vw14) of the MAST archive.

We searched for transit signals among the extracted and corrected
light curves and, after a series of vetting tests, we isolated 33
transiting objects of interest. We extracted the physical parameters
of these objects modelling the transits, and using these information
combined with the stellar properties, we selected 11 candidate
exoplanets orbiting stars of 8 open clusters. One of the youngest open
clusters in our sample, IC\,2602 ($\sim 35$\,Myr), hosts two stars
with candidate Neptune-size exoplanets; while in the 250\,Myr old open
cluster NGC\,2516, we found two warm and one hot Jupiter
candidates. In the remaining clusters were we detected one transiting
candidate exoplanet in each of them: NGC\,2112 ($\sim 2$~Gyr),
NGC\,2323 ($\sim 100$\,Myr), NGC\,2437 ($\sim 220$\,Myr), NGC\,2527
($\sim 830$\,Myr), NGC\,2548 ($\sim 500$\,Myr), and NGC\,3532 ($\sim
400$\,Myr)

We also show that the planet detection is strongly affected by the
bias due to the distance of the open clusters. Comparing the mean
distribution of the photometric precision with the expected depth of
the transits due to exoplanets with different radii and the percentage
of stars that host different size exoplanets tabulated by
\citet{2013ApJ...766...81F}, we expected a null detection of Earth and
super-Earth size exoplanets in the light curves of the members of the
open clusters analysed in this work.

We detected 2 Neptune-size candidate exoplanets around two IC\,2602
members with radii $R_{\star} \lesssim 1.5 R_{\sun}$; consequently, we
estimated that the frequency of this kind of exoplanets is
$f_{\star}=1.34 \pm 0.95$, consistent with that of field stars.

  We also identified 7 Jupiters around stars with $R_{\star} \lesssim
  1.5R_{\sun}$ and estimated a fraction $f_{\star}=0.19 \pm 0.07\,\%$
  for this kind of planets, significantly smaller than what estimated
  for field stars (e.g., \citealt{2013ApJ...766...81F}). Two,
  partially compensating effects (completeness of our detection method
  and false positive rate), make our results still provisional.

The analysis of the light curves of open cluster members in the
northern ecliptic hemisphere, as also the analysis of cluster members
that will be observed during the {\it TESS} extended mission(s), will
be important for a better estimate of planet fraction in cluster stars.
Spectroscopic follow-up, when feasible, is mandatory as well, in order
to confirm the planetary nature and derive the exoplanet masses and
search for a correlation between planet composition and cluster
properties.

\section*{Acknowledgements}
DN acknowledges support from the French Centre National d'Etudes
Spatiales (CNES). DN and GP recognize partial support by UNIPD/DFA
Dipartimental project PIOT\_SID17\_01. GP, VG, LM, MM, DN, acknowledge
support from PLATO ASI-INAF agreements n.2015-019-R0-2015 and
n. 2015-019-R.1-2018. LB, GP, DN acknowledge the funding support from
CHEOPS ASI-INAF agreement n. 2019-29-HH.0.  LRB acknowledges support
by MIUR under PRIN program \#2017Z2HSMF.  The authors warmly thank the
anonymous referee for the prompt and careful reading of our
manuscript.


\bibliographystyle{mnras}
\bibliography{biblio}

\appendix

\section{Light curve modelling and exoplanet parameters estimate}
\label{app:cand}
\begin{figure*}
\includegraphics[bb=77 360 535 691, width=0.33\textwidth]{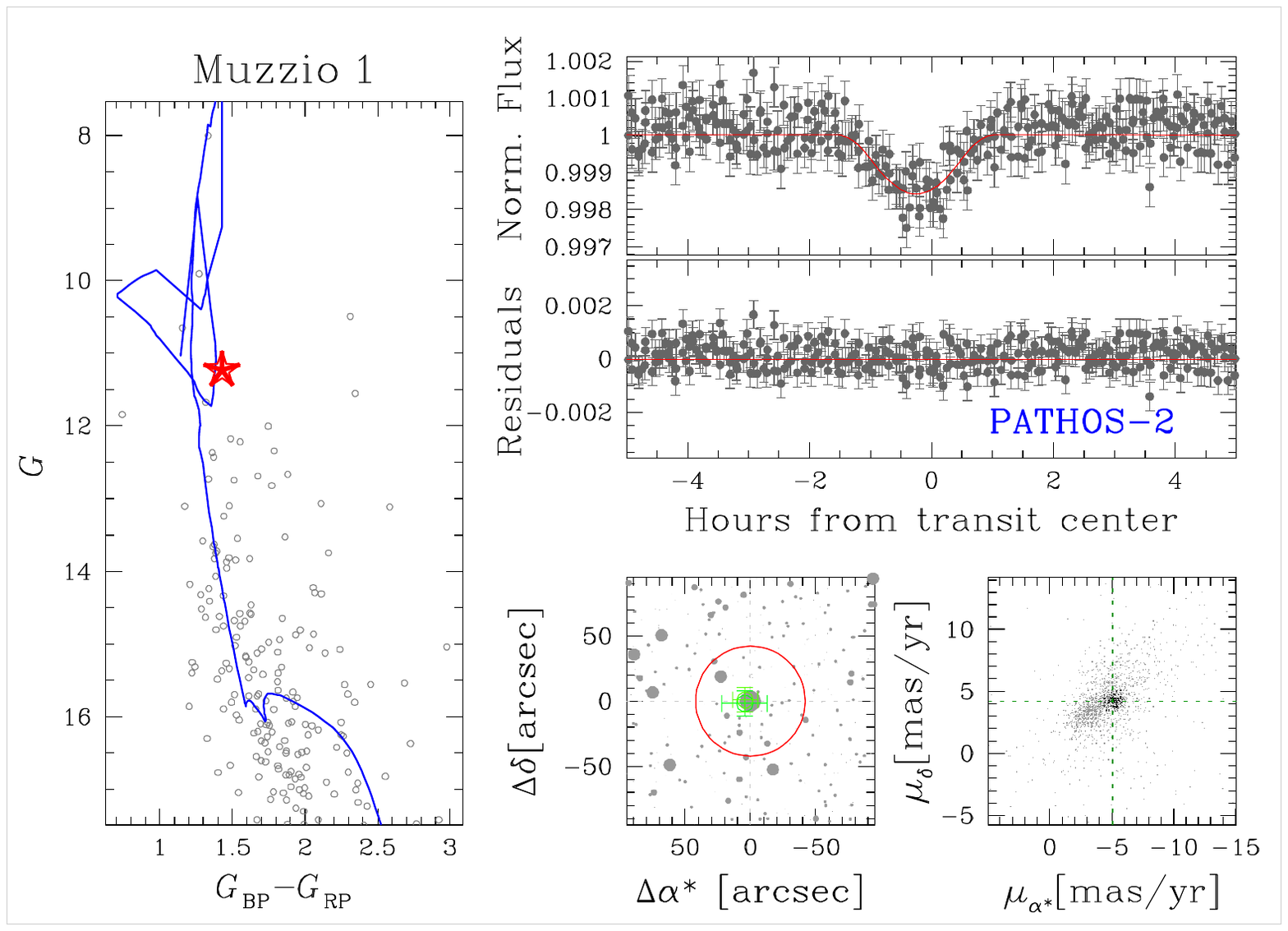}
\includegraphics[bb=77 360 535 691, width=0.33\textwidth]{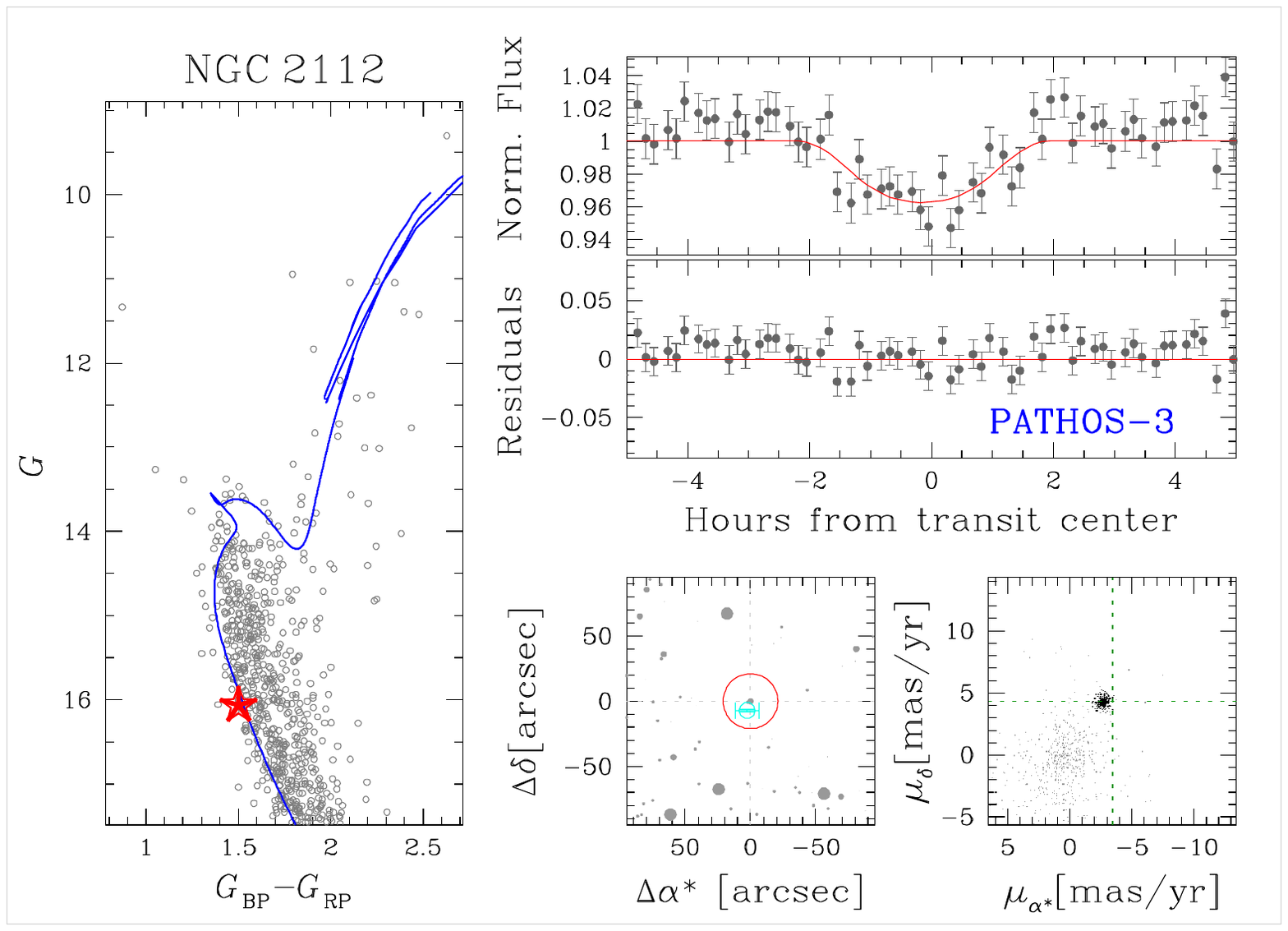}
\includegraphics[bb=77 360 535 691, width=0.33\textwidth]{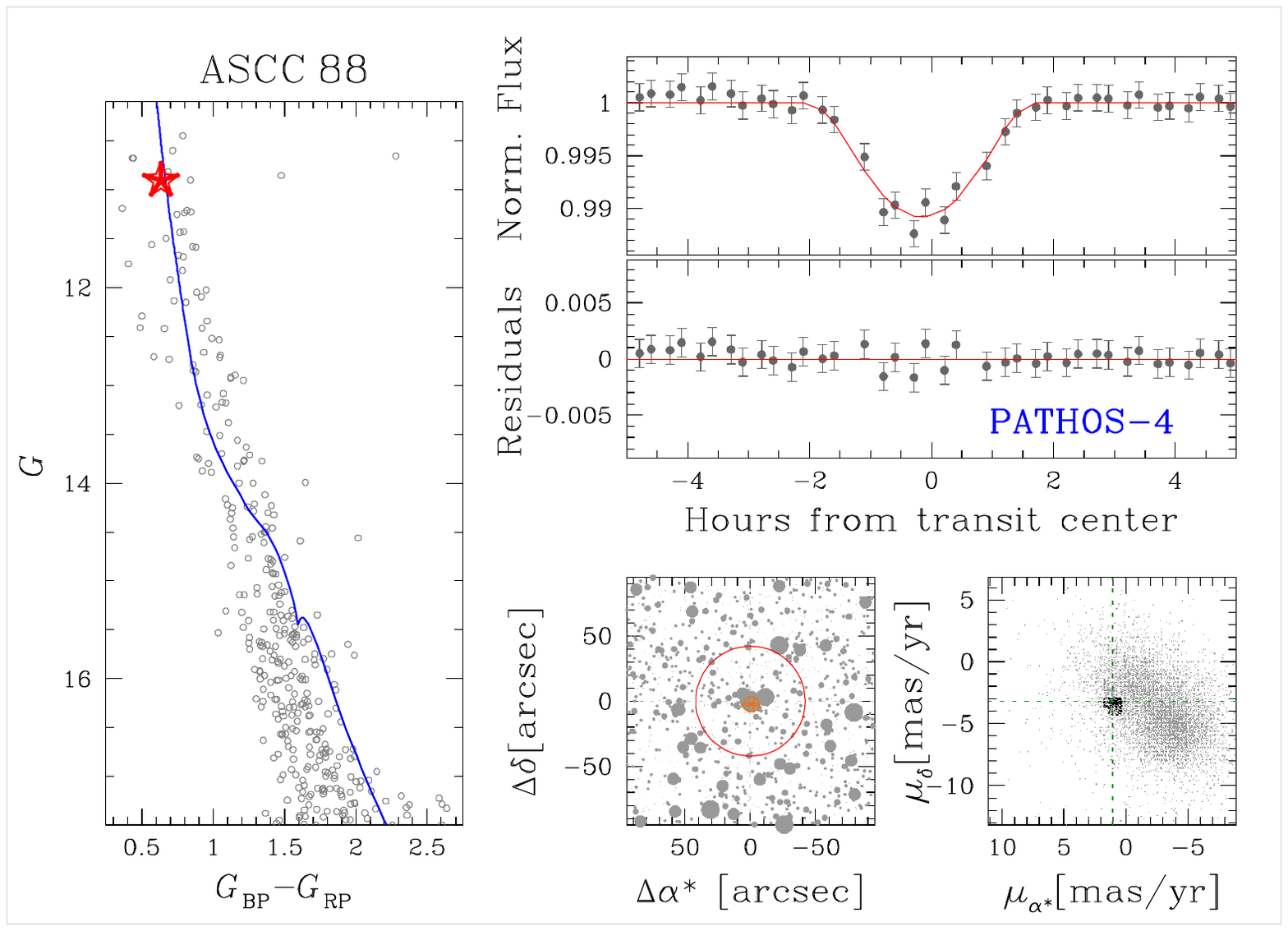} \\
\includegraphics[bb=77 360 535 691, width=0.33\textwidth]{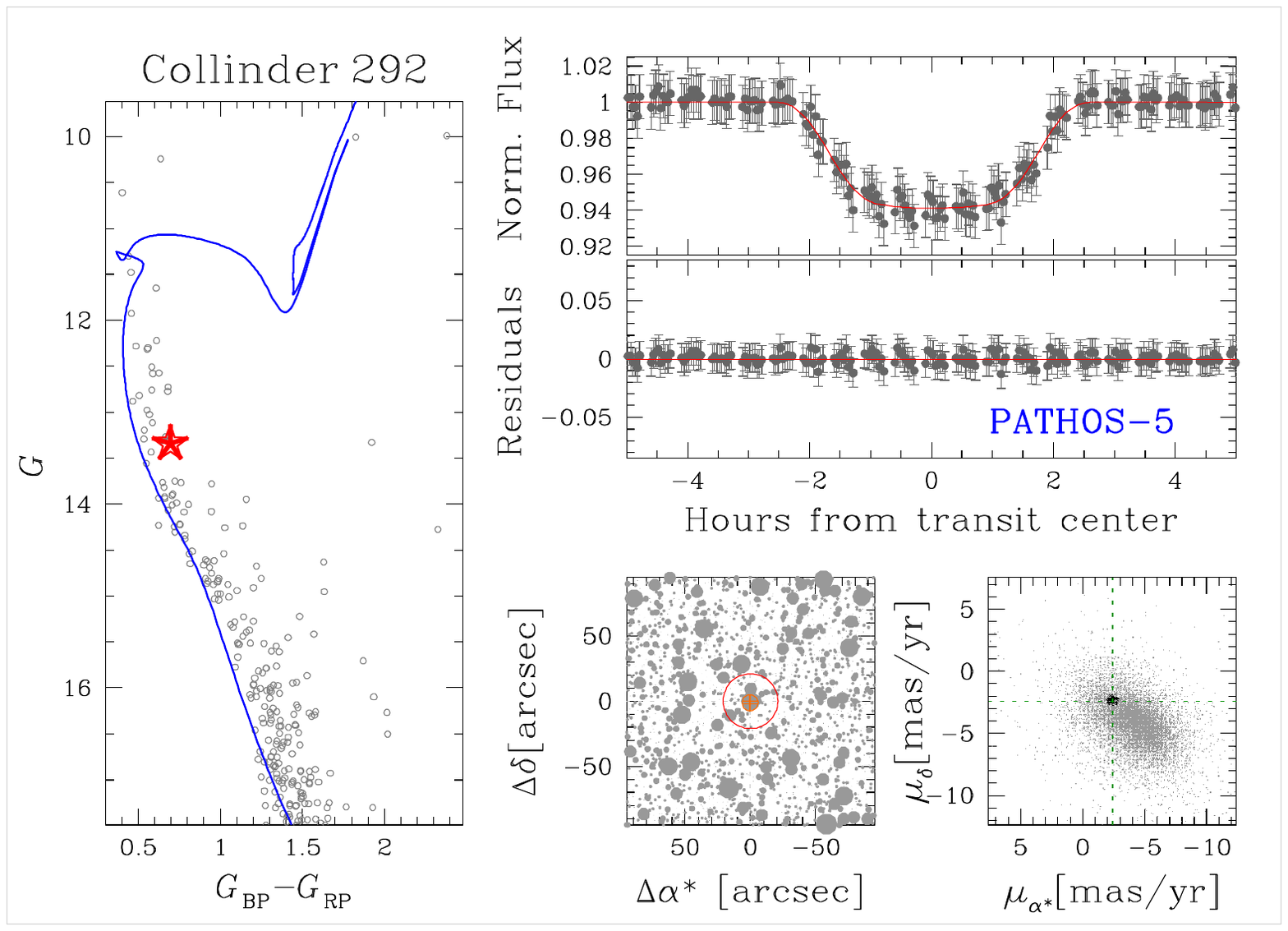}
\includegraphics[bb=77 360 535 691, width=0.33\textwidth]{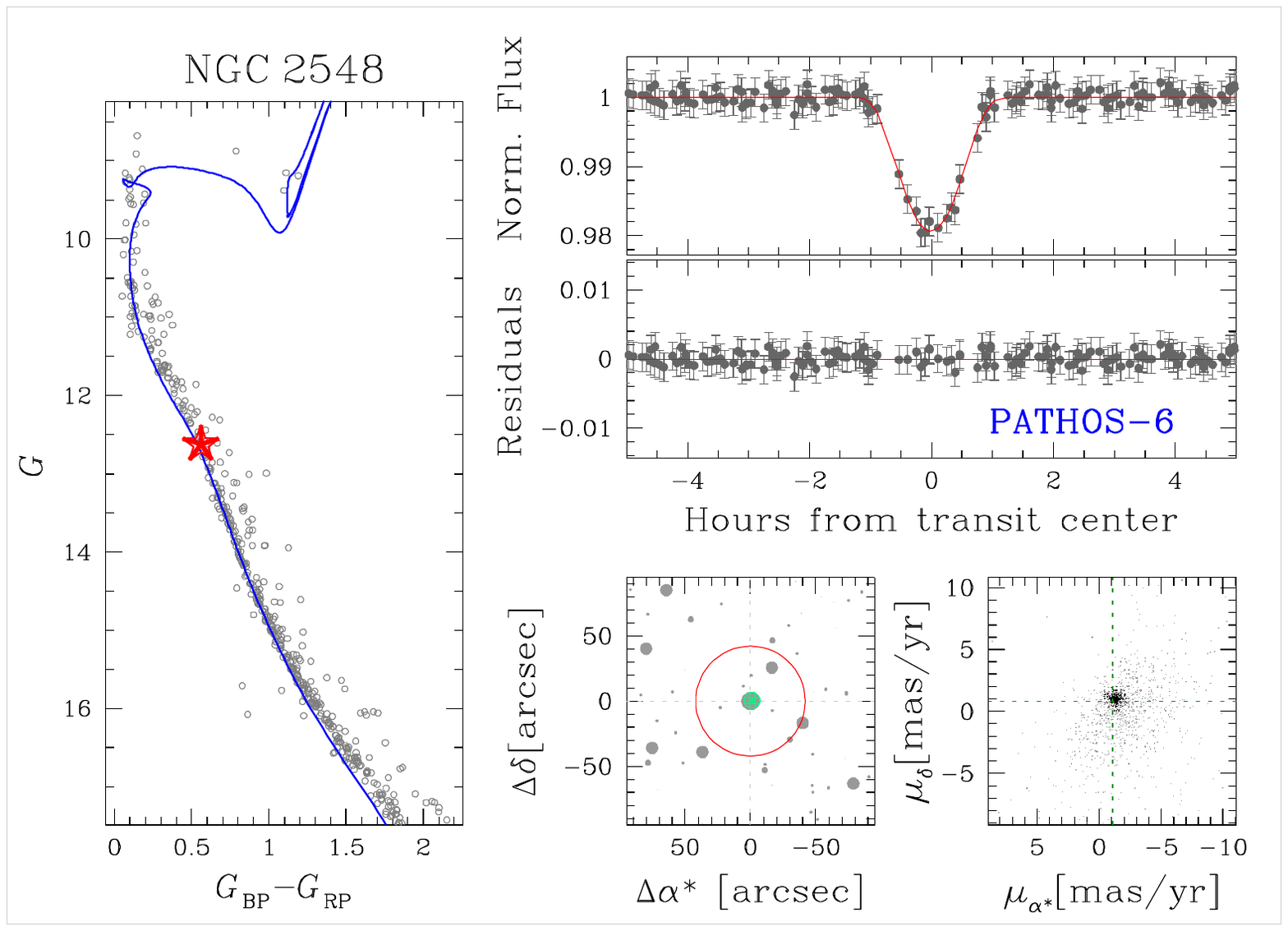}
\includegraphics[bb=77 360 535 691, width=0.33\textwidth]{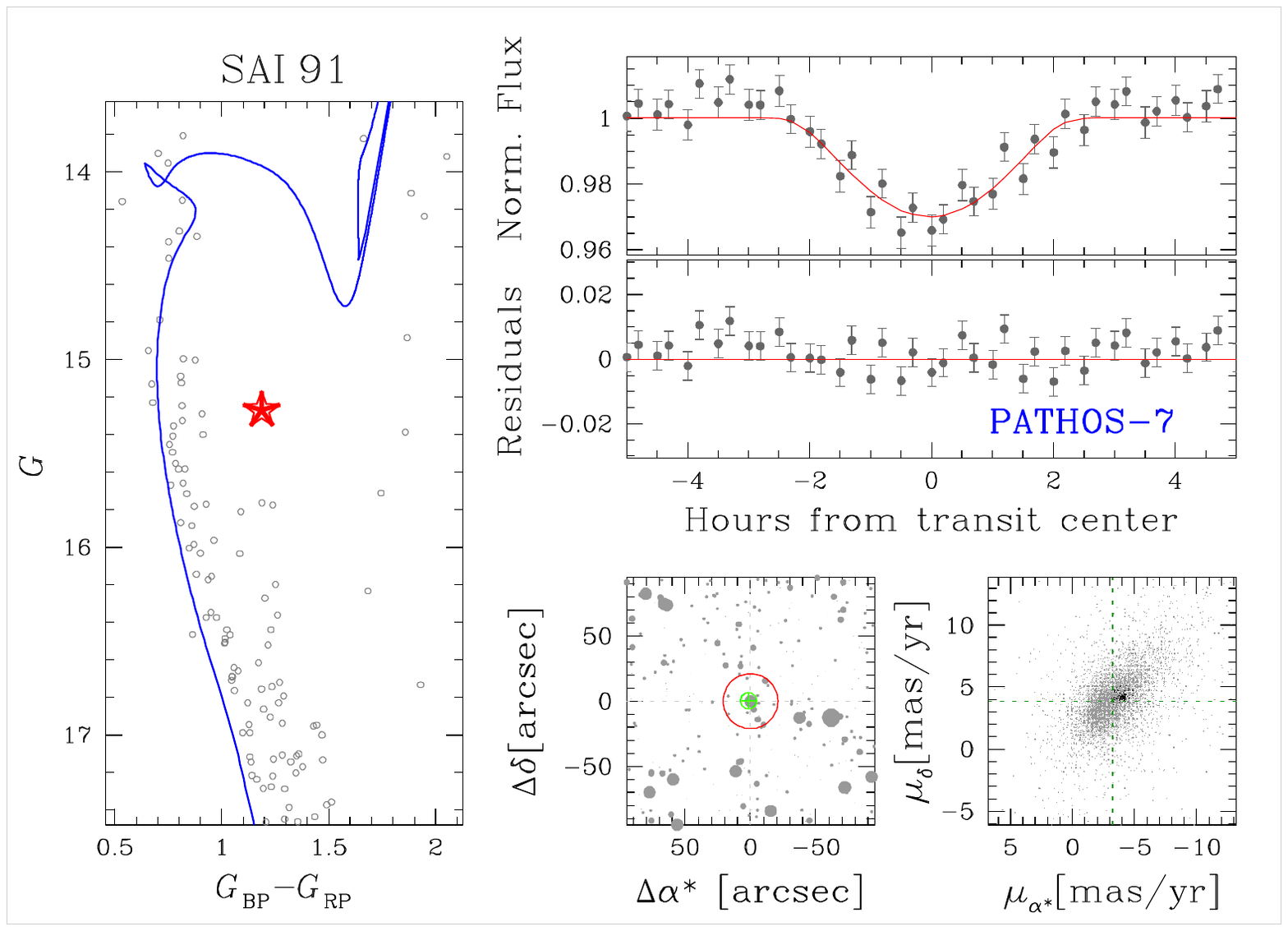} \\
\includegraphics[bb=77 360 535 691, width=0.33\textwidth]{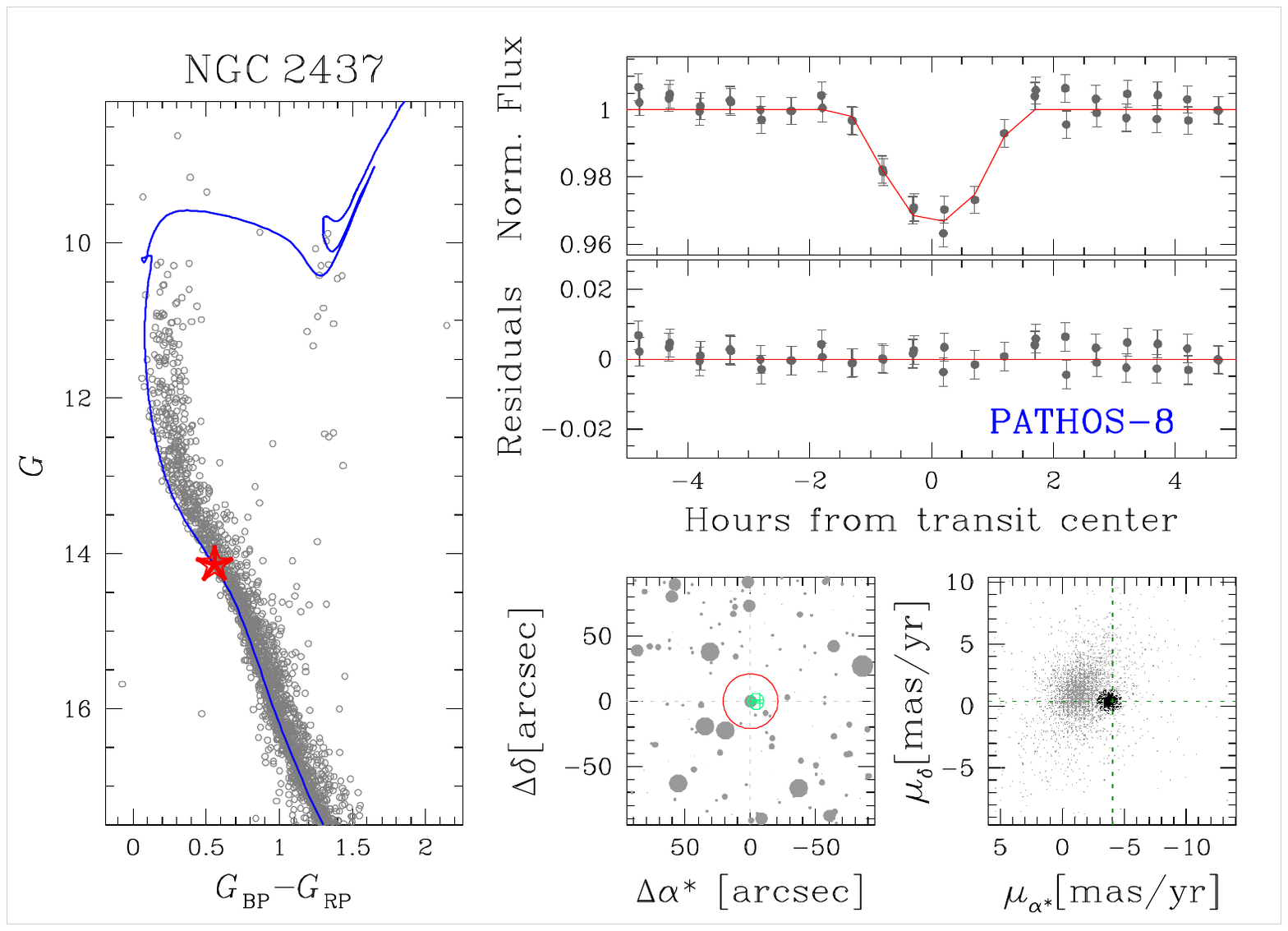}
\includegraphics[bb=77 360 535 691, width=0.33\textwidth]{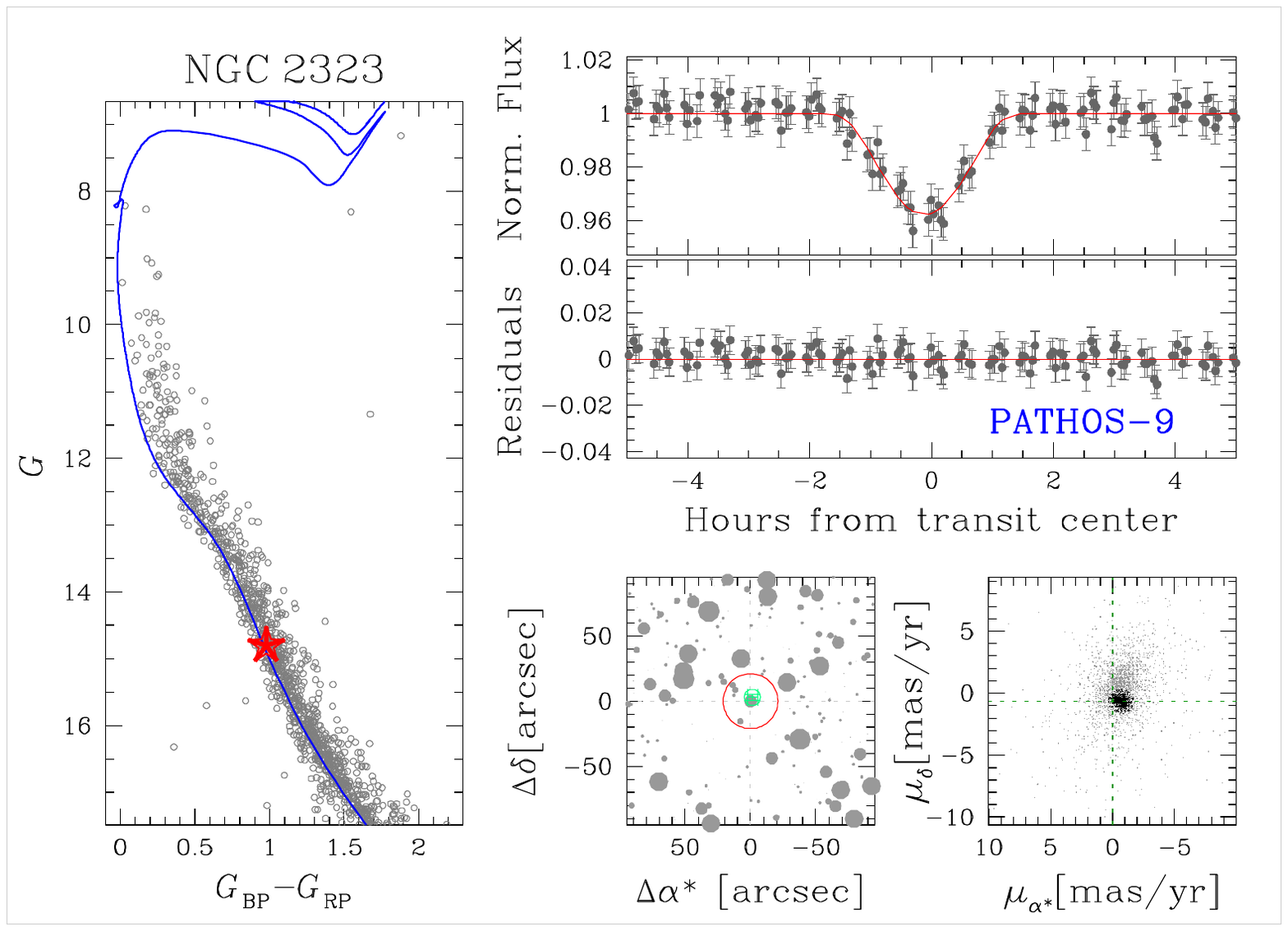}
\includegraphics[bb=77 360 535 691, width=0.33\textwidth]{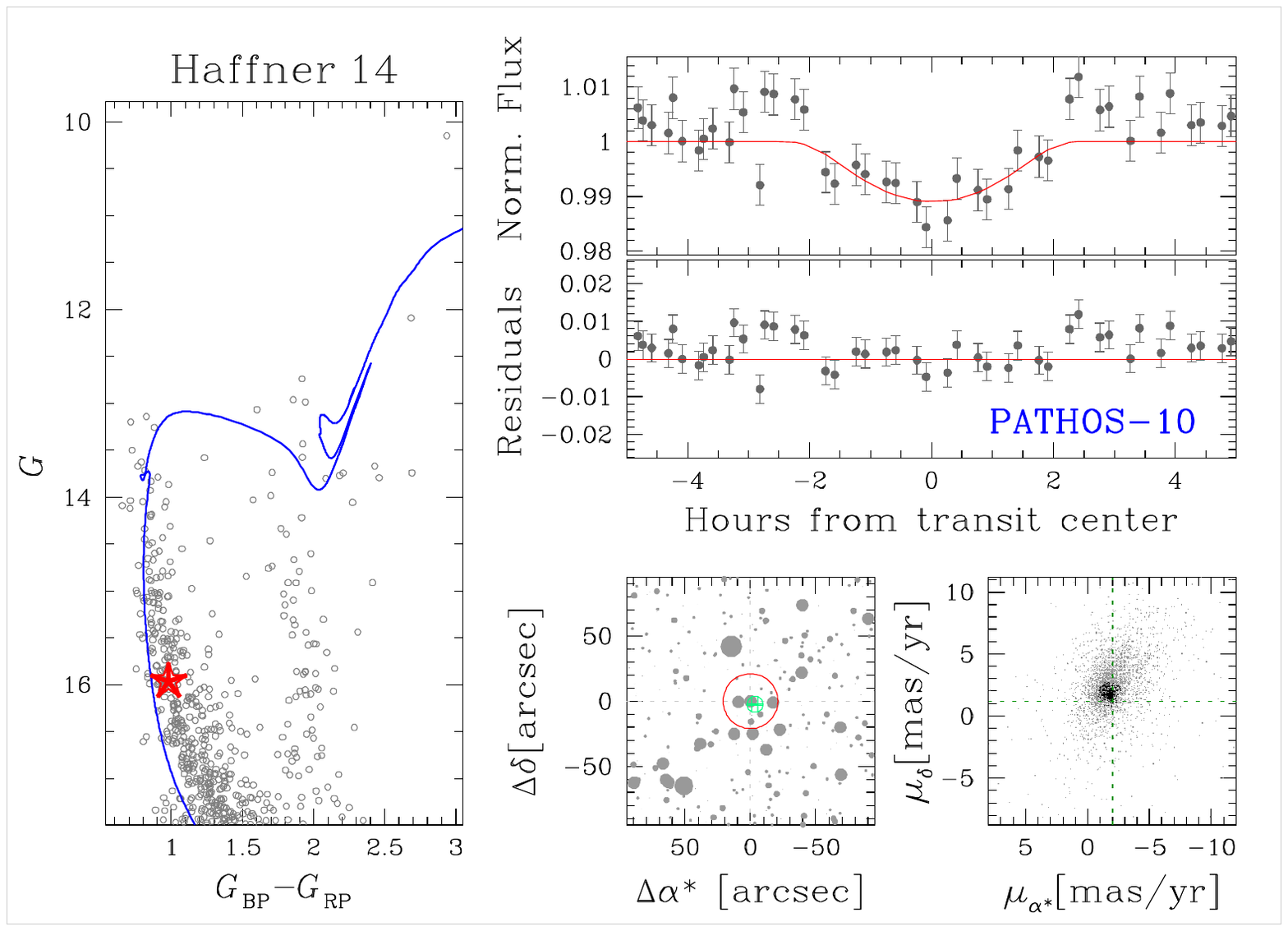} \\
\includegraphics[bb=77 360 535 691, width=0.33\textwidth]{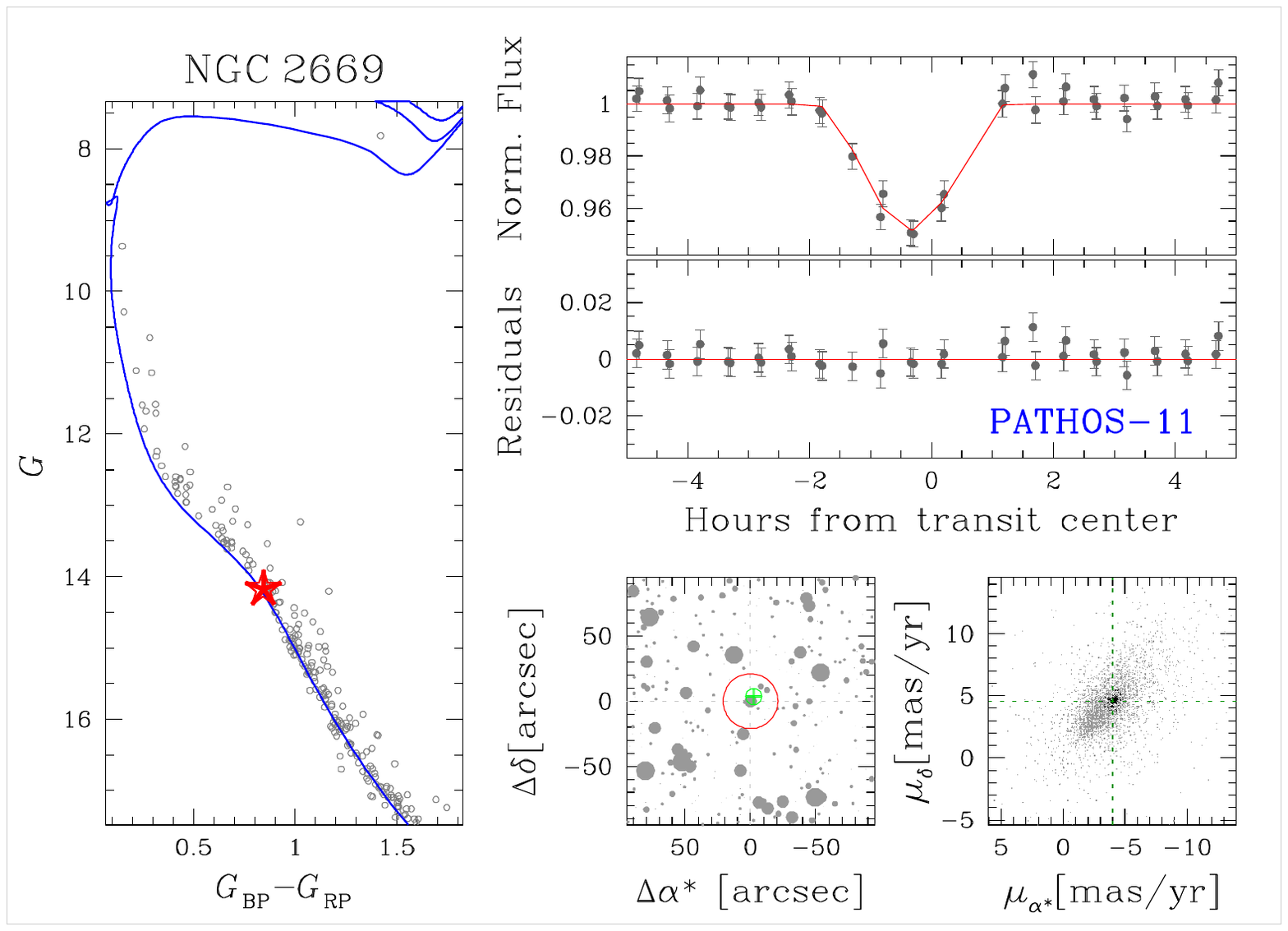}
\includegraphics[bb=77 360 535 691, width=0.33\textwidth]{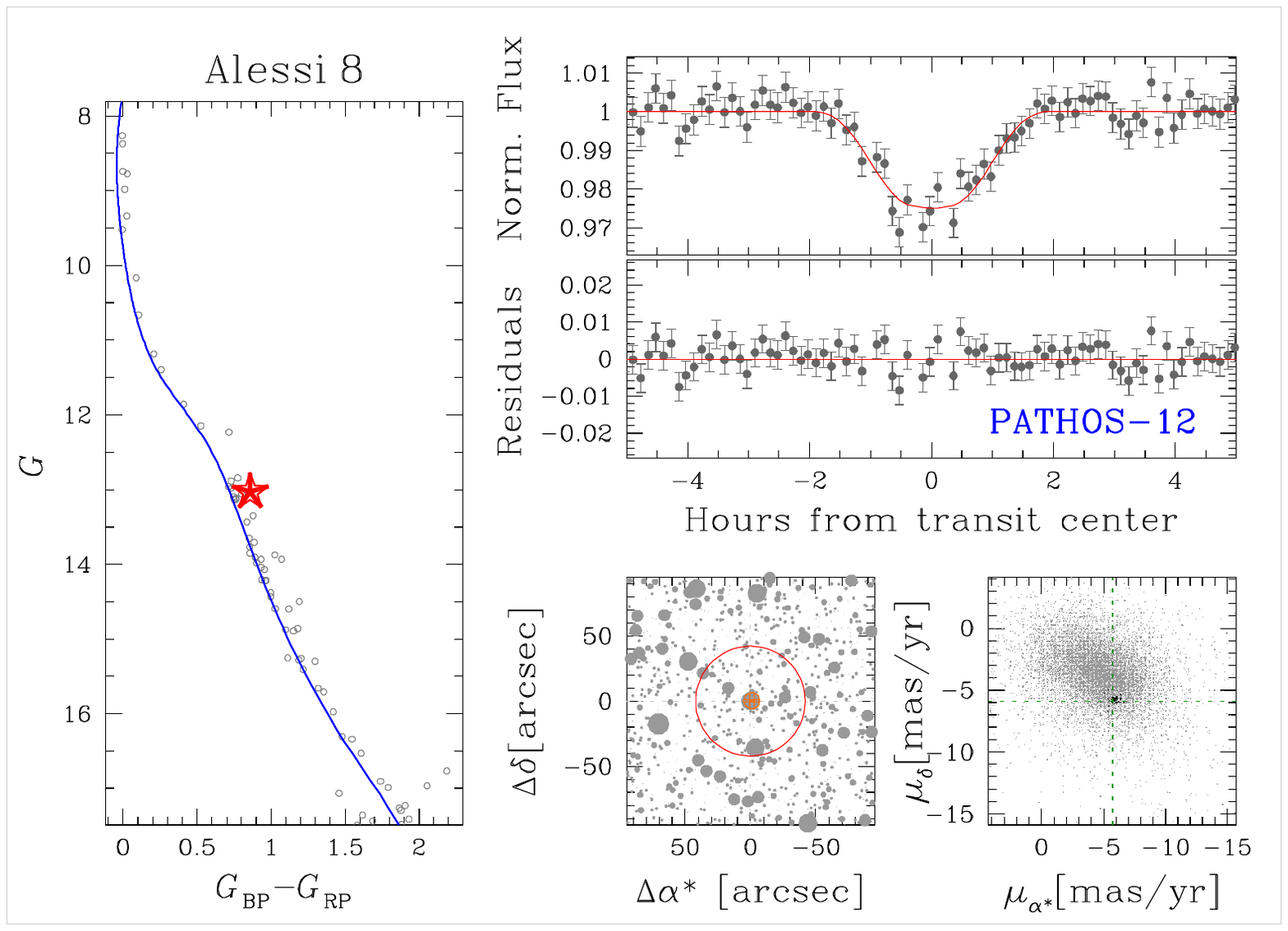}
\includegraphics[bb=77 360 535 691, width=0.33\textwidth]{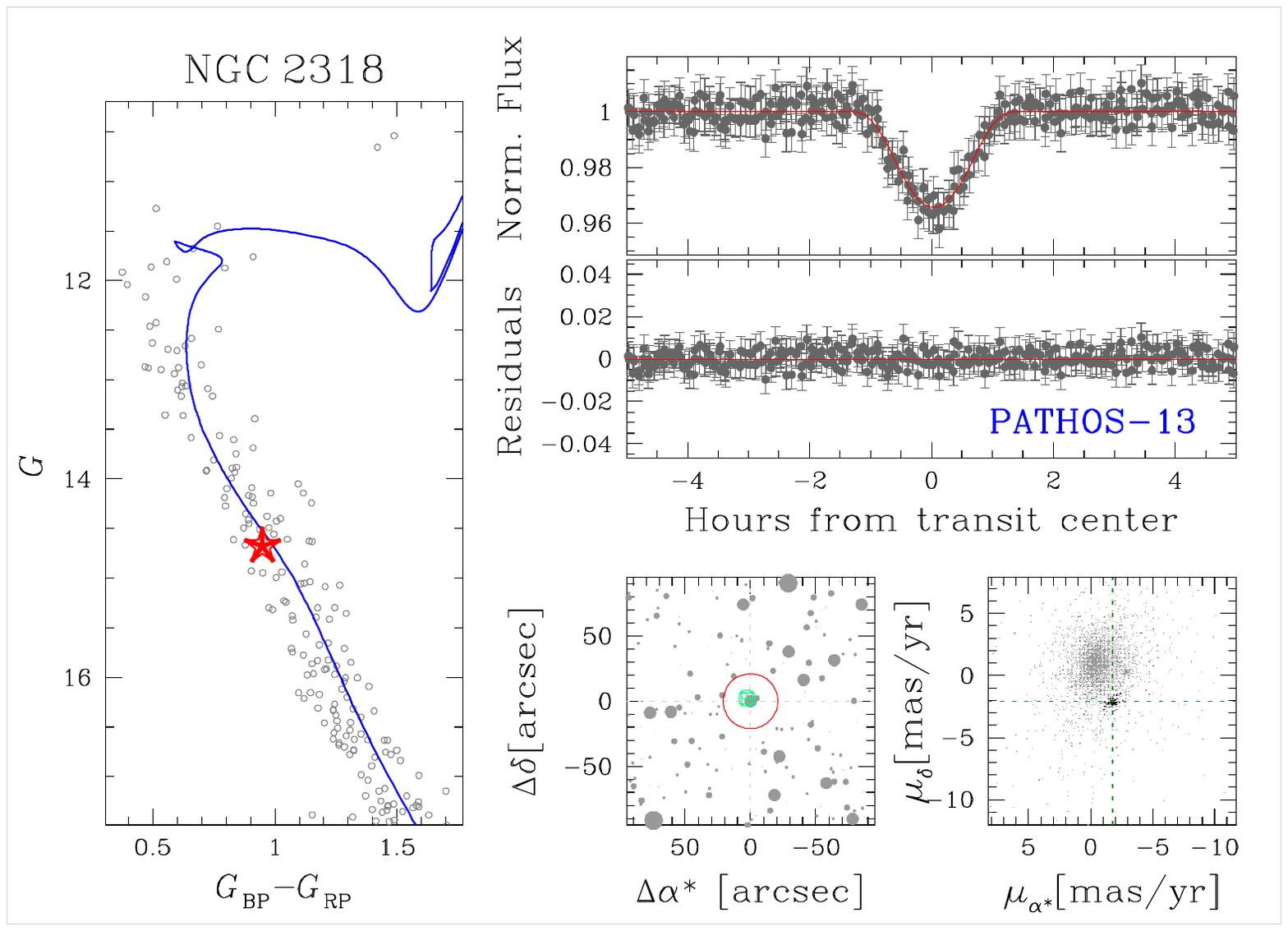} \\
\caption{Overview on the candidate exoplanets PATHOS-2 --
  PATHOS-13. On the left-hand, the $G$ versus $G_{\rm BP}-G_{\rm RP}$
  CMD of the cluster that hosts the target star (red star) and the
  isochrone (blue) fitted with the cluster parameters listed in
  Table~\ref{tab:1}. On the right-hand, top-panel shows the folded
  light curve (grey points) of the candidate and the model (in red)
  found with PyORBIT; middle panel shows the difference between the
  observed points and the model. Bottom left-hand panel shows the $95
  \times 95$\,arcsec$^2$ finding chart centred on the target star;
  red circle is the aperture adopted to extract photometry, crosses
  are the in-/out-of transit difference centroid, colour-coded as in
  Fig.~\ref{fig:6}. Bottom-right panel is the vector-point diagram,
  centred on the target star, for all the stars that are within
  10\,arcmin from the target star; black points are the cluster
  members listed in the catalogue by \citet{2018A&A...618A..93C}.
  \label{fig:7a}}
\end{figure*}
\begin{figure*}
\includegraphics[bb=77 360 535 691, width=0.33\textwidth]{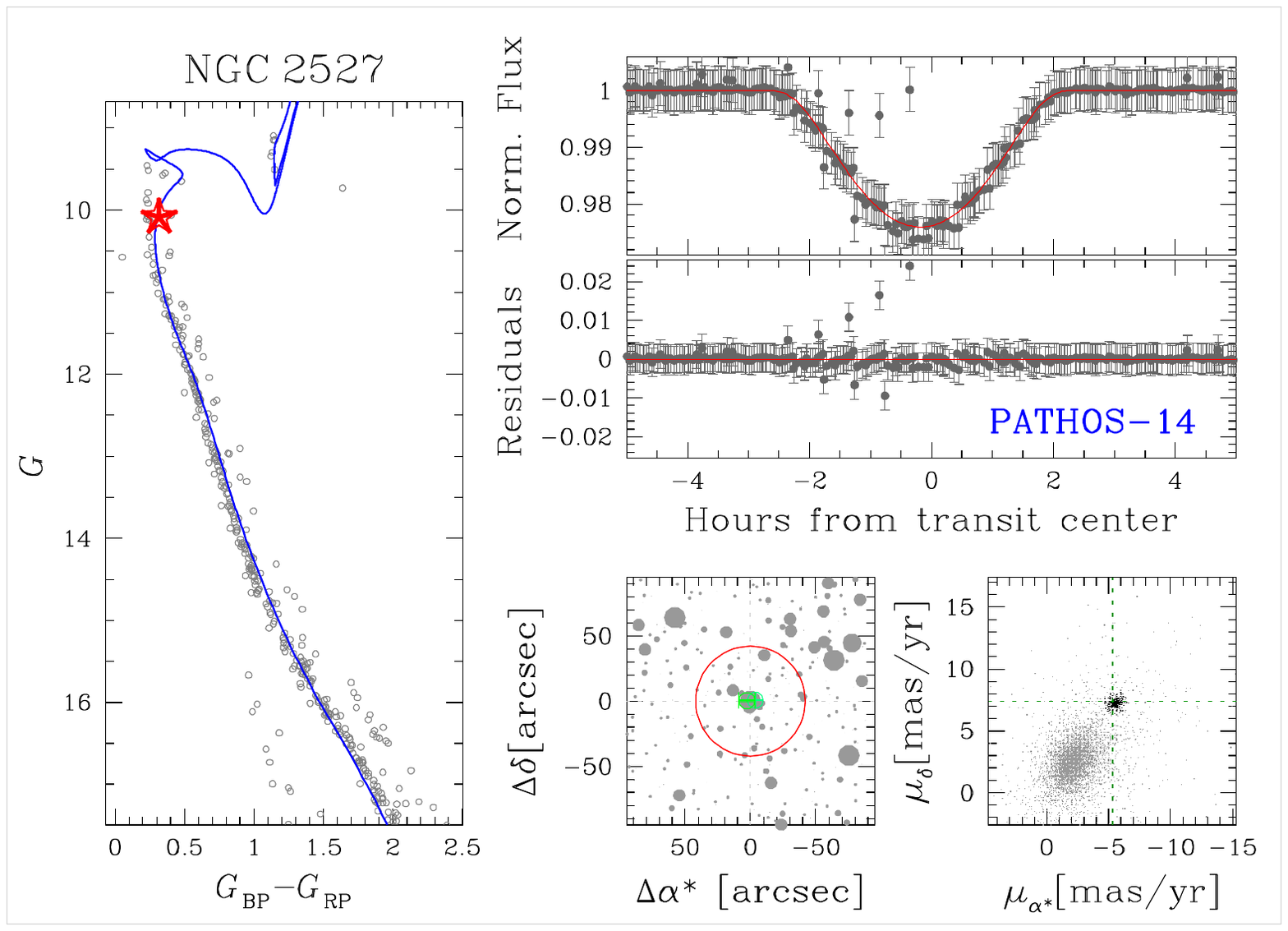}
\includegraphics[bb=77 360 535 691, width=0.33\textwidth]{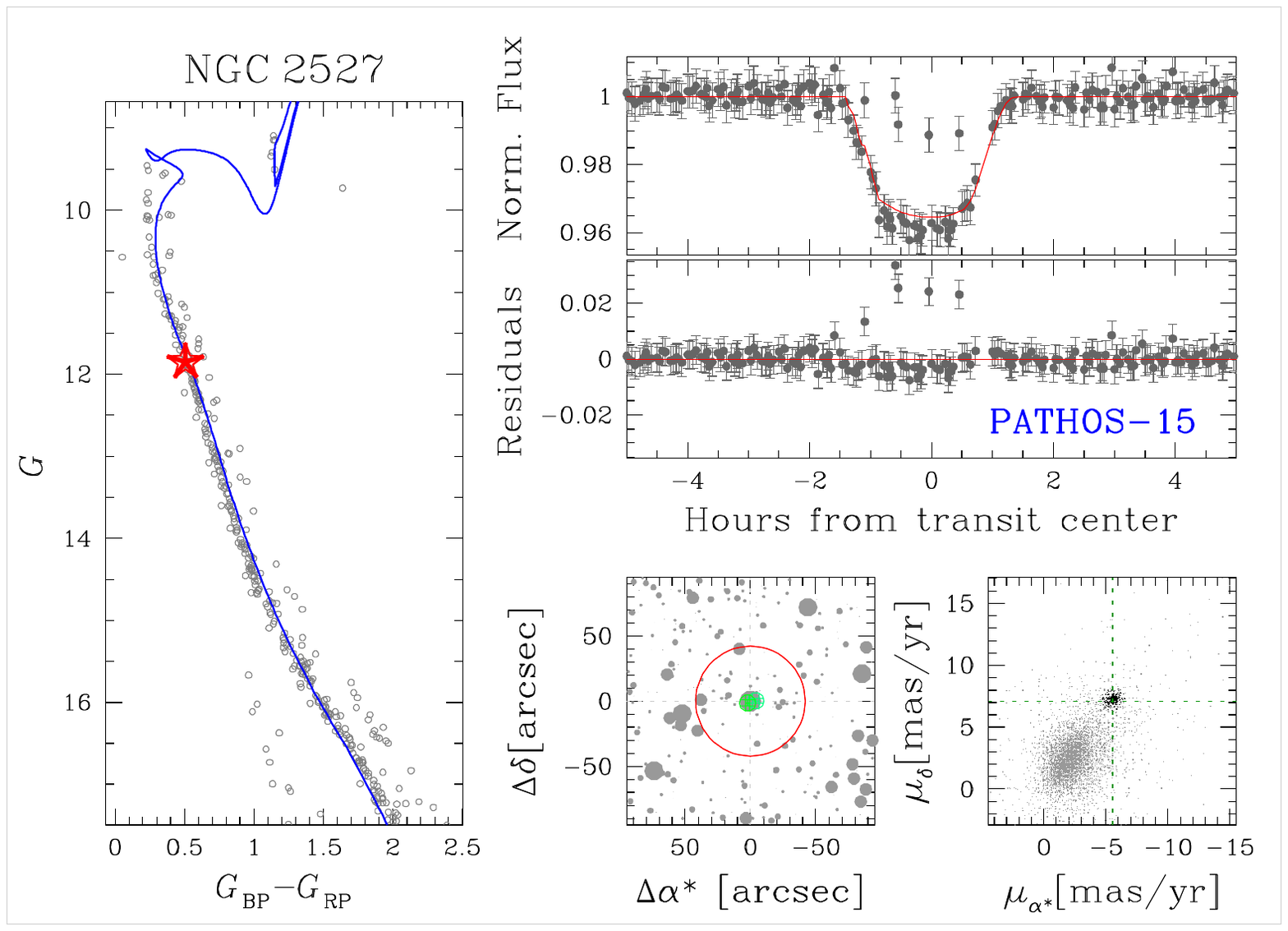}
\includegraphics[bb=77 360 535 691, width=0.33\textwidth]{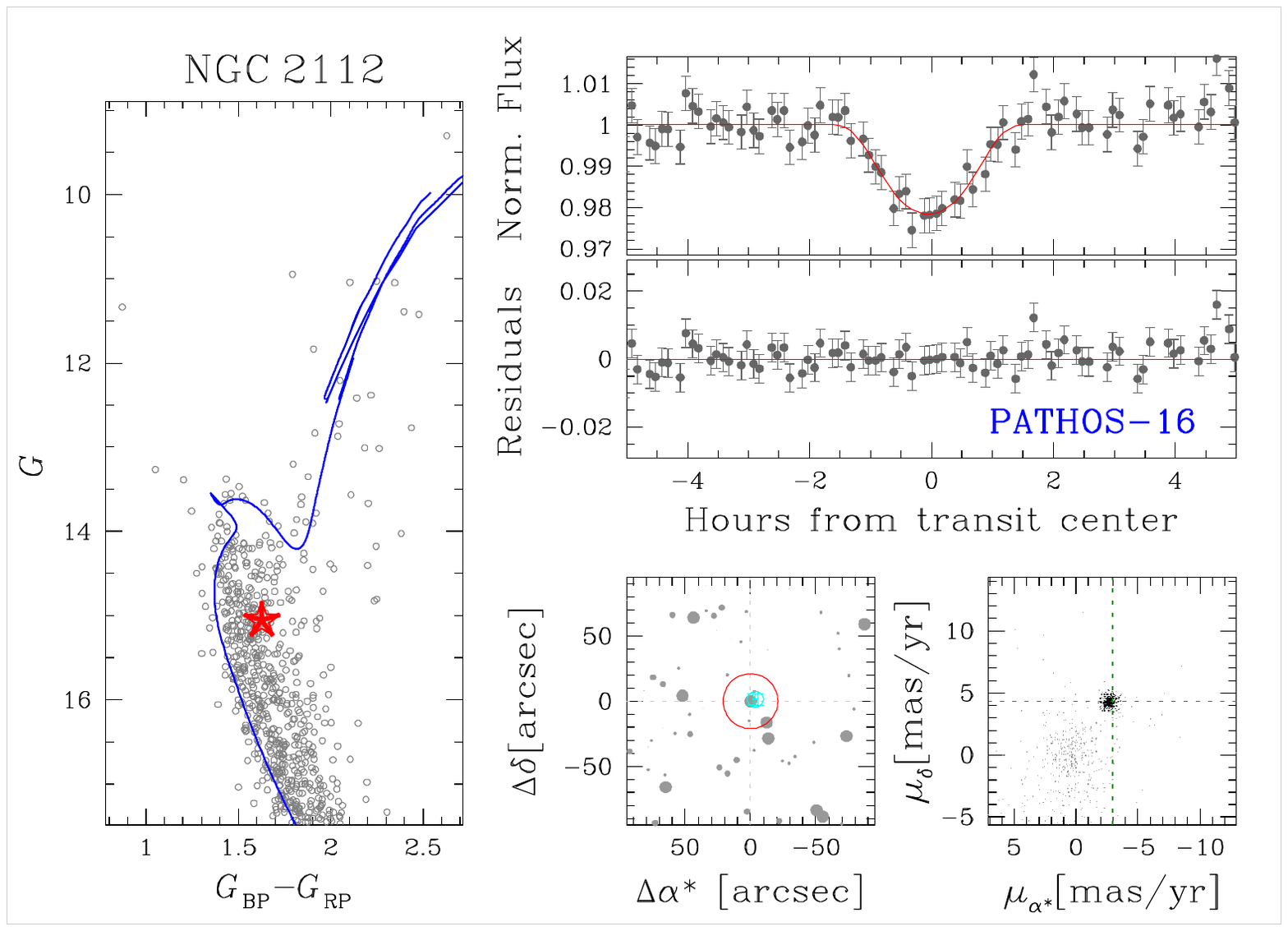} \\
\includegraphics[bb=77 360 535 691, width=0.33\textwidth]{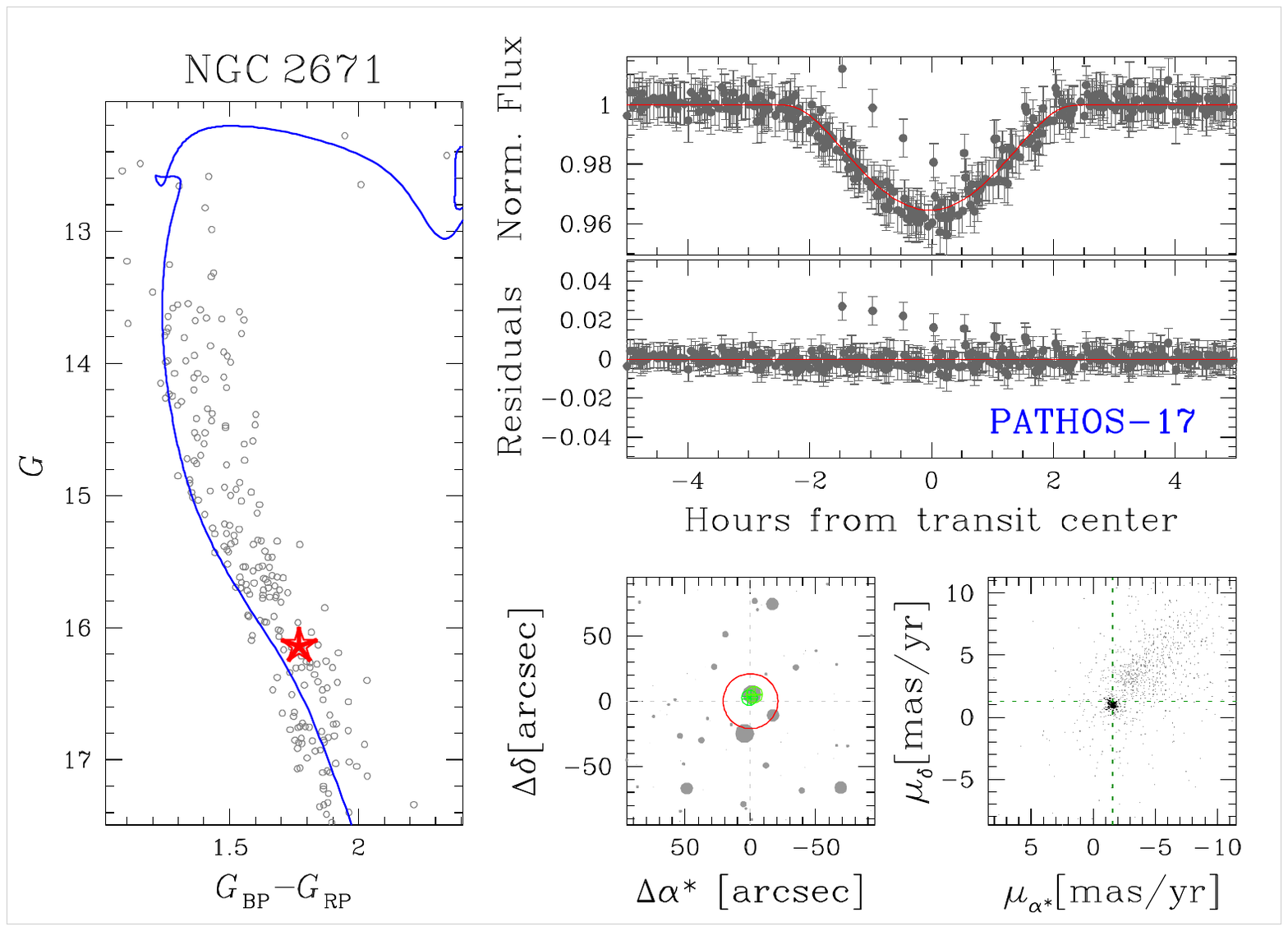}
\includegraphics[bb=77 360 535 691, width=0.33\textwidth]{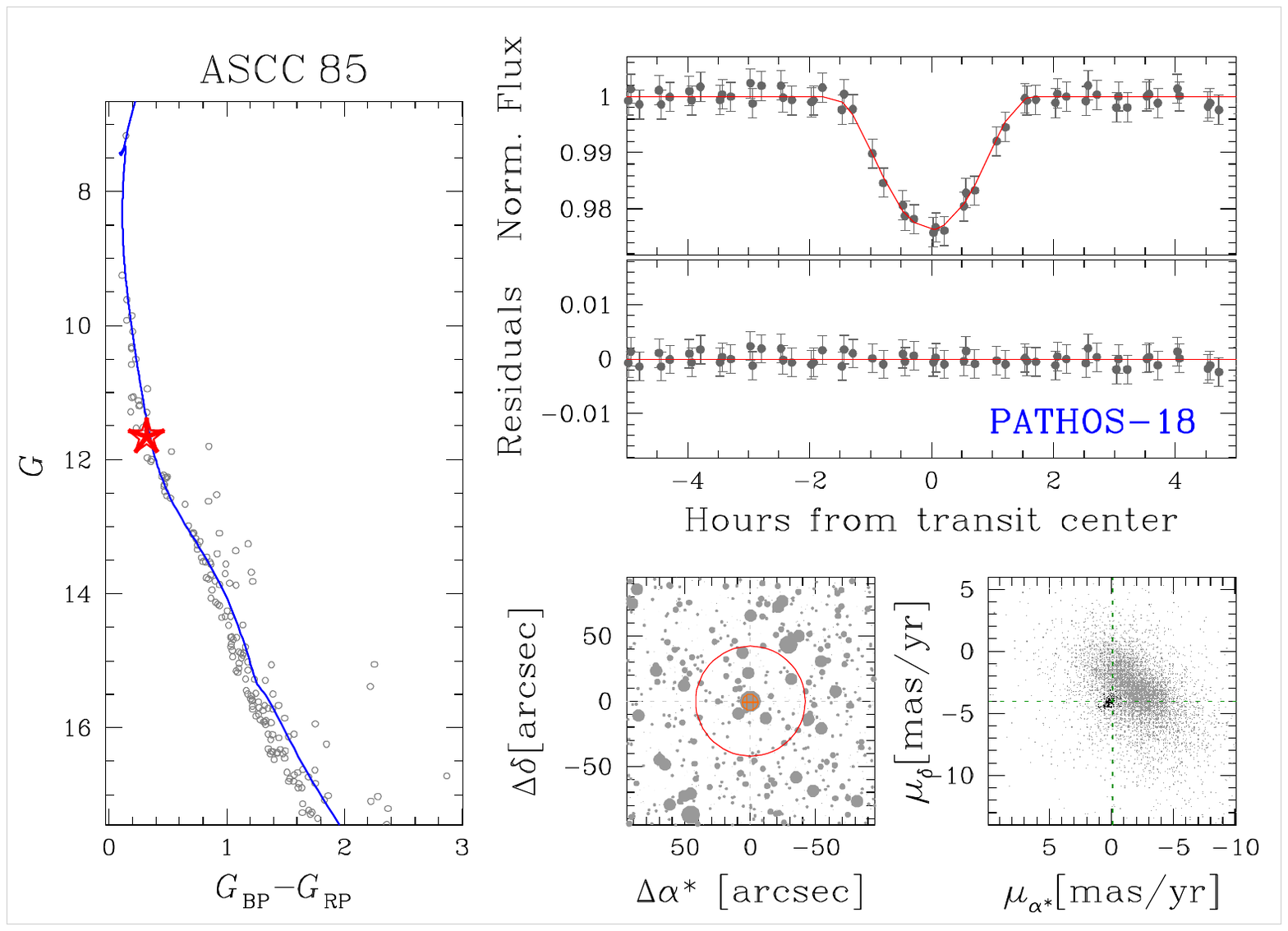}
\includegraphics[bb=77 360 535 691, width=0.33\textwidth]{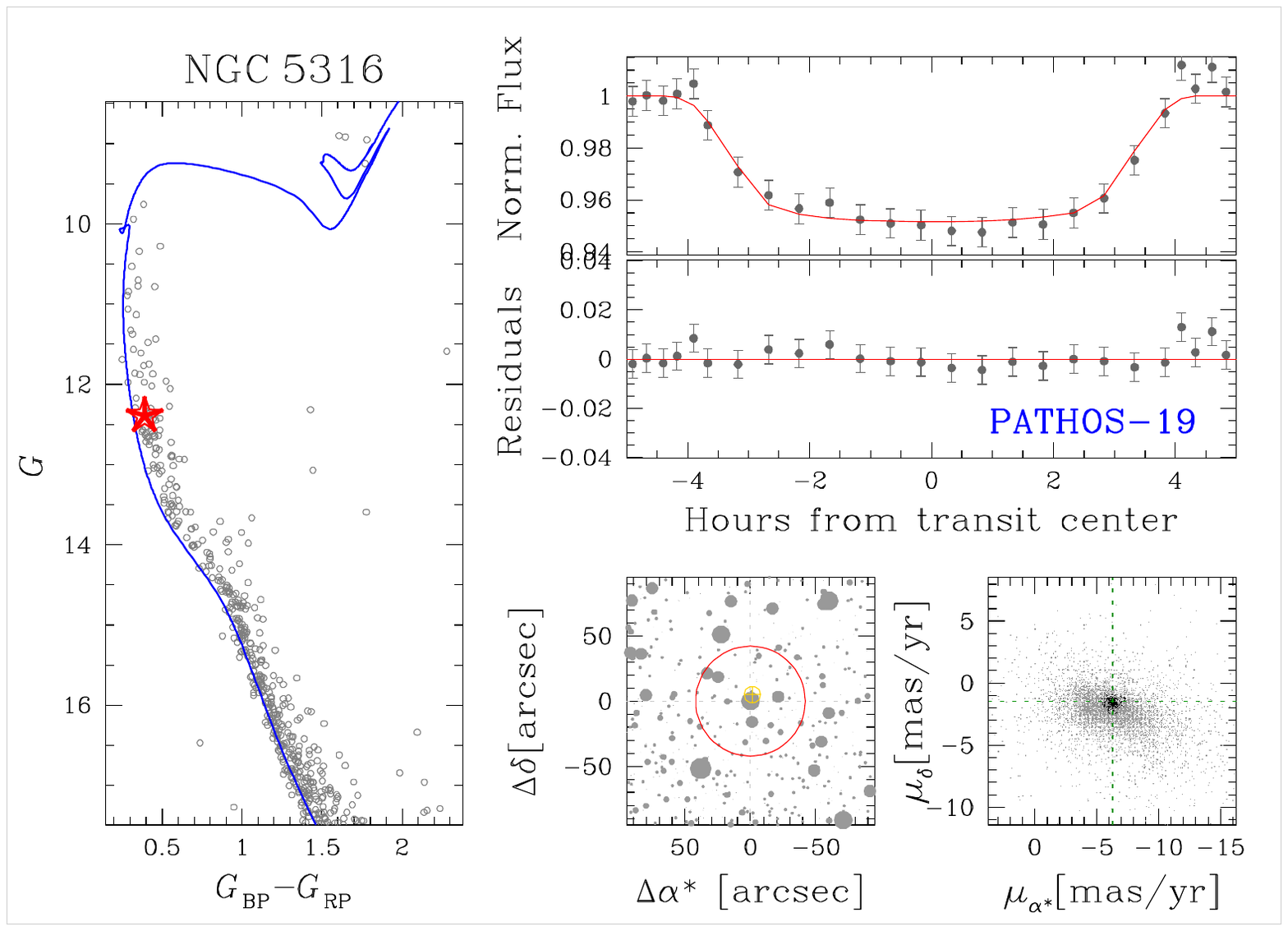} \\
\includegraphics[bb=77 360 535 691, width=0.33\textwidth]{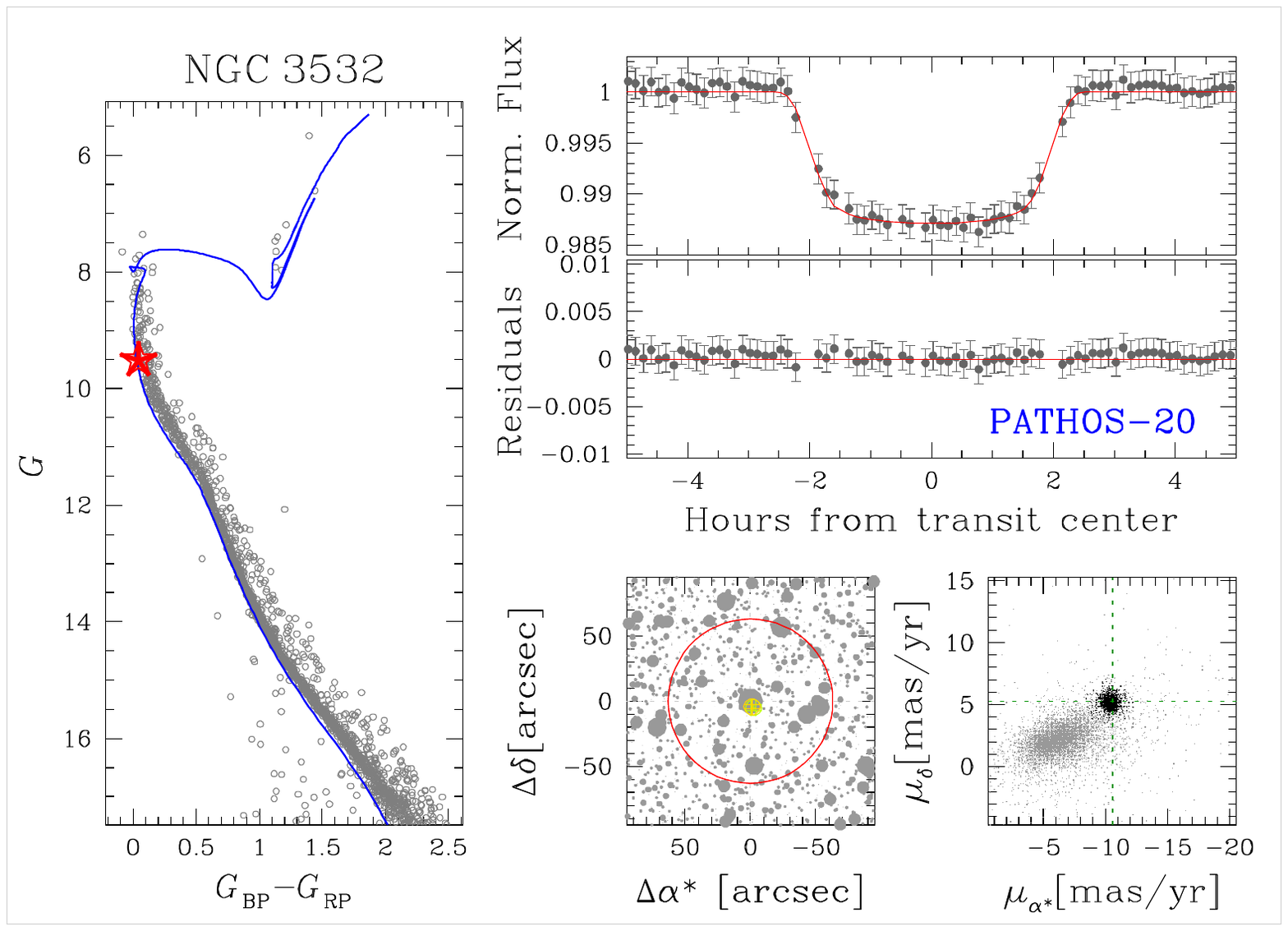}
\includegraphics[bb=77 360 535 691, width=0.33\textwidth]{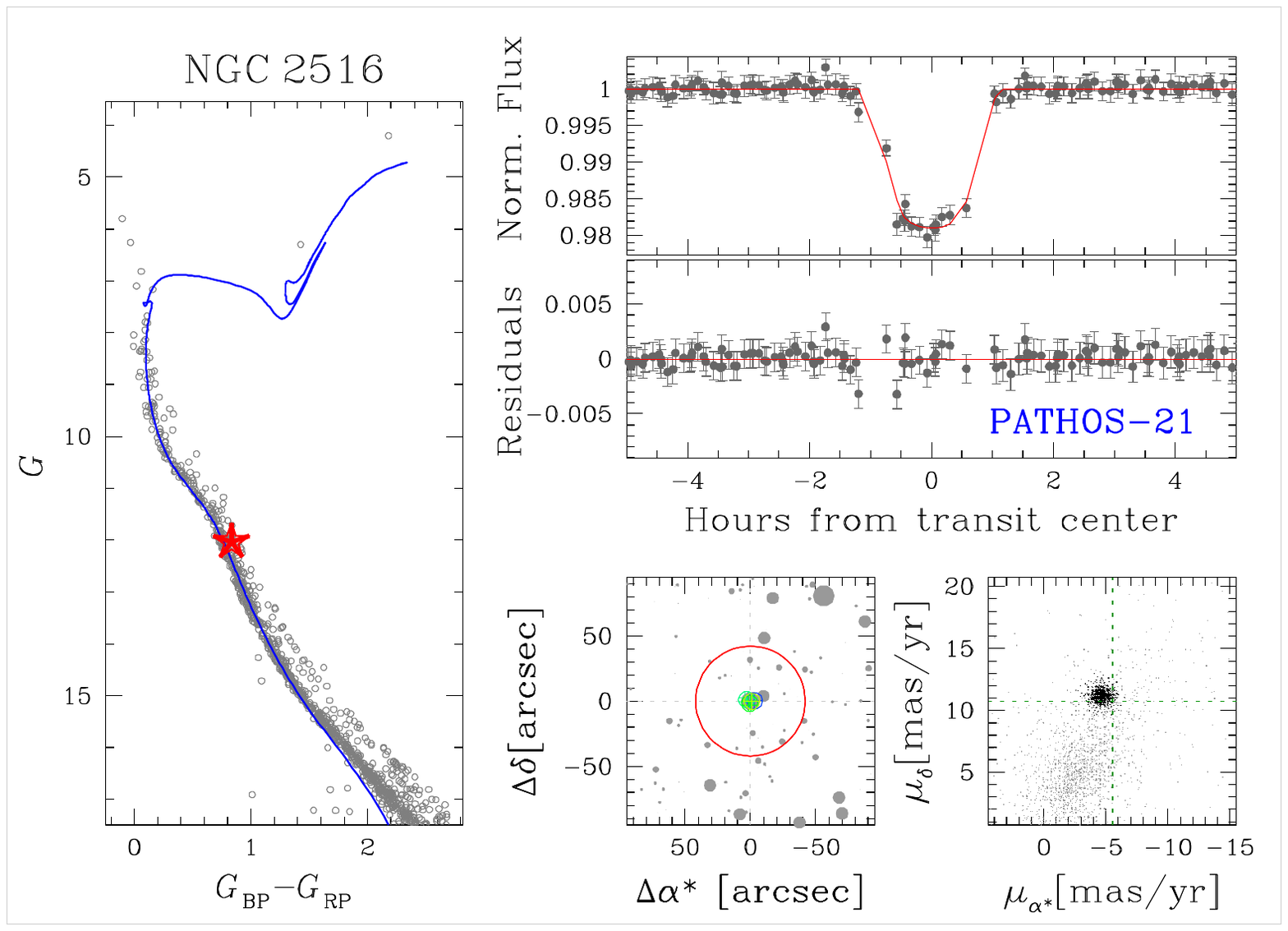}
\includegraphics[bb=77 360 535 691, width=0.33\textwidth]{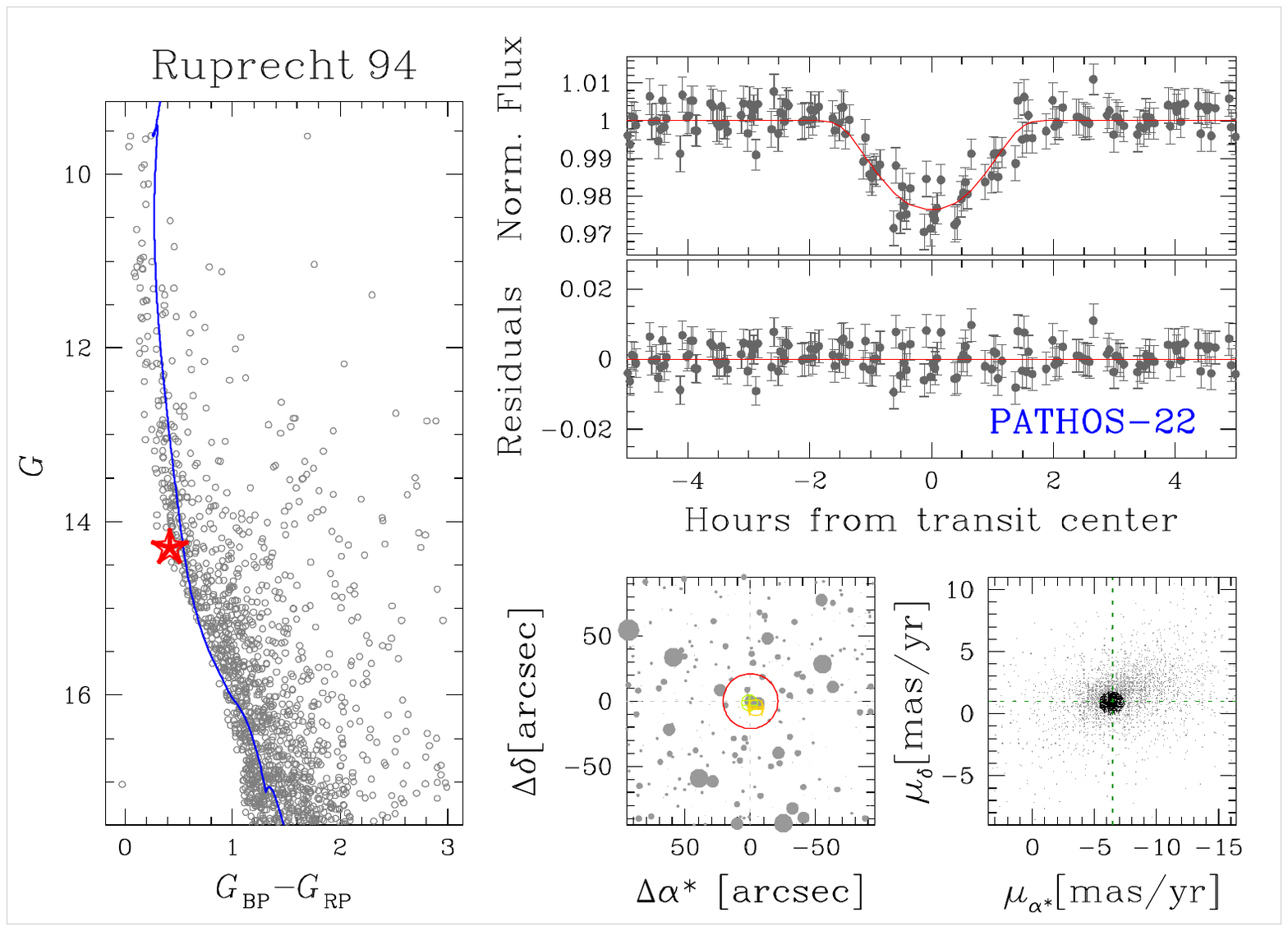} \\
\includegraphics[bb=77 360 535 691, width=0.33\textwidth]{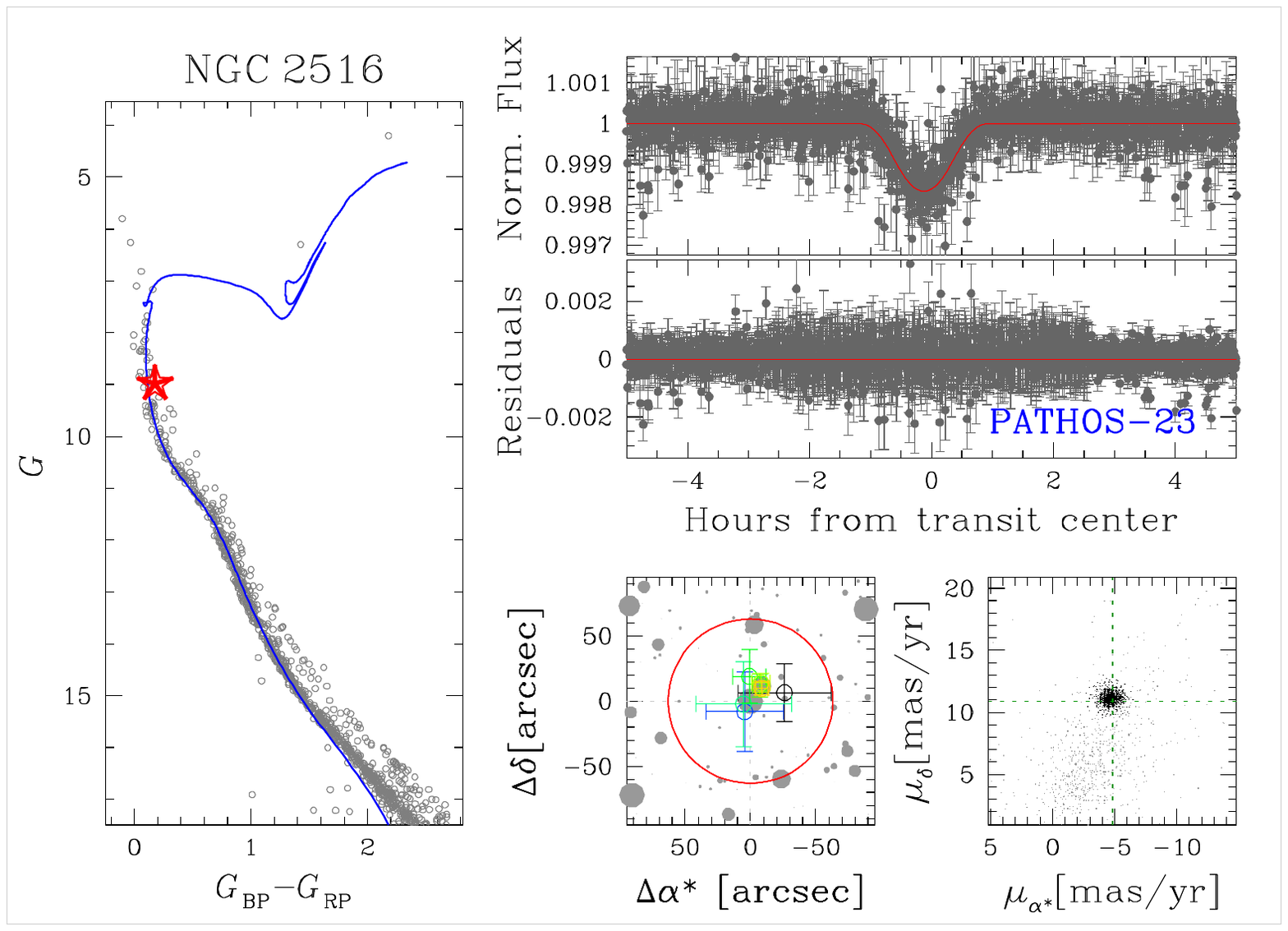}
\includegraphics[bb=77 360 535 691, width=0.33\textwidth]{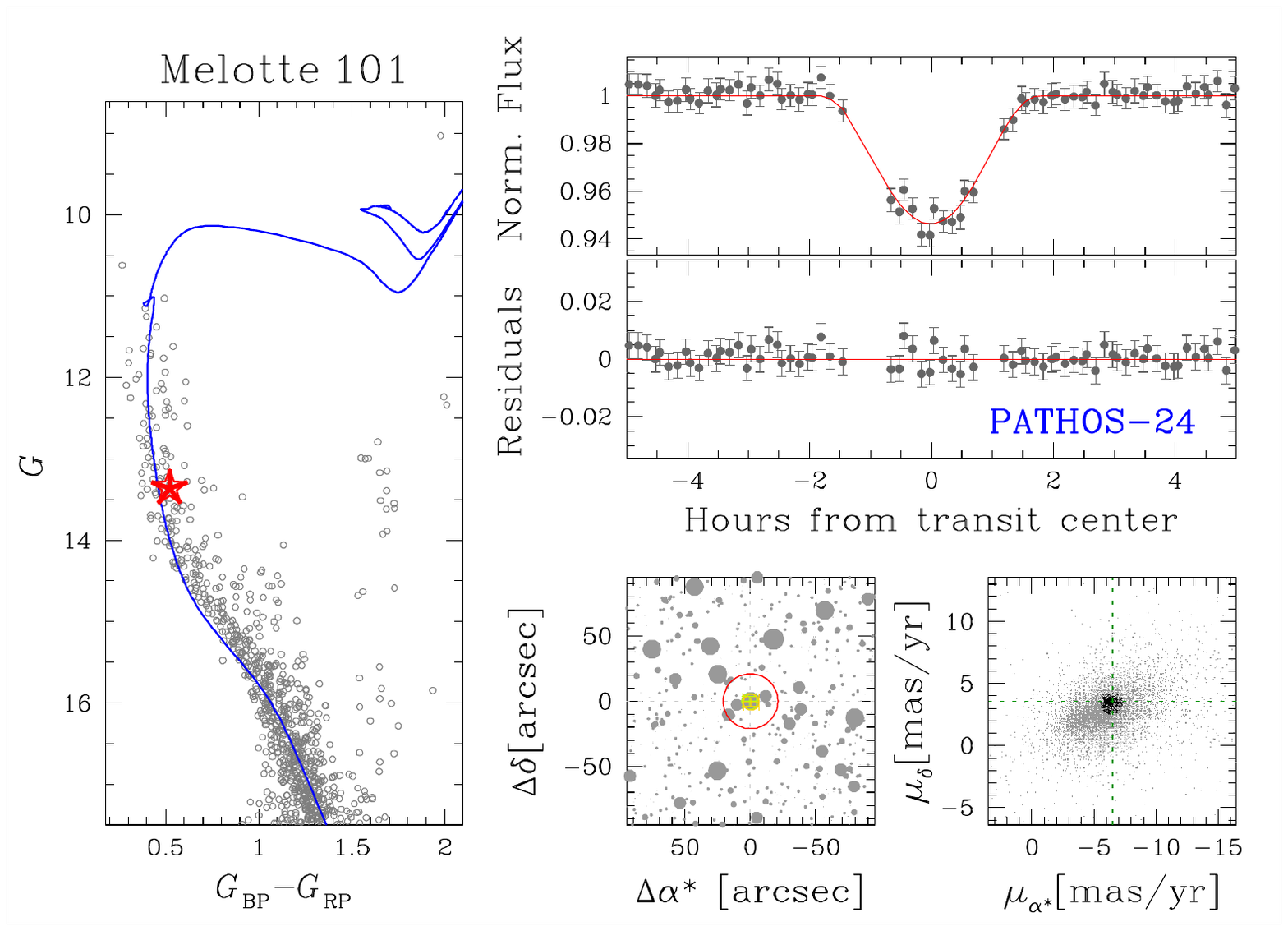}
\includegraphics[bb=77 360 535 691, width=0.33\textwidth]{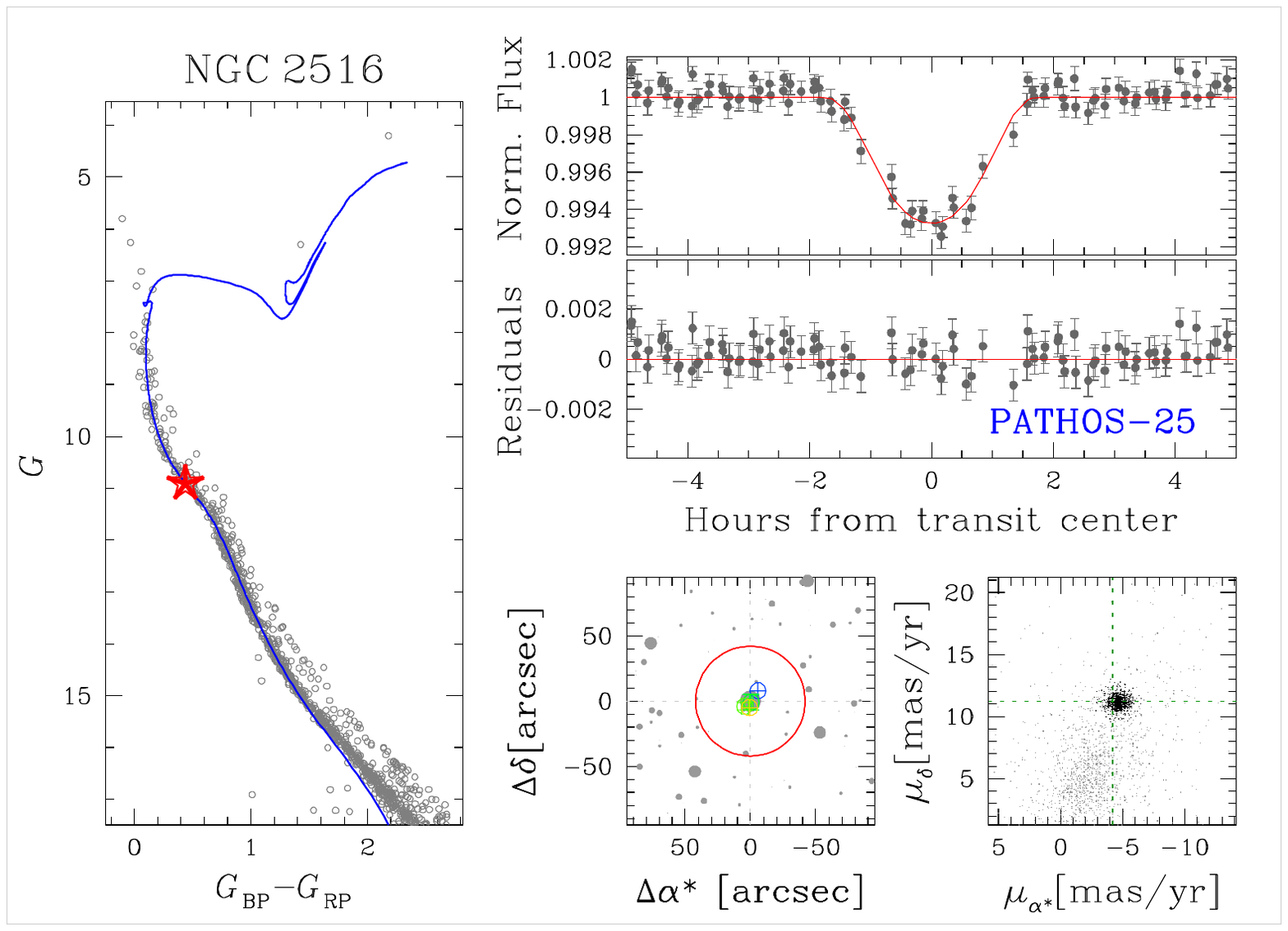} \\
\caption{As in Fig.~\ref{fig:7a}, but for PATHOS-14 -- PATHOS-25.
  \label{fig:7b}}
\end{figure*}
\begin{figure*}
\includegraphics[bb=77 360 535 691, width=0.33\textwidth]{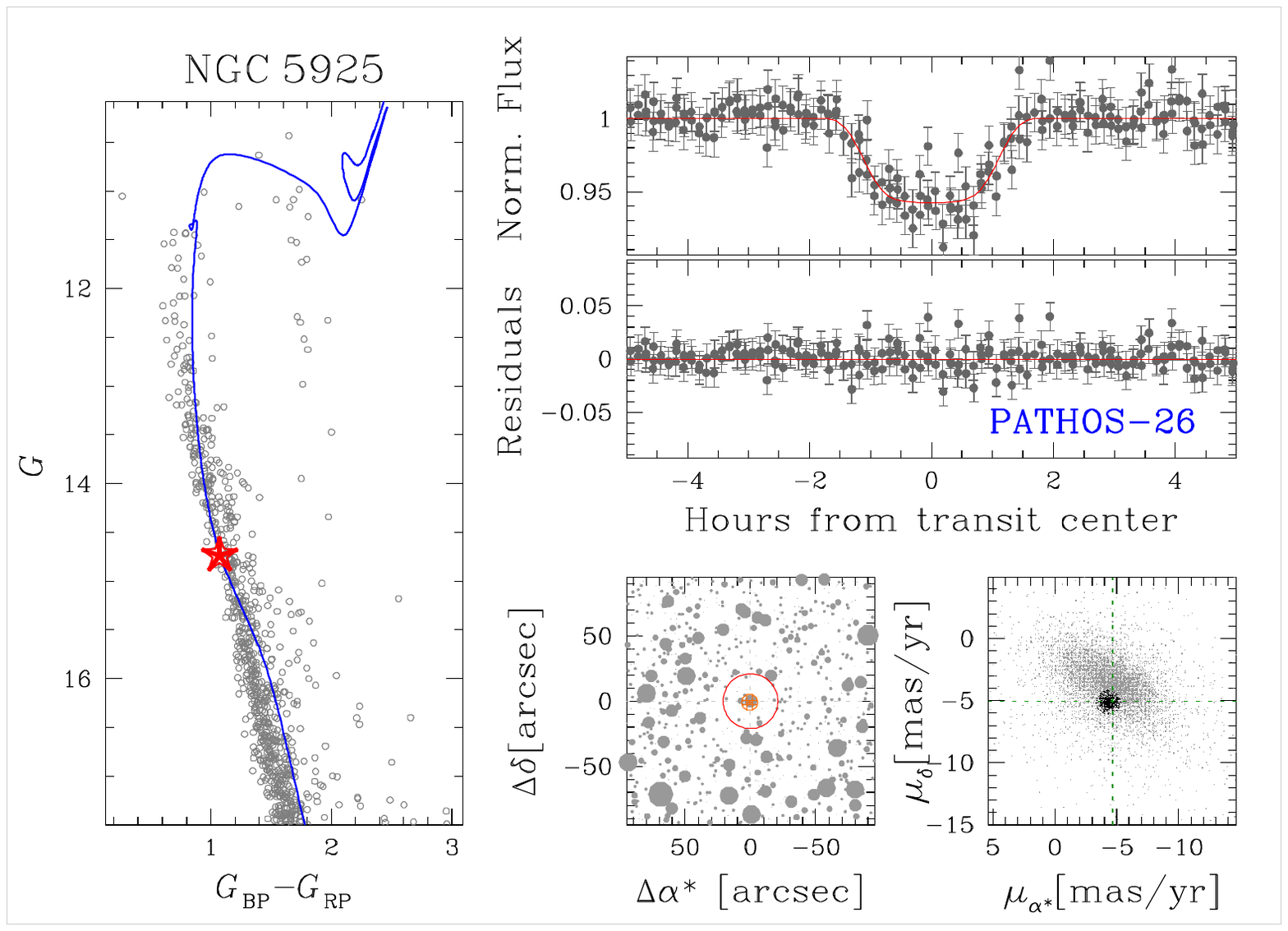}
\includegraphics[bb=77 360 535 691, width=0.33\textwidth]{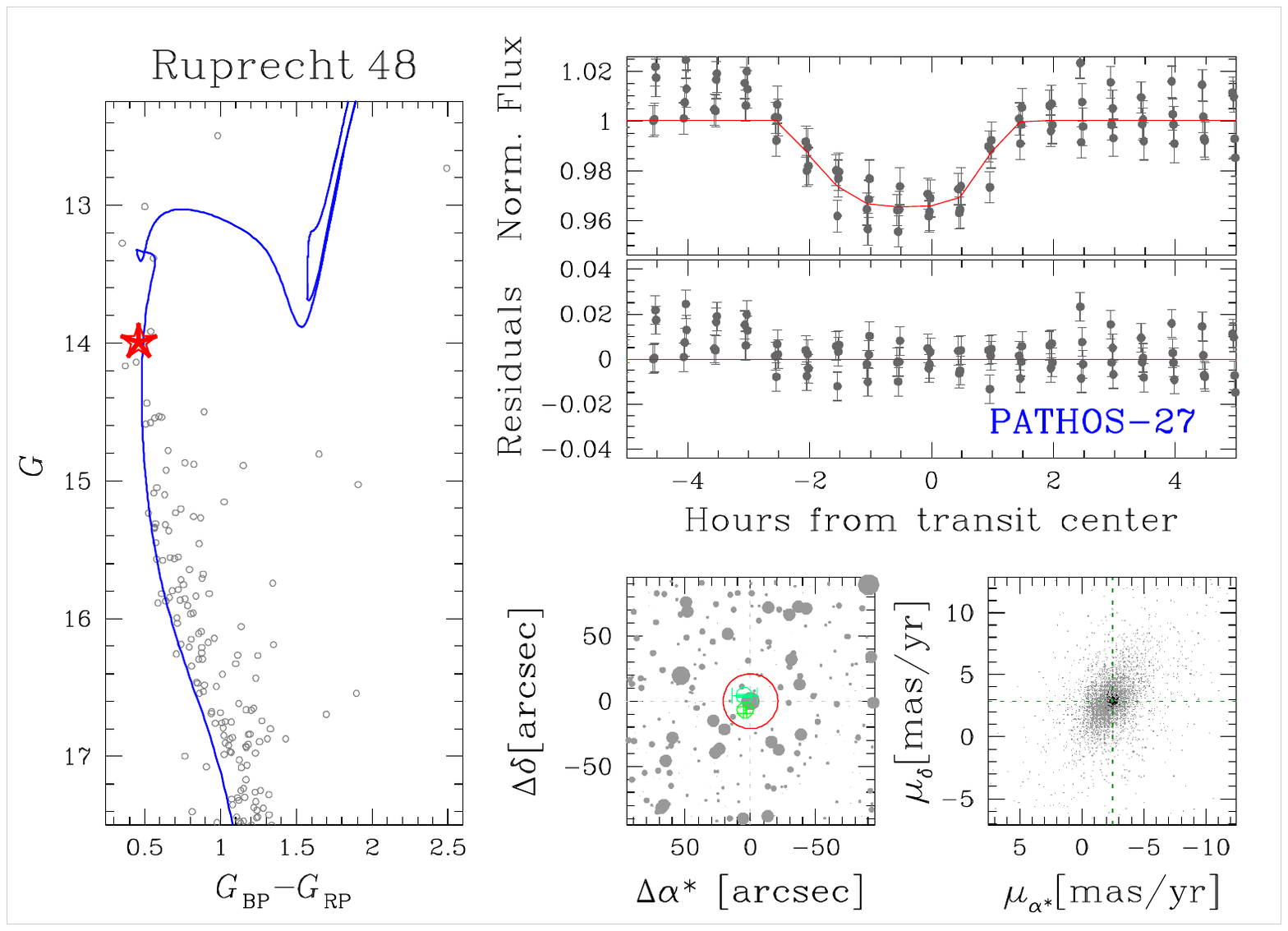}
\includegraphics[bb=77 360 535 691, width=0.33\textwidth]{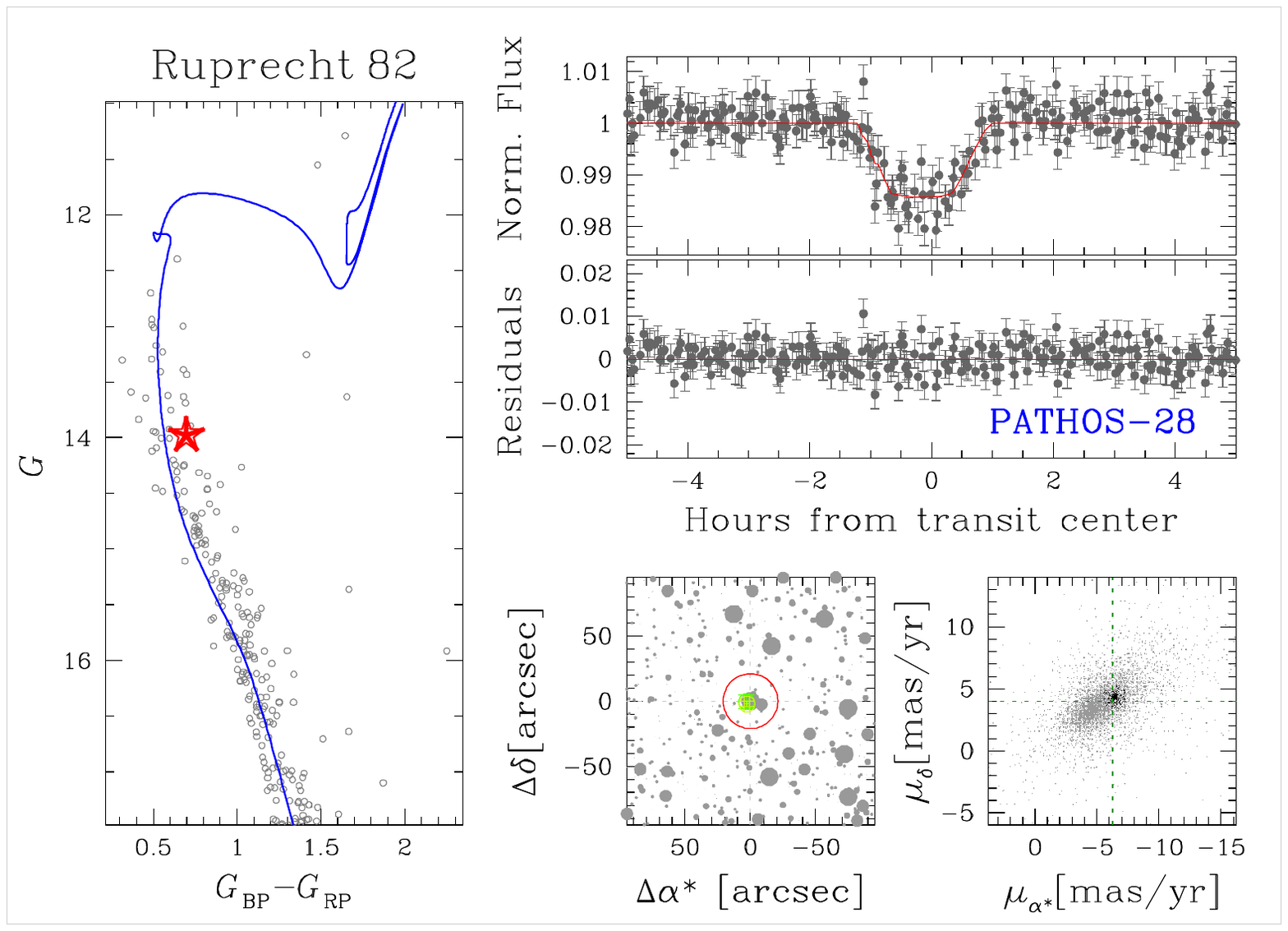} \\
\includegraphics[bb=77 360 535 691, width=0.33\textwidth]{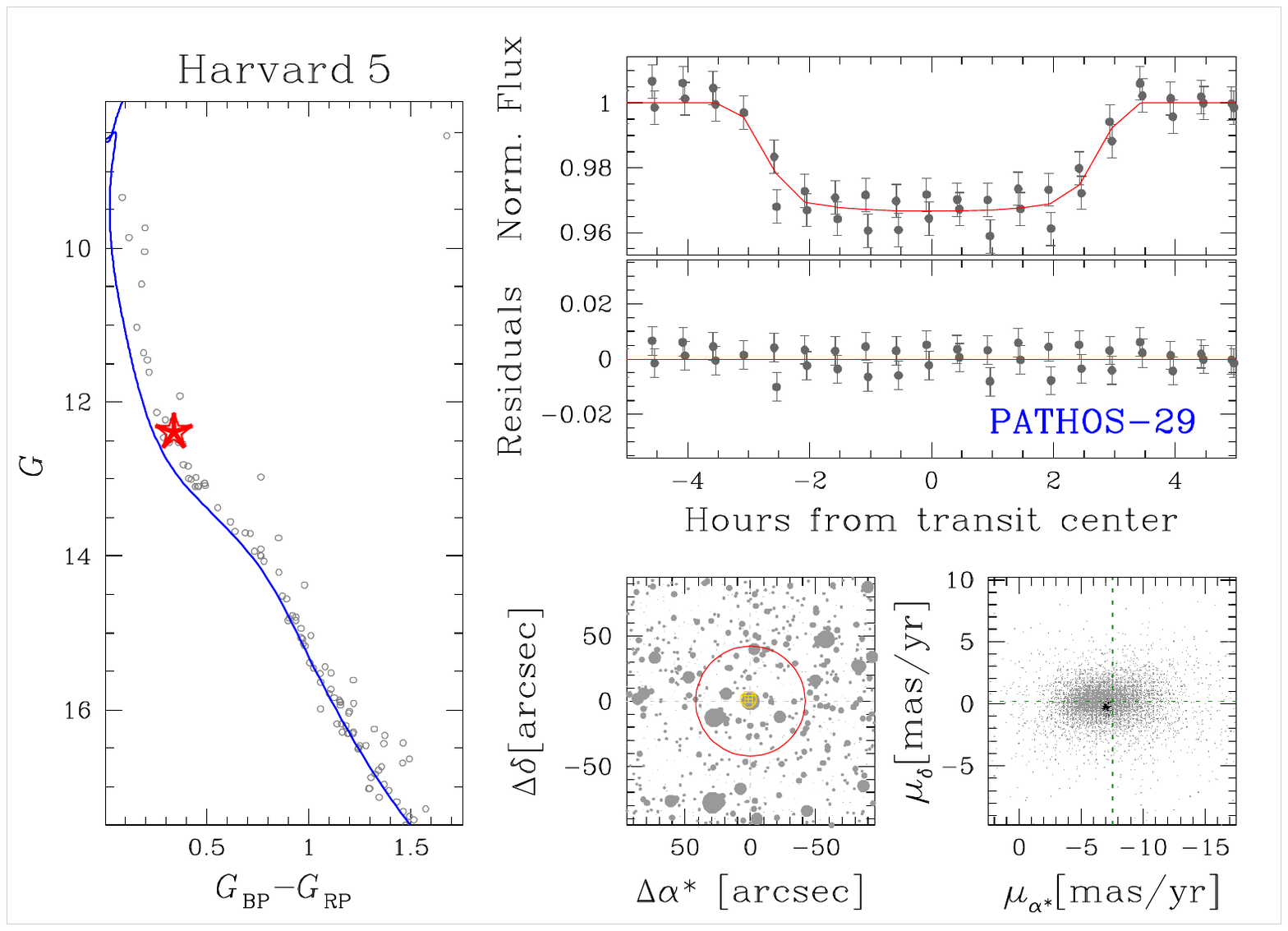}
\includegraphics[bb=77 360 535 691, width=0.33\textwidth]{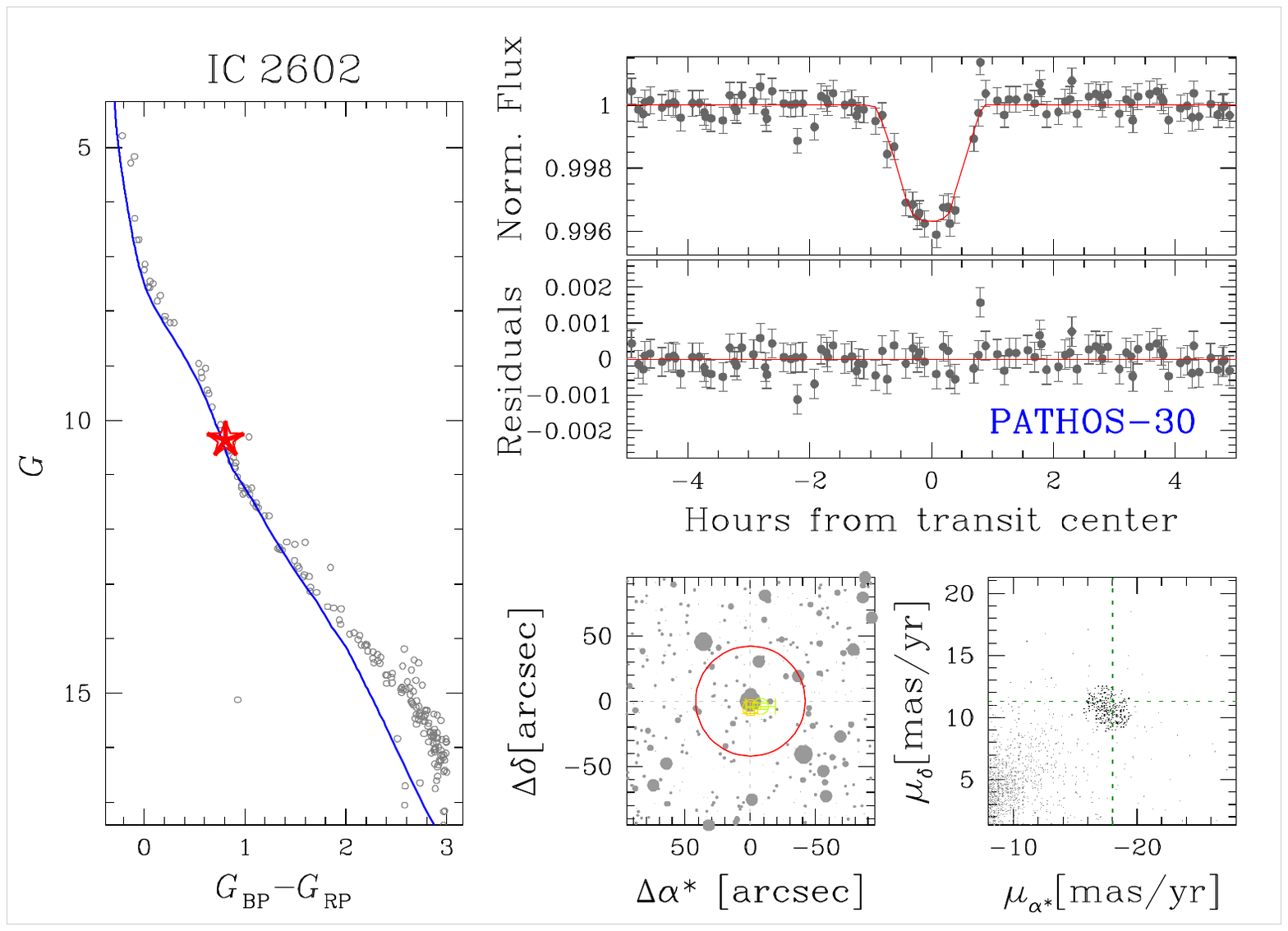}
\includegraphics[bb=77 360 535 691, width=0.33\textwidth]{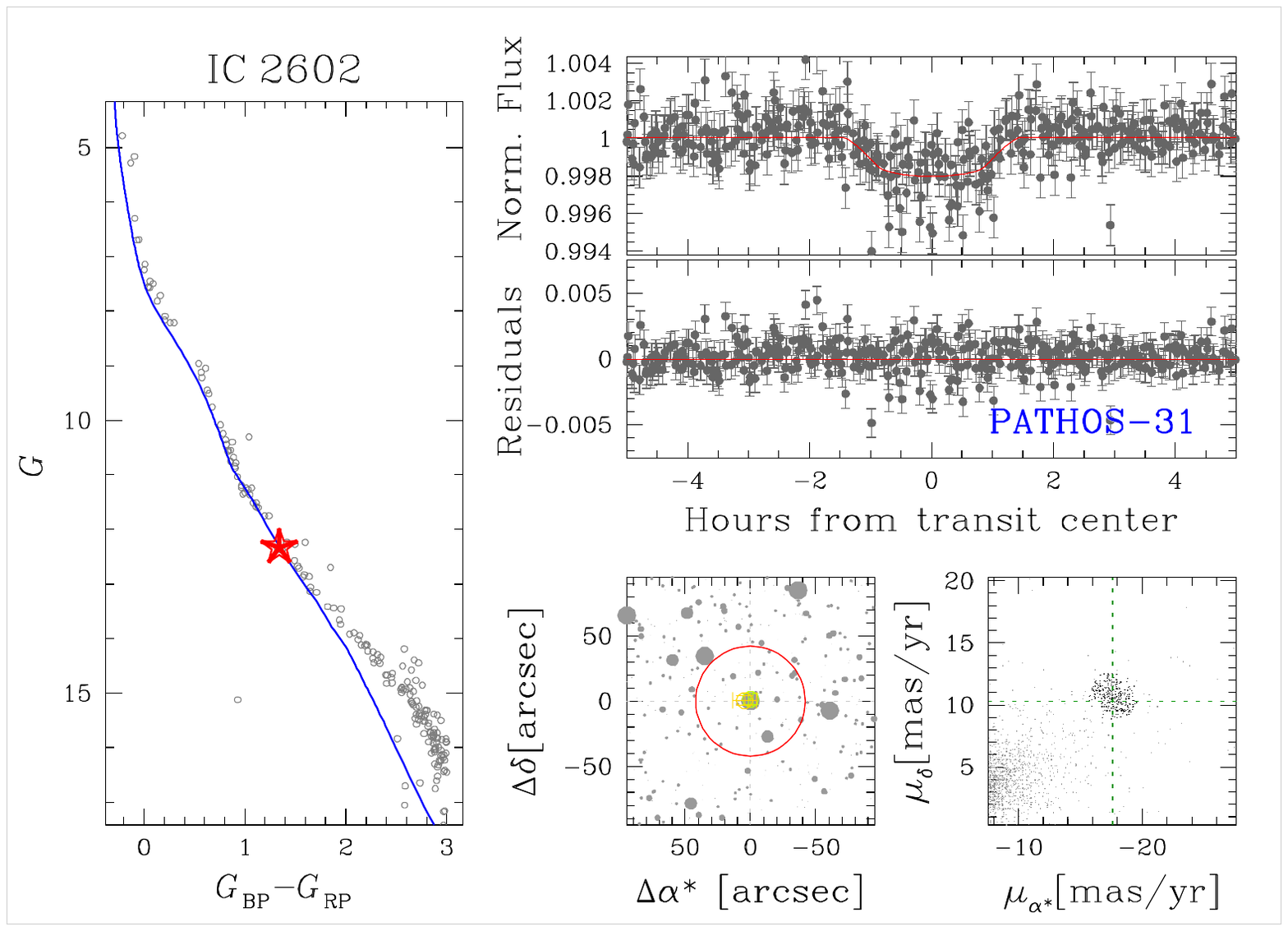} \\
\includegraphics[bb=77 360 535 691, width=0.33\textwidth]{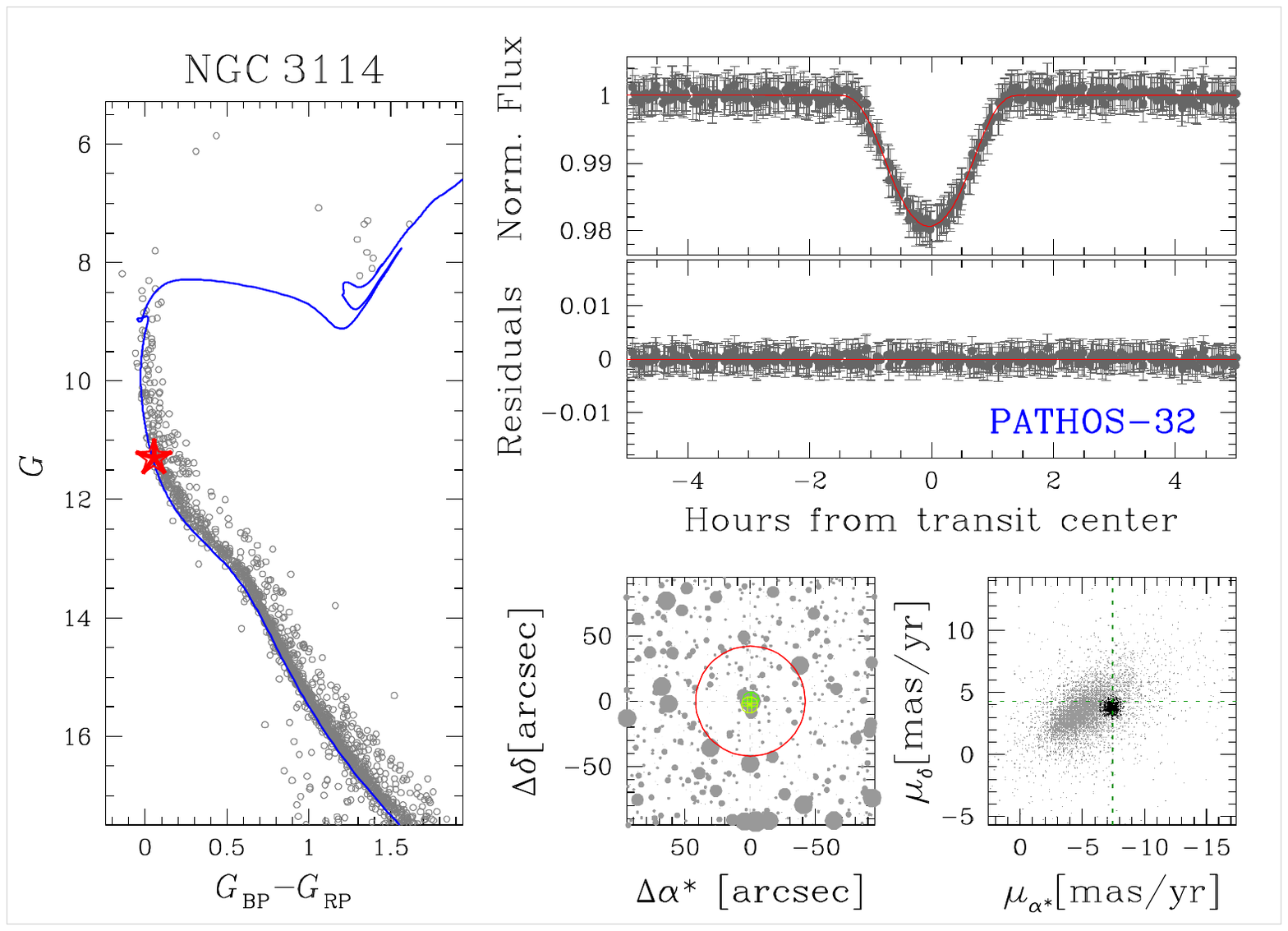}
\includegraphics[bb=77 360 535 691, width=0.33\textwidth]{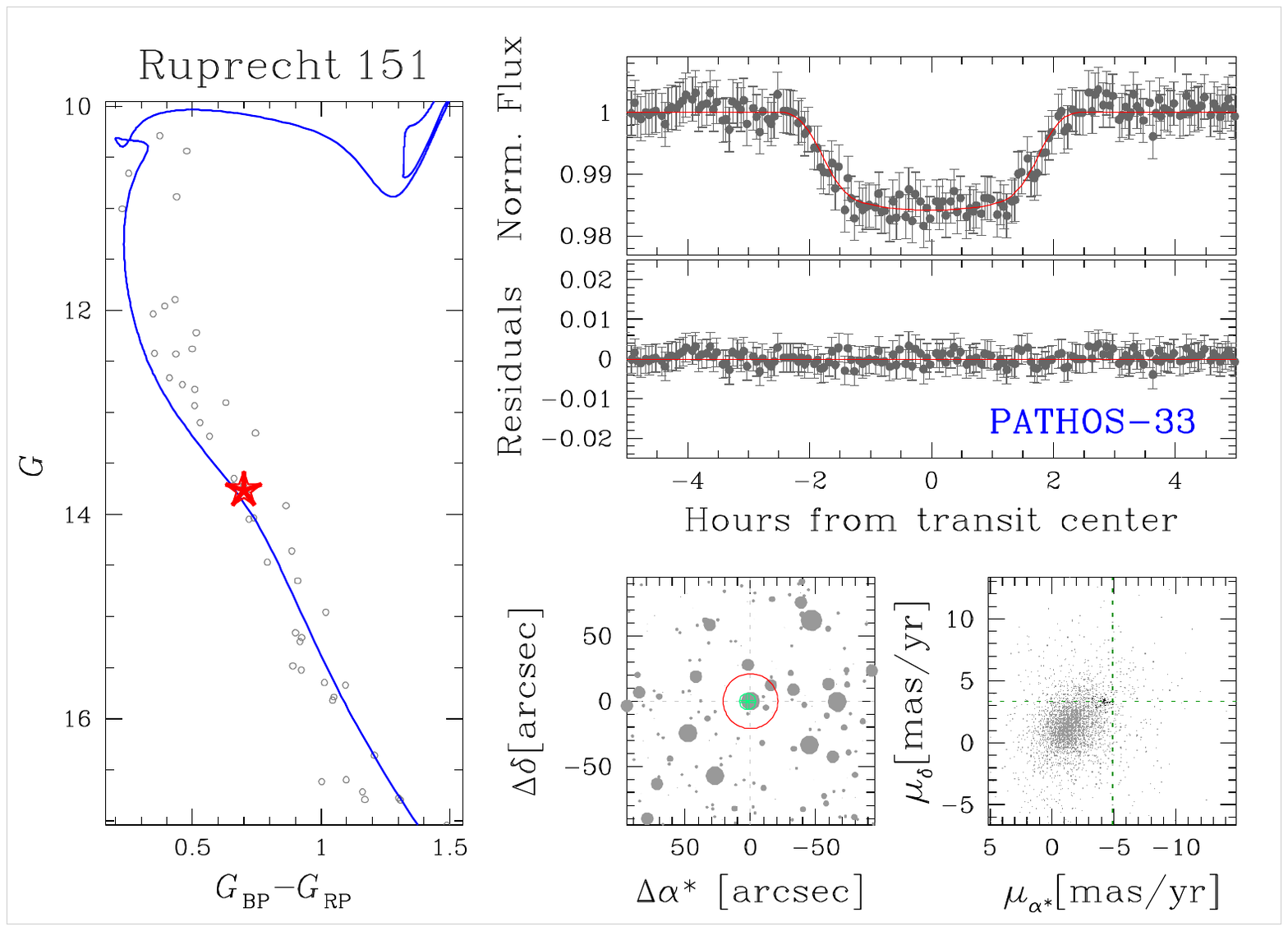}
\includegraphics[bb=77 360 535 691, width=0.33\textwidth]{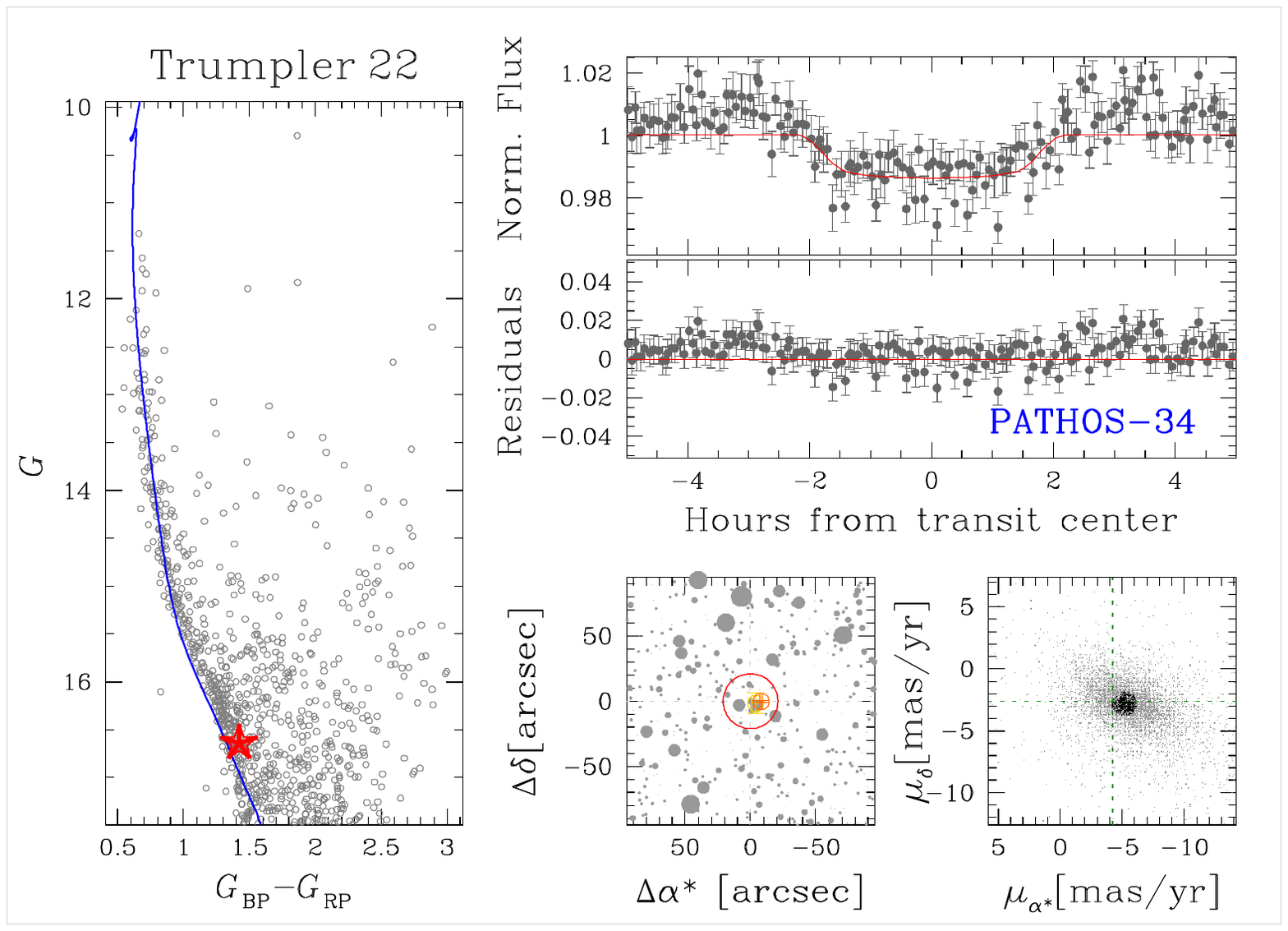} \\
\caption{As in Fig.~\ref{fig:7a}, but for PATHOS-26 -- PATHOS-34.
  \label{fig:7c}}
\end{figure*}

\bsp	
\begin{landscape}
\begin{table}
  \renewcommand{\arraystretch}{1.25}
  \caption{Results of transit modelling}
  \label{tab:3}
    \resizebox{1.29\textwidth}{!}{
      \begin{tabular}{l c l r r r r r r r r r r r l}
  \hline
  TIC & PATHOS & Cluster & \multicolumn{1}{c}{$P$} & \multicolumn{1}{c}{$T_0$}  & \multicolumn{1}{c}{$R_{\rm p}/R_{\star}$} & \multicolumn{1}{c}{$b$} &  \multicolumn{1}{c}{$a/R_{\star}$} &  \multicolumn{1}{c}{$\rho_{\star}$}   &  \multicolumn{1}{c}{LD$_{c1}$} &  \multicolumn{1}{c}{LD$_{c2}$} &  \multicolumn{1}{c}{$i$}   &  \multicolumn{1}{c}{$R_{\rm p}$}  &  \multicolumn{1}{c}{$R_{\rm p}$} &    Note  \\
      &        &         & \multicolumn{1}{c}{(d)} & \multicolumn{1}{c}{(BTJD)} &                    &     &              & ($\rho_{\sun}$)  &          &          & (deg) & ($R_{\rm J}$) & ($R_{\earth}$) &   \\
  \hline
0030654608 &     2     &      Muzzio\,1 & $       2.4955_{     -0.0003}^{+      0.0003} $   & $     1519.662_{      -0.009}^{+       0.009} $  & $        0.048_{      -0.009}^{+       0.014} $   & $         0.96_{       -0.16}^{+        0.03} $  & $          3.7_{        -0.6}^{+         0.6} $  & $         0.11_{       -0.04}^{+        0.06} $   & $         0.38_{       -0.26}^{+        0.35} $   & $         0.18_{       -0.28}^{+        0.32} $    & $         71.7_{       -13.6}^{+         5.2} $    &  $          2.5_{        -0.7}^{+         1.0} $   &   $         28.6_{        -8.0}^{+        10.8} $        &  \\
0039291805 &     3     &      NGC\,2112 & $       7.1719_{     -0.0059}^{+      0.0057} $   & $     1469.033_{      -0.009}^{+       0.009} $  & $        0.190_{      -0.018}^{+       0.064} $   & $         0.69_{       -0.40}^{+        0.21} $  & $         14.5_{        -0.6}^{+         0.6} $  & $         0.80_{       -0.10}^{+        0.10} $   & $         0.31_{       -0.05}^{+        0.05} $   & $         0.29_{       -0.05}^{+        0.05} $    & $         87.1_{        -0.8}^{+         1.2} $    &  $          2.1_{        -0.2}^{+         0.7} $   &   $         23.5_{        -2.6}^{+         7.9} $        &  \\
0042524156 &     4     &       ASCC\,88 & $      10.0130_{     -0.0026}^{+      0.0025} $   & $     1631.474_{      -0.006}^{+       0.005} $  & $        0.125_{      -0.019}^{+       0.024} $   & $         0.94_{       -0.06}^{+        0.04} $  & $         12.0_{        -1.8}^{+         1.4} $  & $         0.23_{       -0.09}^{+        0.09} $   & $         0.18_{       -0.11}^{+        0.12} $   & $         0.16_{       -0.12}^{+        0.13} $    & $         84.0_{        -3.8}^{+         1.6} $    &  $          3.5_{        -0.7}^{+         0.8} $   &   $         39.1_{        -7.3}^{+         9.5} $        &  \\
0080317933 &     5     & Collinder\,292 & $       3.2098_{     -0.0006}^{+      0.0006} $   & $     1625.529_{      -0.003}^{+       0.003} $  & $        0.232_{      -0.005}^{+       0.006} $   & $         0.39_{       -0.24}^{+        0.17} $  & $          6.3_{        -0.6}^{+         0.5} $  & $         0.32_{       -0.09}^{+        0.09} $   & $         0.19_{       -0.05}^{+        0.05} $   & $         0.23_{       -0.05}^{+        0.05} $    & $         86.3_{        -1.8}^{+         2.2} $    &  $          4.1_{        -0.3}^{+         0.5} $   &   $         45.8_{        -3.8}^{+         5.4} $        &  \\
0088977253 &     6     &      NGC\,2548 & $       2.7677_{     -0.0004}^{+      0.0004} $   & $     1492.708_{      -0.004}^{+       0.004} $  & $        0.146_{      -0.016}^{+       0.048} $   & $         0.85_{       -0.31}^{+        0.11} $  & $          6.6_{        -0.5}^{+         0.4} $  & $         0.50_{       -0.10}^{+        0.10} $   & $         0.27_{       -0.05}^{+        0.05} $   & $         0.28_{       -0.05}^{+        0.05} $    & $         78.1_{        -4.3}^{+         2.7} $    &  $          2.1_{        -0.3}^{+         0.6} $   &   $         23.0_{        -3.2}^{+         7.1} $        & TOI-496 \\
0092835691 &     7     &        SAI\,91 & $      22.3661_{     -0.0069}^{+      0.0067} $   & $     1527.825_{      -0.007}^{+       0.007} $  & $        0.214_{      -0.033}^{+       0.034} $   & $         0.92_{       -0.07}^{+        0.05} $  & $         20.0_{        -3.1}^{+         2.5} $  & $         0.22_{       -0.09}^{+        0.09} $   & $         0.19_{       -0.05}^{+        0.05} $   & $         0.23_{       -0.05}^{+        0.05} $    & $         85.4_{        -2.9}^{+         1.3} $    &  $          4.4_{        -0.8}^{+         1.0} $   &   $         49.8_{        -9.3}^{+        11.2} $        &  \\
0094589619 &     8     &      NGC\,2437 & $      12.0829_{     -0.0047}^{+      0.0045} $   & $     1496.198_{      -0.004}^{+       0.004} $  & $        0.187_{      -0.019}^{+       0.062} $   & $         0.81_{       -0.39}^{+        0.14} $  & $         17.6_{        -1.2}^{+         1.1} $  & $         0.50_{       -0.10}^{+        0.10} $   & $         0.24_{       -0.05}^{+        0.05} $   & $         0.26_{       -0.05}^{+        0.05} $    & $         84.9_{        -1.9}^{+         1.3} $    &  $          2.7_{        -0.4}^{+         0.8} $   &   $         30.4_{        -4.0}^{+         9.4} $        &  \\
0125414447 &     9     &      NGC\,2323 & $       3.7069_{     -0.0008}^{+      0.0008} $   & $     1493.200_{      -0.004}^{+       0.004} $  & $        0.229_{      -0.035}^{+       0.052} $   & $         0.88_{       -0.10}^{+        0.09} $  & $         10.4_{        -0.3}^{+         0.3} $  & $         1.09_{       -0.10}^{+        0.10} $   & $         0.32_{       -0.05}^{+        0.05} $   & $         0.30_{       -0.05}^{+        0.05} $    & $         85.4_{        -0.8}^{+         0.5} $    &  $          2.2_{        -0.4}^{+         0.5} $   &   $         25.1_{        -4.0}^{+         5.7} $        &  \\
0126600730 &    10     &    Haffner\,14 & $       6.2991_{     -0.0067}^{+      0.0067} $   & $     1497.305_{      -0.009}^{+       0.011} $  & $        0.102_{      -0.010}^{+       0.027} $   & $         0.75_{       -0.46}^{+        0.18} $  & $          9.3_{        -1.2}^{+         1.0} $  & $         0.27_{       -0.09}^{+        0.10} $   & $         0.15_{       -0.05}^{+        0.05} $   & $         0.20_{       -0.05}^{+        0.05} $    & $         84.7_{        -2.4}^{+         1.9} $    &  $          2.1_{        -0.3}^{+         0.7} $   &   $         23.8_{        -3.7}^{+         8.0} $        &  \\
0144995073 &    11     &      NGC\,2669 & $      20.0207_{     -0.0050}^{+      0.0051} $   & $     1518.823_{      -0.005}^{+       0.005} $  & $        0.240_{      -0.038}^{+       0.066} $   & $         0.83_{       -0.34}^{+        0.12} $  & $         25.6_{        -1.6}^{+         1.4} $  & $         0.56_{       -0.10}^{+        0.10} $   & $         0.27_{       -0.05}^{+        0.05} $   & $         0.28_{       -0.05}^{+        0.05} $    & $         84.9_{        -2.3}^{+         1.4} $    &  $          3.2_{        -0.5}^{+         0.8} $   &   $         36.0_{        -6.1}^{+         9.5} $        &  \\
0147069011 &    12     &      Alessi\,8 & $       7.3270_{     -0.0018}^{+      0.0019} $   & $     1628.559_{      -0.004}^{+       0.004} $  & $        0.448_{      -0.045}^{+       0.035} $   & $         0.38_{       -0.25}^{+        0.22} $  & $         13.9_{        -0.7}^{+         0.6} $  & $         0.67_{       -0.10}^{+        0.10} $   & $         0.28_{       -0.05}^{+        0.05} $   & $         0.29_{       -0.05}^{+        0.05} $    & $         86.8_{        -1.6}^{+         1.9} $    &  $          5.4_{        -0.6}^{+         0.5} $   &   $         61.0_{        -6.5}^{+         5.6} $        &  \\
0147426828 &    13     &      NGC\,2318 & $       1.4618_{     -0.0002}^{+      0.0002} $   & $     1492.845_{      -0.004}^{+       0.004} $  & $        0.221_{      -0.037}^{+       0.047} $   & $         0.89_{       -0.12}^{+        0.08} $  & $          4.2_{        -0.3}^{+         0.3} $  & $         0.47_{       -0.09}^{+        0.09} $   & $         0.25_{       -0.05}^{+        0.05} $   & $         0.27_{       -0.05}^{+        0.05} $    & $         76.2_{        -3.9}^{+         2.2} $    &  $          3.2_{        -0.6}^{+         0.7} $   &   $         35.9_{        -6.4}^{+         7.6} $        &  \\
0153734545 &    14     &      NGC\,2527 & $       3.2196_{     -0.0003}^{+      0.0003} $   & $     1493.665_{      -0.008}^{+       0.008} $  & $        0.187_{      -0.025}^{+       0.034} $   & $         0.92_{       -0.06}^{+        0.06} $  & $          4.5_{        -0.6}^{+         0.8} $  & $         0.11_{       -0.04}^{+        0.07} $   & $         0.22_{       -0.05}^{+        0.05} $   & $         0.25_{       -0.05}^{+        0.05} $    & $         78.1_{        -5.1}^{+         3.1} $    &  $          4.8_{        -0.9}^{+         1.2} $   &   $         53.7_{       -10.4}^{+        13.7} $        &  \\
0153735144 &    15     &      NGC\,2527 & $       3.7785_{     -0.0003}^{+      0.0003} $   & $     1493.406_{      -0.002}^{+       0.002} $  & $        0.178_{      -0.004}^{+       0.006} $   & $         0.37_{       -0.25}^{+        0.24} $  & $          7.8_{        -0.6}^{+         0.5} $  & $         0.44_{       -0.10}^{+        0.10} $   & $         0.26_{       -0.05}^{+        0.05} $   & $         0.27_{       -0.05}^{+        0.05} $    & $         84.4_{        -2.6}^{+         3.5} $    &  $          2.6_{        -0.2}^{+         0.2} $   &   $         29.0_{        -2.1}^{+         2.7} $        &  \\
0159059181 &    16     &      NGC\,2112 & $       5.2167_{     -0.0020}^{+      0.0020} $   & $     1471.198_{      -0.005}^{+       0.005} $  & $        0.162_{      -0.025}^{+       0.048} $   & $         0.87_{       -0.31}^{+        0.09} $  & $          9.3_{        -0.8}^{+         0.7} $  & $         0.40_{       -0.09}^{+        0.10} $   & $         0.28_{       -0.05}^{+        0.05} $   & $         0.28_{       -0.05}^{+        0.05} $    & $         82.5_{        -3.3}^{+         1.8} $    &  $          2.4_{        -0.4}^{+         0.7} $   &   $         27.0_{        -4.8}^{+         7.6} $        &  \\
0181602717 &    17     &      NGC\,2671 & $       2.4557_{     -0.0002}^{+      0.0002} $   & $     1517.489_{      -0.009}^{+       0.009} $  & $        0.246_{      -0.034}^{+       0.030} $   & $         0.93_{       -0.06}^{+        0.05} $  & $          5.9_{        -0.5}^{+         0.4} $  & $         0.46_{       -0.10}^{+        0.10} $   & $         0.25_{       -0.05}^{+        0.05} $   & $         0.25_{       -0.05}^{+        0.05} $    & $         83.2_{        -1.5}^{+         0.9} $    &  $          3.6_{        -0.5}^{+         0.5} $   &   $         40.2_{        -5.8}^{+         6.0} $        &  \\
0236084210 &    18     &       ASCC\,85 & $       5.3465_{     -0.0016}^{+      0.0016} $   & $     1631.974_{      -0.004}^{+       0.004} $  & $        0.182_{      -0.028}^{+       0.035} $   & $         0.91_{       -0.07}^{+        0.06} $  & $          9.8_{        -0.8}^{+         0.6} $  & $         0.44_{       -0.09}^{+        0.09} $   & $         0.15_{       -0.05}^{+        0.05} $   & $         0.18_{       -0.05}^{+        0.05} $    & $         83.9_{        -1.8}^{+         0.9} $    &  $          3.3_{        -0.5}^{+         0.6} $   &   $         36.6_{        -5.8}^{+         7.2} $        &  \\
0300362600 &    19     &      NGC\,5316 & $      11.6556_{     -0.0075}^{+      0.0073} $   & $     1602.169_{      -0.004}^{+       0.003} $  & $        0.213_{      -0.005}^{+       0.005} $   & $         0.20_{       -0.14}^{+        0.18} $  & $         13.7_{        -1.4}^{+         1.5} $  & $         0.25_{       -0.07}^{+        0.09} $   & $         0.15_{       -0.05}^{+        0.05} $   & $         0.19_{       -0.05}^{+        0.05} $    & $         89.2_{        -0.8}^{+         0.6} $    &  $          4.6_{        -0.5}^{+         0.5} $   &   $         51.3_{        -5.2}^{+         6.0} $        &  \\
0306385801 &    20     &      NGC\,3532 & $      14.0367_{     -0.0018}^{+      0.0020} $   & $     1575.028_{      -0.003}^{+       0.003} $  & $        0.110_{      -0.002}^{+       0.003} $   & $         0.48_{       -0.30}^{+        0.19} $  & $         15.2_{        -2.3}^{+         1.8} $  & $         0.24_{       -0.09}^{+        0.10} $   & $         0.17_{       -0.05}^{+        0.05} $   & $         0.21_{       -0.05}^{+        0.05} $    & $         86.9_{        -2.0}^{+         1.6} $    &  $          2.3_{        -0.3}^{+         0.4} $   &   $         25.5_{        -2.8}^{+         4.5} $        &  \\
0308538095 &    21     &      NGC\,2516 & $      11.6929_{     -0.0002}^{+      0.0002} $   & $     1420.294_{      -0.002}^{+       0.002} $  & $        0.130_{      -0.003}^{+       0.004} $   & $         0.40_{       -0.27}^{+        0.24} $  & $         18.6_{        -1.0}^{+         0.9} $  & $         0.63_{       -0.10}^{+        0.10} $   & $         0.28_{       -0.05}^{+        0.05} $   & $         0.29_{       -0.05}^{+        0.05} $    & $         84.6_{        -2.0}^{+         3.2} $    &  $          1.6_{        -0.1}^{+         0.1} $   &   $         18.2_{        -1.0}^{+         1.2} $        &  \\
0317536999 &    22     &   Ruprecht\,94 & $       5.8557_{     -0.0007}^{+      0.0007} $   & $     1576.007_{      -0.004}^{+       0.004} $  & $        0.178_{      -0.025}^{+       0.037} $   & $         0.90_{       -0.08}^{+        0.06} $  & $         10.4_{        -0.8}^{+         0.7} $  & $         0.44_{       -0.09}^{+        0.09} $   & $         0.15_{       -0.05}^{+        0.05} $   & $         0.17_{       -0.05}^{+        0.05} $    & $         84.5_{        -1.5}^{+         0.8} $    &  $          3.2_{        -0.5}^{+         0.7} $   &   $         36.1_{        -5.7}^{+         7.9} $        &  \\
0372913337 &    23     &      NGC\,2516 & $       1.7860_{     -0.0001}^{+      0.0001} $   & $     1325.341_{      -0.006}^{+       0.006} $  & $        0.045_{      -0.004}^{+       0.011} $   & $         0.96_{       -0.10}^{+        0.03} $  & $          3.7_{        -0.5}^{+         0.5} $  & $         0.22_{       -0.07}^{+        0.09} $   & $         0.15_{       -0.05}^{+        0.05} $   & $         0.19_{       -0.05}^{+        0.05} $    & $         71.1_{       -16.1}^{+         4.9} $    &  $          1.1_{        -0.2}^{+         0.3} $   &   $         11.8_{        -2.4}^{+         3.5} $        &  \\
0389927567 &    24     &   Melotte\,101 & $       6.9408_{     -0.0010}^{+      0.0009} $   & $     1575.805_{      -0.006}^{+       0.006} $  & $        0.261_{      -0.036}^{+       0.058} $   & $         0.85_{       -0.14}^{+        0.10} $  & $         10.0_{        -1.3}^{+         1.1} $  & $         0.28_{       -0.09}^{+        0.10} $   & $         0.15_{       -0.05}^{+        0.05} $   & $         0.17_{       -0.05}^{+        0.05} $    & $         80.1_{        -5.0}^{+         2.7} $    &  $          5.7_{        -1.0}^{+         1.3} $   &   $         64.1_{       -11.0}^{+        14.6} $        &  \\
0410450228 &    25     &      NGC\,2516 & $      15.7784_{     -0.0004}^{+      0.0004} $   & $     1420.250_{      -0.004}^{+       0.004} $  & $        0.095_{      -0.010}^{+       0.022} $   & $         0.93_{       -0.04}^{+        0.04} $  & $         21.0_{        -1.5}^{+         1.2} $  & $         0.50_{       -0.10}^{+        0.09} $   & $         0.22_{       -0.05}^{+        0.05} $   & $         0.25_{       -0.05}^{+        0.05} $    & $         87.1_{        -0.9}^{+         0.4} $    &  $          1.4_{        -0.2}^{+         0.3} $   &   $         15.7_{        -2.0}^{+         3.7} $        &  TOI-681 \\
0413809436 &    26     &      NGC\,5925 & $       2.7030_{     -0.0006}^{+      0.0006} $   & $     1627.336_{      -0.003}^{+       0.003} $  & $        0.233_{      -0.007}^{+       0.011} $   & $         0.52_{       -0.31}^{+        0.18} $  & $          6.4_{        -0.5}^{+         0.4} $  & $         0.49_{       -0.10}^{+        0.10} $   & $         0.18_{       -0.05}^{+        0.05} $   & $         0.23_{       -0.05}^{+        0.05} $    & $         84.3_{        -1.9}^{+         2.8} $    &  $          3.6_{        -0.3}^{+         0.4} $   &   $         40.2_{        -2.9}^{+         4.4} $        &  \\
0419091401 &    27     &   Ruprecht\,48 & $      12.7071_{     -0.0023}^{+      0.0024} $   & $     1499.685_{      -0.007}^{+       0.007} $  & $        0.188_{      -0.011}^{+       0.047} $   & $         0.75_{       -0.28}^{+        0.16} $  & $         11.9_{        -2.2}^{+         2.0} $  & $         0.14_{       -0.06}^{+        0.08} $   & $         0.16_{       -0.09}^{+        0.10} $   & $         0.21_{       -0.10}^{+        0.10} $    & $         81.4_{        -7.6}^{+         3.3} $    &  $          5.2_{        -0.9}^{+         1.5} $   &   $         57.9_{       -10.6}^{+        17.2} $        &  \\
0432564189 &    28     &   Ruprecht\,82 & $       3.7611_{     -0.0003}^{+      0.0003} $   & $     1546.783_{      -0.004}^{+       0.004} $  & $        0.120_{      -0.005}^{+       0.008} $   & $         0.64_{       -0.42}^{+        0.20} $  & $          6.5_{        -0.8}^{+         0.7} $  & $         0.27_{       -0.09}^{+        0.09} $   & $         0.16_{       -0.05}^{+        0.05} $   & $         0.21_{       -0.05}^{+        0.05} $    & $         76.1_{        -9.3}^{+         5.6} $    &  $          2.4_{        -0.3}^{+         0.4} $   &   $         27.3_{        -3.1}^{+         5.0} $        &  \\
0450610413 &    29     &     Harvard\,5 & $      10.6780_{     -0.0027}^{+      0.0027} $   & $     1600.682_{      -0.004}^{+       0.004} $  & $        0.176_{      -0.004}^{+       0.004} $   & $         0.21_{       -0.15}^{+        0.19} $  & $         15.3_{        -1.2}^{+         1.2} $  & $         0.42_{       -0.09}^{+        0.10} $   & $         0.15_{       -0.05}^{+        0.05} $   & $         0.20_{       -0.05}^{+        0.05} $    & $         89.2_{        -0.7}^{+         0.6} $    &  $          3.0_{        -0.2}^{+         0.3} $   &   $         33.9_{        -2.5}^{+         2.9} $        &  \\
0460205581 &    30     &       IC\,2602 & $       8.3252_{     -0.0006}^{+      0.0006} $   & $     1574.271_{      -0.002}^{+       0.002} $  & $        0.059_{      -0.002}^{+       0.005} $   & $         0.57_{       -0.35}^{+        0.26} $  & $         16.9_{        -0.6}^{+         0.6} $  & $         0.94_{       -0.10}^{+        0.10} $   & $         0.31_{       -0.05}^{+        0.05} $   & $         0.30_{       -0.05}^{+        0.05} $    & $         83.4_{        -2.4}^{+         2.7} $    &  $          0.6_{        -0.1}^{+         0.1} $   &   $          6.9_{        -0.4}^{+         0.6} $        &  TOI-837\\
0460950389 &    31     &       IC\,2602 & $       2.8622_{     -0.0004}^{+      0.0004} $   & $     1572.224_{      -0.004}^{+       0.004} $  & $        0.042_{      -0.002}^{+       0.002} $   & $         0.30_{       -0.20}^{+        0.26} $  & $          9.4_{        -0.2}^{+         0.2} $  & $         1.37_{       -0.10}^{+        0.10} $   & $         0.43_{       -0.05}^{+        0.05} $   & $         0.38_{       -0.05}^{+        0.05} $    & $         88.1_{        -1.3}^{+         1.3} $    &  $          0.3_{        -0.1}^{+         0.1} $   &   $          3.8_{        -0.2}^{+         0.2} $        &  \\
0462004618 &    32     &      NGC\,3114 & $       2.3669_{     -0.0001}^{+      0.0001} $   & $     1544.599_{      -0.004}^{+       0.004} $  & $        0.172_{      -0.024}^{+       0.027} $   & $         0.93_{       -0.05}^{+        0.04} $  & $          5.1_{        -0.5}^{+         0.4} $  & $         0.32_{       -0.09}^{+        0.08} $   & $         0.15_{       -0.05}^{+        0.05} $   & $         0.20_{       -0.05}^{+        0.05} $    & $         78.2_{        -3.9}^{+         1.9} $    &  $          3.3_{        -0.5}^{+         0.6} $   &   $         37.3_{        -5.9}^{+         6.6} $        &  \\
0748919024 &    33     &  Ruprecht\,151 & $       2.5395_{     -0.0006}^{+      0.0007} $   & $     1494.064_{      -0.004}^{+       0.004} $  & $        0.118_{      -0.003}^{+       0.004} $   & $         0.37_{       -0.25}^{+        0.26} $  & $          6.0_{        -0.5}^{+         0.4} $  & $         0.46_{       -0.11}^{+        0.10} $   & $         0.25_{       -0.05}^{+        0.05} $   & $         0.27_{       -0.05}^{+        0.05} $    & $         86.9_{        -1.7}^{+         2.1} $    &  $          1.7_{        -0.1}^{+         0.2} $   &   $         19.2_{        -1.4}^{+         1.9} $        &  \\
1036769612 &    34     &   Trumpler\,22 & $       7.4732_{     -0.0013}^{+      0.0014} $   & $     1599.298_{      -0.005}^{+       0.005} $  & $        0.111_{      -0.004}^{+       0.005} $   & $         0.37_{       -0.25}^{+        0.24} $  & $         13.4_{        -1.6}^{+         1.3} $  & $         0.57_{       -0.19}^{+        0.19} $   & $         0.24_{       -0.05}^{+        0.05} $   & $         0.27_{       -0.05}^{+        0.05} $    & $         88.1_{        -1.2}^{+         1.2} $    &  $          1.5_{        -0.2}^{+         0.2} $   &   $         17.0_{        -1.7}^{+         2.6} $        &  \\
  \hline
\end{tabular}

      }
\end{table}
\end{landscape}

\label{lastpage}
\end{document}